\documentclass[%
reprint,
superscriptaddress,
amsmath,amssymb,
aps,longbibliography,
]{revtex4-1}

\usepackage{graphicx}%
\usepackage{verbatim}
\usepackage[version=4]{mhchem}
\usepackage{xcolor}
\usepackage{physics}
\usepackage[normalem]{ulem}
\usepackage{pdfpages}
\usepackage{pgffor}
\usepackage{multirow}

\makeatletter
\AtBeginDocument{\let\LS@rot\@undefined}
\makeatother

\makeatletter
\def\@fnsymbol#1{\ensuremath{\ifcase#1\or \dagger\or *\or \ddagger\or
   \mathsection\or \mathparagraph\or \|\or **\or \dagger\dagger
   \or \ddagger\ddagger \else\@ctrerr\fi}}
    \makeatother

\newcommand{\Othree}{{O(3)}}
\newcommand{\SOthree}{{SO(3)}}

\newcommand{\mbf}[1]{\ensuremath{\mathbf{#1}}}

\NewDocumentCommand\te{s}{\tilde{\e}\IfBooleanTF{#1}{'}{}}
\NewDocumentCommand\tn{s}{\tilde{n}\IfBooleanTF{#1}{'}{}}
\NewDocumentCommand\tl{s}{\tilde{l}\IfBooleanTF{#1}{'}{}}
\NewDocumentCommand\tm{s}{\tilde{m}\IfBooleanTF{#1}{'}{}}
\NewDocumentCommand\tlm{s}{\IfBooleanTF{#1}{\tl*\tm*}{\tl\tm}}
\NewDocumentCommand\tnlm{s}{\IfBooleanTF{#1}{\tnl*\tm*}{\tnl\tm}}
\NewDocumentCommand\tnl{s}{\IfBooleanTF{#1}{\tn*\tl*}{\tn\tl}}

\begin{document}

\title{Electronic excited states from physically-constrained machine learning}

\author{Edoardo Cignoni}
\thanks{These authors contributed equally to this work }
\affiliation{Dipartimento di Chimica e Chimica Industriale, Università di Pisa, Pisa, Italy}

\author{Divya Suman}
\thanks{These authors contributed equally to this work }
\affiliation{Laboratory of Computational Science and Modeling, Institut des Mat\'eriaux, \'Ecole Polytechnique F\'ed\'erale de Lausanne, 1015 Lausanne, Switzerland}

\author{Jigyasa Nigam}
\affiliation{Laboratory of Computational Science and Modeling, Institut des Mat\'eriaux, \'Ecole Polytechnique F\'ed\'erale de Lausanne, 1015 Lausanne, Switzerland}

\author{Lorenzo Cupellini}
\affiliation{Dipartimento di Chimica e Chimica Industriale, Università di Pisa, Pisa, Italy}

\author{Benedetta Mennucci}
\affiliation{Dipartimento di Chimica e Chimica Industriale, Università di Pisa, Pisa, Italy}

\author{Michele Ceriotti}
\email{michele.ceriotti@epfl.ch}
\affiliation{Laboratory of Computational Science and Modeling, Institut des Mat\'eriaux, \'Ecole Polytechnique F\'ed\'erale de Lausanne, 1015 Lausanne, Switzerland}
\affiliation{Division of Chemistry and Chemical Engineering, California Institute of Technology, Pasadena, CA, USA}

\newcommand{\mc}[1]{{\color{blue}#1}}

\newcommand{\ds}[1]{{\color{red}#1}}

\newcommand{\ec}[1]{{\color{purple}#1}}

\newcommand{\jn}[1]{{\color{cyan}#1}}

\date{\today}%

\begin{abstract}
Data-driven techniques are increasingly used to replace electronic-structure calculations of matter. 
In this context, a relevant question is whether machine learning (ML) should be applied directly to predict the desired properties or be combined explicitly with physically-grounded operations.
We present an example of an integrated modeling approach, in which a symmetry-adapted ML model of an effective Hamiltonian is trained to reproduce electronic excitations from a quantum-mechanical calculation.   
The resulting model can make predictions for molecules that are much larger and more complex than those that it is trained on, and allows for dramatic computational savings by indirectly targeting the outputs of well-converged calculations while using a parameterization corresponding to a minimal atom-centered basis. 
These results emphasize the merits of intertwining data-driven techniques with physical approximations, improving the transferability and interpretability of ML models without affecting their accuracy and computational efficiency, and providing a blueprint for developing ML-augmented electronic-structure methods.

\end{abstract}

\maketitle

\section{Introduction}

\begin{figure*}[htb]
    \centering
    \includegraphics[width=0.9\textwidth]{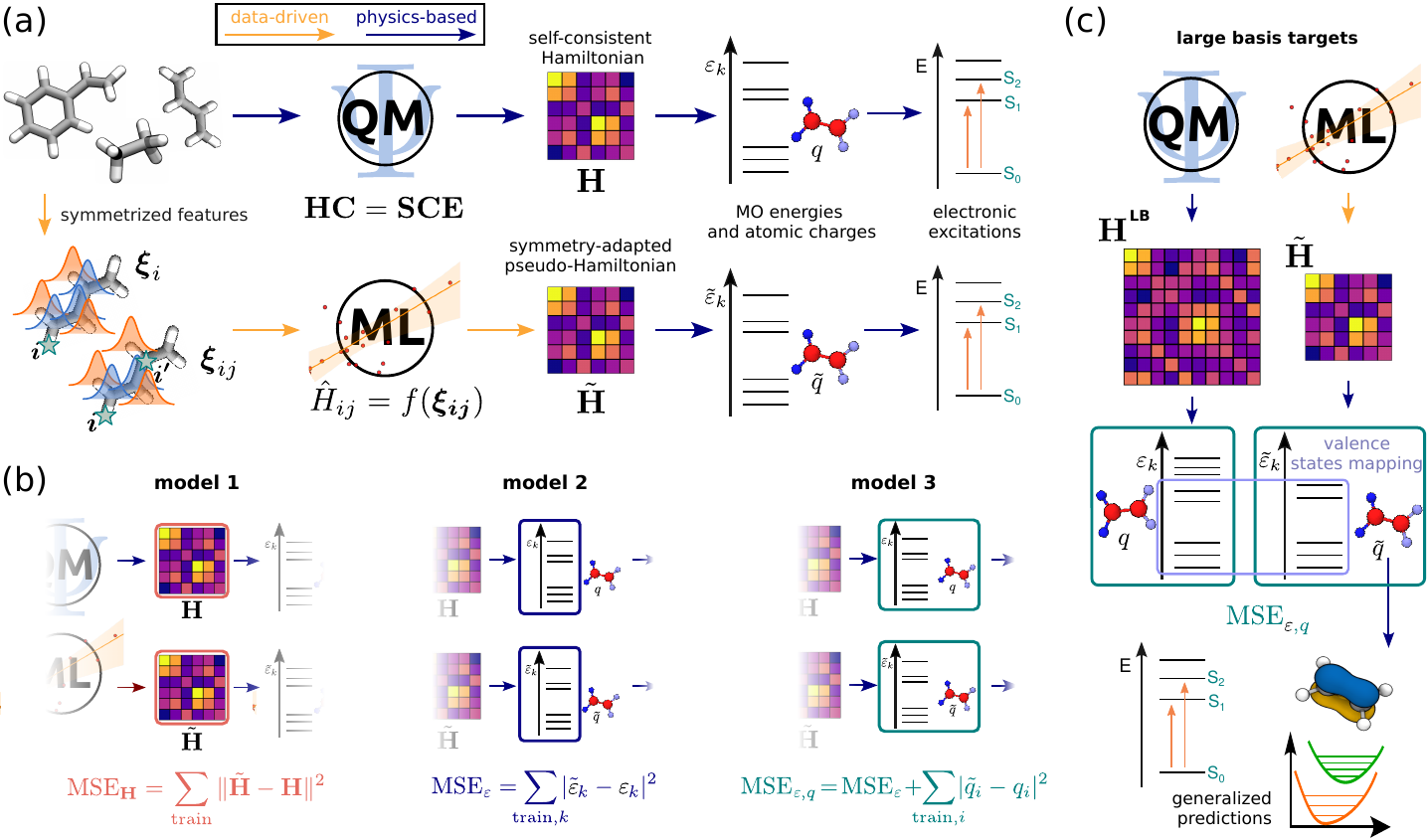}
    \caption{\textbf{Schematic representation of the indirect learning framework.}
    (a) An ML model designed to target the molecular orbital (MO) energies $\varepsilon_n$ and the L\"owdin charges $q$ by using an equivariant model of the single-particle Hamiltonian as an intermediate layer. Cross-hatching shows the Hamiltonian blocks for each pair of atoms $(i,j)$ in the structure that is modeled by learnable functions $f$ with corresponding input features $\boldsymbol{\xi}$.  Arrows are color-coded to indicate data-driven predictions and physics-based approximations.
    (b) We compare different training protocols, targeting the elements of the Hamiltonian (model 1), the minimal-basis eigenvalues (model 2), and both the eigenvalues and L\"{o}wdin charges (model 3). 
    (c) Illustration of a model that is trained on charges and a selection of MO energies computed with a larger basis. Once the effective Hamiltonian is learned, it can be used to compute other electronic-structure quantities, beyond those used during training.
    }
    \label{fig:workflow}
\end{figure*}

Machine learning (ML) techniques have been applied very successfully to circumvent the complexity and computational cost of physics-based modelling techniques.\cite{carl+19rmp,ceri+21cr}
For example, ML interatomic potentials trained on quantum mechanical (QM) calculations have become ubiquitous, making it possible to simulate the structure and stability of complex systems in realistic thermodynamic conditions\cite{chen+19pnas,deri+21nature,zhou+23ne}. 
ML approaches are increasingly applied to a broader array of quantum mechanical properties\cite{ceri22mrsb}, from the ground-state electron density\cite{broc+17nc,gris+19acscs,Shao2023} to electronic excitations\cite{Dral2021,Cignoni2023,Westermayr2020ChemRev,dral+18jpcl,chen_exploiting_2020}, the latter of which are the main focus of the present study. %
In QM calculations, these properties are often the result of a sequence of computational steps, that operate on intermediate quantities describing the electronic structure of a molecule. 
For instance, mean-field methods such as Hartree-Fock and Kohn-Sham density functional theory (DFT)\cite{szabo_MQC,Becke1993} evaluate a self-consistent, effective single-particle Hamiltonian, which can be diagonalized to obtain numerous properties of the system.
Methods based on this single-particle picture also form the basis of more accurate \emph{ab-initio} methods such as complete active space (CAS) or coupled-cluster (CC) theories, as well as of less demanding semi-empirical methods\cite{Baird1969, Dewar1977, Stewart2012}, which parametrize the Hamiltonian using \emph{ab-initio} and empirical data.
In addition to the parametrization, which avoids computing many expensive integrals, semi-empirical schemes usually work in a ``minimal basis''\cite{wats-chan16jctc,schu-vand18jctc} and consider only valence electrons, discarding core electrons and high-energy virtual orbitals\cite{chan+15jcp}, to further speed up the calculations.
Significant work has been devoted to exploring these approximations,
which can be used as inspiration to design ML schemes for electronic properties\cite{Fedik2023}.

When the single-particle wavefunction is expressed in terms of an atom-centered basis, the elements of the Hamiltonian matrix
are indexed by two atoms and determined by interactions with their neighbors, which makes them very well suited as the target of an ML model built on geometric and chemical information.\cite{niga+22jcp} 
Over the past few years, several works have discussed the prediction of single-particle electronic Hamiltonians of molecular systems~\cite{schu+19nc, hegd-bowe17sr, li2022deep, deephe3, yu2023efficient, zhang2022equivariant}.
Physical symmetries constrain their electronic structure, which can be exploited by constructing equivariant ML schemes that ensure that the data-driven model conforms to these constraints.\cite{unke+21nips,niga+22jcp,zhan+22npjcm} 
Once the Hamiltonian matrix is obtained, all sorts of ground-state properties such as the electron density can be obtained with simple, inexpensive manipulations. Furthermore, excited states can also be predicted, at least approximately, by post-processing the ground-state single-particle Hamiltonian.
As a matter of fact, an ML algorithm could be built to directly target the property one wants to predict, e.g. electronic excitation energies\cite{Dral2021,ramakrishnan2015electronic,westermayr2020neural}, HOMO-LUMO gaps\cite{mazouin2022selected}.
Here we pursue an alternative strategy, which integrates data-driven modelling and QM calculations more closely.

We build a symmetry-adapted ML model with an intermediate layer that mimics a minimal-basis, single-particle electronic Hamiltonian, which is then used to compute the molecular orbital (MO) energy levels and atomic charges. 
We then train this model against the MOs obtained from quantum chemical calculations with a richer basis set.
The resulting architecture inherits the accuracy of QM calculations, as well as the transferability to larger, more complex molecules, while being orders of magnitude faster.
This approach also enables predictions of molecular excited states, which we demonstrate using the simplified Tamm-Dancoff Approximation (sTDA)~\cite{Grimme2013,Bannwarth2014} to calculate valence excitation energies.

We analyze the resulting model for a dataset of hydrocarbons, training on a few small molecules and assessing predictions on much larger systems. Our model not only can reproduce the energies and shapes of MOs of previously unseen molecules but can also predict their excitation energy with remarkable accuracy. We finally showcase an example of calculating the vibronic spectra of a molecule not present in the training set.
Our observations have relevance beyond the specific type of excited-state calculations we apply, as they indicate that models combining data-driven steps with physically motivated manipulations and constraints combine the advantages of both approaches and deserve to be more widely adopted as a tool for accurate and affordable atomistic modeling of the electronic properties of molecules and materials.

\begin{table*}[htb!]
    \centering
    \begin{tabular}{ccccccc}
    \hline
    \multirow{2}{*}{\textbf{Molecule}} & \multicolumn{2}{c}{\textbf{$\mathcal{L} = MSE_{\rm \mathbf{H}}$}} & \multicolumn{2}{c}{\textbf{$\mathcal{L} = MSE_{\varepsilon}$}} & \multicolumn{2}{c}{\textbf{$\mathcal{L} = MSE_{\varepsilon,q}$}} \\
    
    & $MAE_{\varepsilon}$ (meV) & $MAE_{q}$ (e) & $MAE_{\varepsilon}$ (meV) & $MAE_{q}$(e) & $MAE_{\varepsilon}$ (meV) & $MAE_{q}$ (e) \\
    \hline
    \textbf{Ethane} & 41.82 & 1.4 x 10\textsuperscript{-3} & 9.96 & 0.11 & 16.15 & 4.1 x 10\textsuperscript{-4}  \\
    \textbf{Ethene} & 68.45 & 1.6 x 10\textsuperscript{-3} & 7.99 & 0.15 & 12.37 & 4.9 x 10\textsuperscript{-4} \\
    \textbf{Butadiene} & 76.92 & 4.8 x 10\textsuperscript{-3} & 32.60 & 0.13 & 44.50 & 1.0 x 10\textsuperscript{-3} \\
    \textbf{Hexane} & 81.10 & 4.7 x 10\textsuperscript{-3} & 53.33 & 0.08 & 63.04 & 1.7 x 10\textsuperscript{-3} \\
    \textbf{Hexatriene} & 93.08 & 6.9 x 10\textsuperscript{-3} & 53.17 & 0.12 & 64.16 & 1.9 x 10\textsuperscript{-3} \\
    \textbf{Styrene} & 78.39 & 7.3 x 10\textsuperscript{-3} & 44.07 & 0.12 & 55.44 & 1.5 x 10\textsuperscript{-3} \\
    \textbf{Isoprene} & 96.61 & 7.3 x 10\textsuperscript{-3} & 52.07 & 0.11 & 69.04 & 2.0 x 10\textsuperscript{-3} \\
    \hline
    \end{tabular}
    \caption{\textbf{Performance comparison of the three different ML models.} The errors on MO energies and the L\"owdin charges obtained from different models 1, 2 and 3 for minimal basis targets. $\mathcal{L} = MSE_{H}$ is the mean squared loss on the Hamiltonian used in model 1, $\mathcal{L} = MSE_{\varepsilon}$ is the mean squared loss on the MO energies used in model 2 and $\mathcal{L} = MSE_{\varepsilon,q}$ is the sum of mean squared losses on the MO energies  and the L\"owdin charges used in model 3.}
    \label{tab:diff_ML}
\end{table*}

\section{Results}

The details of our framework are described in the Materials and Methods section and the Supplementary Materials. However, to better appreciate the result, %
we outline the motivation behind some of our technical choices. 
Our overarching goal is to demonstrate how a hybrid ML scheme can achieve transferability on different axes. On one hand, we aim to show that it can be trained on small, simple molecules, and reach useful accuracy when making predictions on more complicated, larger compounds. 
On the other hand, we want to demonstrate that the model can predict quantities other than those that the model has been trained on: specifically, electronic excitations and their coupling with nuclear vibrations. 
With these goals in mind, we train and validate our model on small hydrocarbon systems (specifically ethane, ethene, butadiene, hexane, hexatriene, styrene and isoprene), which provide a concise but representative palette of saturated, unsaturated, and aromatic motifs.
These structures were generated from high-temperature replica-exchange molecular dynamics (REMD) simulations, which contain multiple conformers and distorted configurations.
We then use the trained model for predictions on larger and more complex molecules, from azulene to beta-carotene. We select classes of compounds with interesting yet well-understood physical effects (e.g. the dependence of band gap on the extent of a conjugated system), and use a simple approximation to compute electronic excitations, allowing for a comparison with explicit quantum mechanical calculations for large molecules and more subtle properties, such as vibronic spectra.
Even though our architecture is fully compatible with any deep-learning scheme, we use an equivariant model based on linear regression to emphasize the impact of the coupling between the ML scheme and the physical approximations over the fine-tuning of the architecture of a non-linear ML model. 

\subsection{Hybrid ML architecture}

Most of the ML frameworks that predict the Hamiltonian directly target the elements of a single-particle electronic matrix $\mathbf{H}$ (the Fock, or Kohn-Sham matrix) obtained from a quantum mechanical calculation\cite{unke+21nips,niga+22jcp,zhan+22npjcm,deephe3}, which describes the interactions between a suitable set of basis functions centered on the different atoms. 
The molecular orbitals (MO) and their energies $\{\varepsilon_n\}$ are obtained by diagonalizing the predicted (or quantum mechanical) $\mathbf{H}$.
A notable exception is given by the SchNet+H approach\cite{west-maur21cs}, in which the target is a ``pseudo-Hamiltonian'' $\tilde{\mathbf{H}}$ that is invariant to the molecular orientation, which is diagonalized to obtain eigenvalues $\{\tilde{\varepsilon}_n\}$ that are then compared with the reference calculation. 
This was shown to provide better accuracy (and a much-simplified architecture) than targeting directly the matrix elements, at the cost of losing the natural symmetries of the physical Hamiltonian. 
Ref.~\citenum{niga+22jcp} also discusses the construction of a symmetry-adapted projected Hamiltonian that reproduces the MO energies from a converged calculation using a smaller set of orbitals, and was then used as the target of the ML model.
This simplifies the calculation (the cost of diagonalizing a matrix scales cubically with the number of orbitals), but introduces a non-physical  training target, as there is no unique definition of the reduced matrix. 

\begin{figure}[tbh]
    \centering
    \includegraphics[width=.4\textwidth]{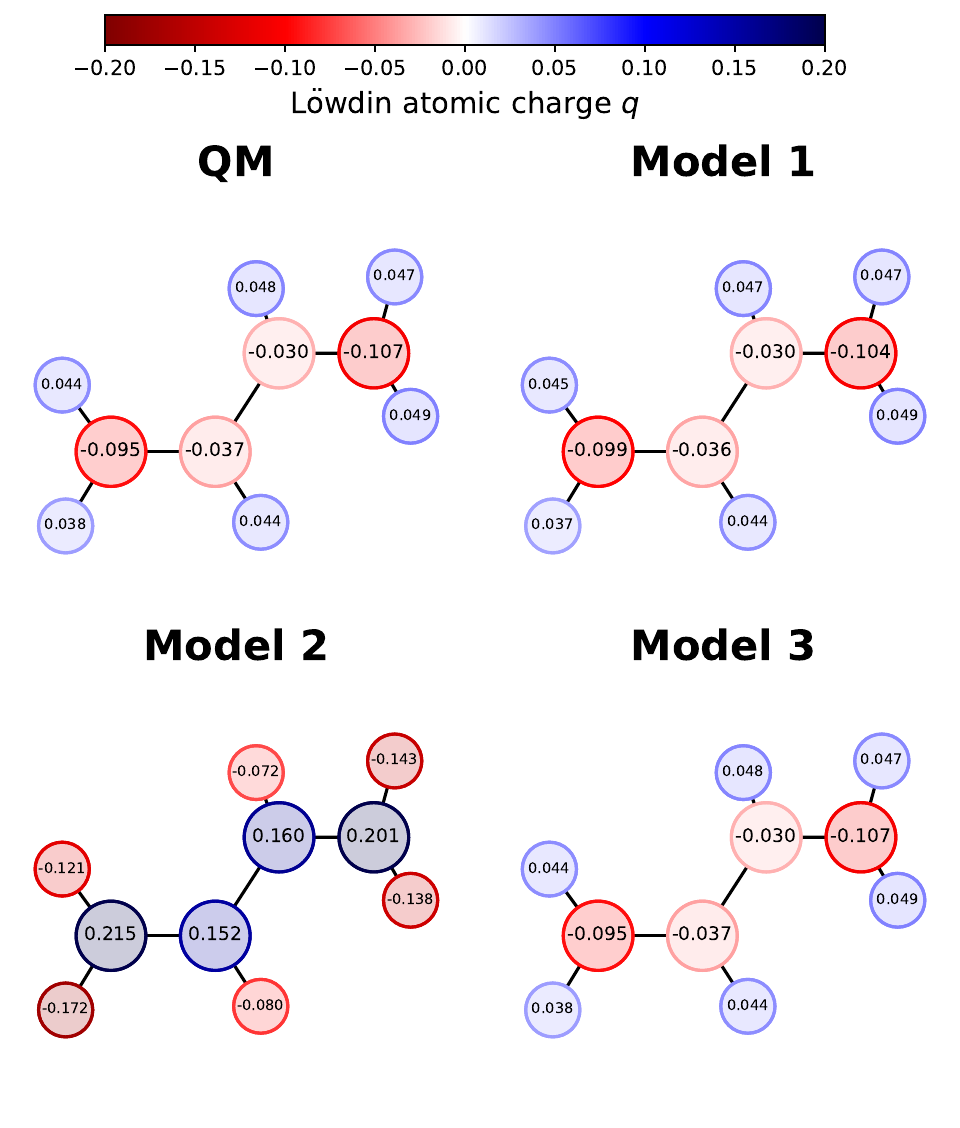}
    \caption{\textbf{Comparison of atomic L\"{o}wdin charges for the three ML models.} The QM target is computed at B3LYP/STO-3G level of theory on a randomly selected geometry of butadiene.
    Model 1 is trained directly on the STO-3G Hamiltonian. Model 2 is trained indirectly using only MO energies as target. Model 3 is trained indirectly using MO energies and L\"{o}wdin charges.}
    \label{fig:buta_charges}
\end{figure}

\begin{figure*}[htb]
    \centering
    \includegraphics[width=\textwidth]{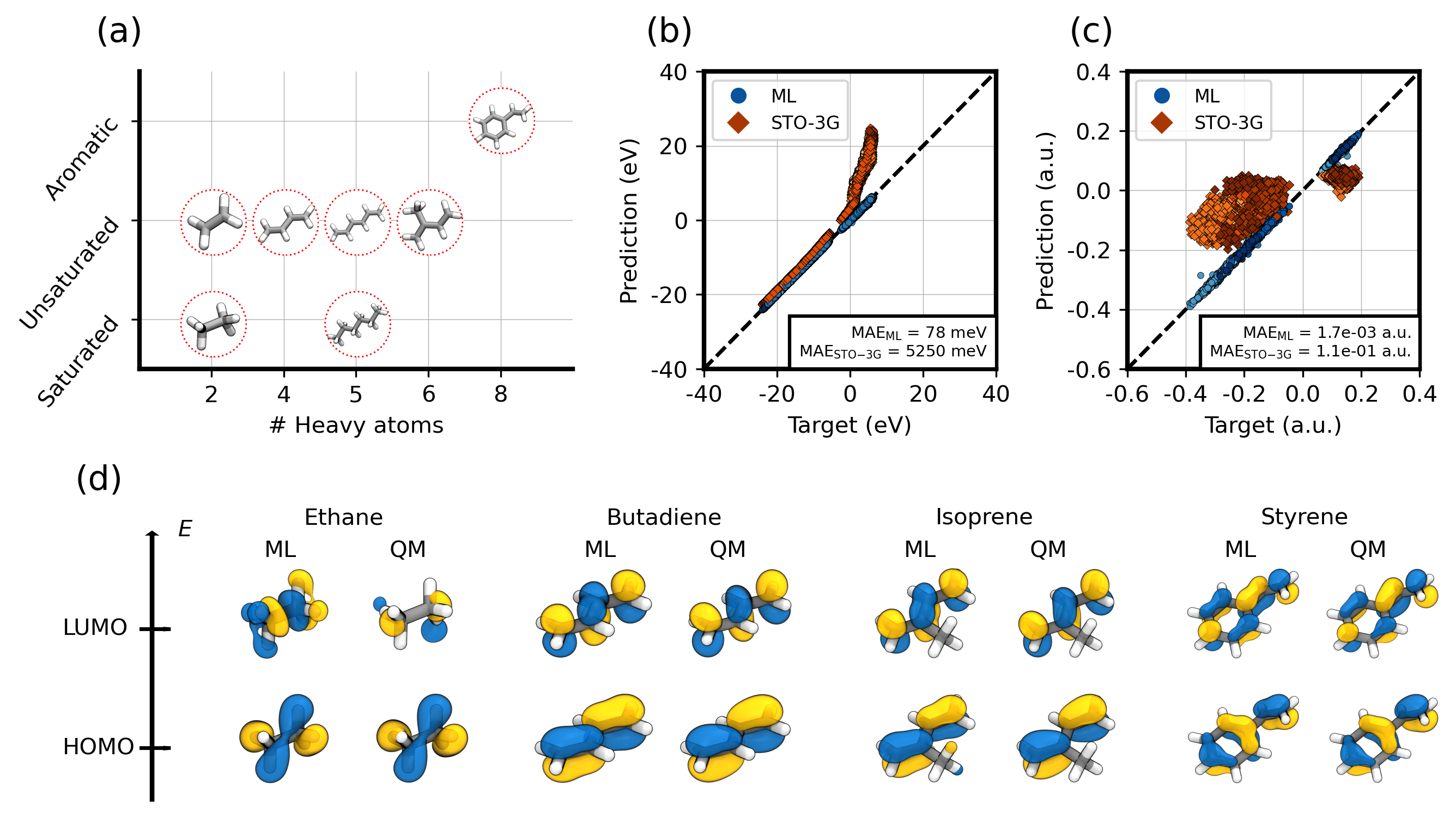}
    \caption{\textbf{Performance of the ML model on large basis targets.}
    (a) The hydrocarbons in the training dataset: ethane, ethene, butadiene, hexane, hexatriene, isoprene, and styrene.
    The two panels on the right show the performance of the ML model on test geometries of the training molecules, for the MO energy (b) and the L\"{o}wdin charges (c).
    The target is always computed with B3LYP/def2-TZVP.
    The prediction is either the ML one (blue points) or the STO-3G calculation (orange points). The mean absolute error (MAE) is reported for both.
    Different shades of blue and orange correspond to the different molecules (panel (a)). A detailed plot showing the prediction for each molecule separately is reported in Figure~\ref{sifig:mo_lowdin}.
    (d) Comparison of HOMO and LUMO molecular orbitals between the ML prediction in a minimal pseudo-basis and the target B3LYP/def2-TZVP.
    }
    \label{fig:mo_lowdin_allin}
\end{figure*}

The hybrid architecture we propose here is shown schematically in Fig.~\ref{fig:workflow}, and combines the most desirable features of these earlier schemes. 
We use a symmetry-adapted ridge regression model based on equivariant 2-center-1-neighbor atom-density correlation features\cite{niga+22jcp} to parameterize the matrix elements of an effective minimal-basis Hamiltonian $\tilde{\mathbf{H}}$. 
This matrix has a structure and the \Othree{} symmetries corresponding to an atom-centered STO-3G basis, containing 1s orbitals for H and 1s, 2s and 2p for C atoms.
$\tilde{\mathbf{H}}$ can be used in different ways, corresponding to different strategies to incorporate data-driven techniques in an electronic structure calculation, and to the different models and loss functions depicted in Fig.~\ref{fig:workflow} (b).

In model 1, we take the effective Hamiltonian matrix as a literal prediction of the results of a self-consistent STO-3G calculation and compute the loss as the $\mathcal{L}^2$ norm of the difference between the predicted $\tilde{\mathbf{H}}$ and the target  ${\mathbf{H}}$.
In model 2, we compute the eigenvalues $\tilde{\varepsilon}_n$ by diagonalizing  $\tilde{\mathbf{H}}$, and define the loss as the error in reproducing the eigenvalues ${\varepsilon}_n$ of the STO-3G calculation. In this case, $\tilde{\mathbf{H}}$ plays the role of a ``pseudo-Hamiltonian'', that (in contrast to SchNet+H) has the correct symmetry properties, but is not bound to be equal to the matrix elements computed in a specified minimal basis. The imposed symmetry properties help the model to recover the correct shape of MOs\cite{niga+22jcp}.
Finally, in model 3 we supplement the MO energies with other quantities computed from the electronic-structure calculations and compute a combined loss that measures the errors in reproducing all the physical constraints. In our case, we use the L\"owdin atomic charges from the QM STO-3G calculation. This constraint is meant to guide the model towards a correct prediction of the QM density on each atom.
We stress that in all cases but model 1, minimizing the model loss is a non-convex optimization problem despite the fact that we use a linear expression for the relation between the matrix elements of $\tilde{\mbf{H}}$ and the structural descriptors.

As shown in Table~\ref{tab:diff_ML},  targeting the matrix elements of the Hamiltonian in a direct learning setup (model 1) leads to consistently larger prediction errors for the single-particle energy levels than using the ML model of the Hamiltonian as an intermediate step in the calculation of the eigenvalues (model 2). 
However, the indirect optimization leads to dramatic errors on other quantities that can be computed from the Hamiltonian matrix. Model 2 gives unphysical predictions for atomic charges, such as negative charges on hydrogen atoms (Fig.~\ref{fig:buta_charges}). %
Combining multiple targets (model 3) achieves a better-balanced model, that improves upon direct Hamiltonian learning for both eigenvalues and atomic charges.

\begin{figure*}[htb]
    \centering
    \includegraphics[width=\textwidth]{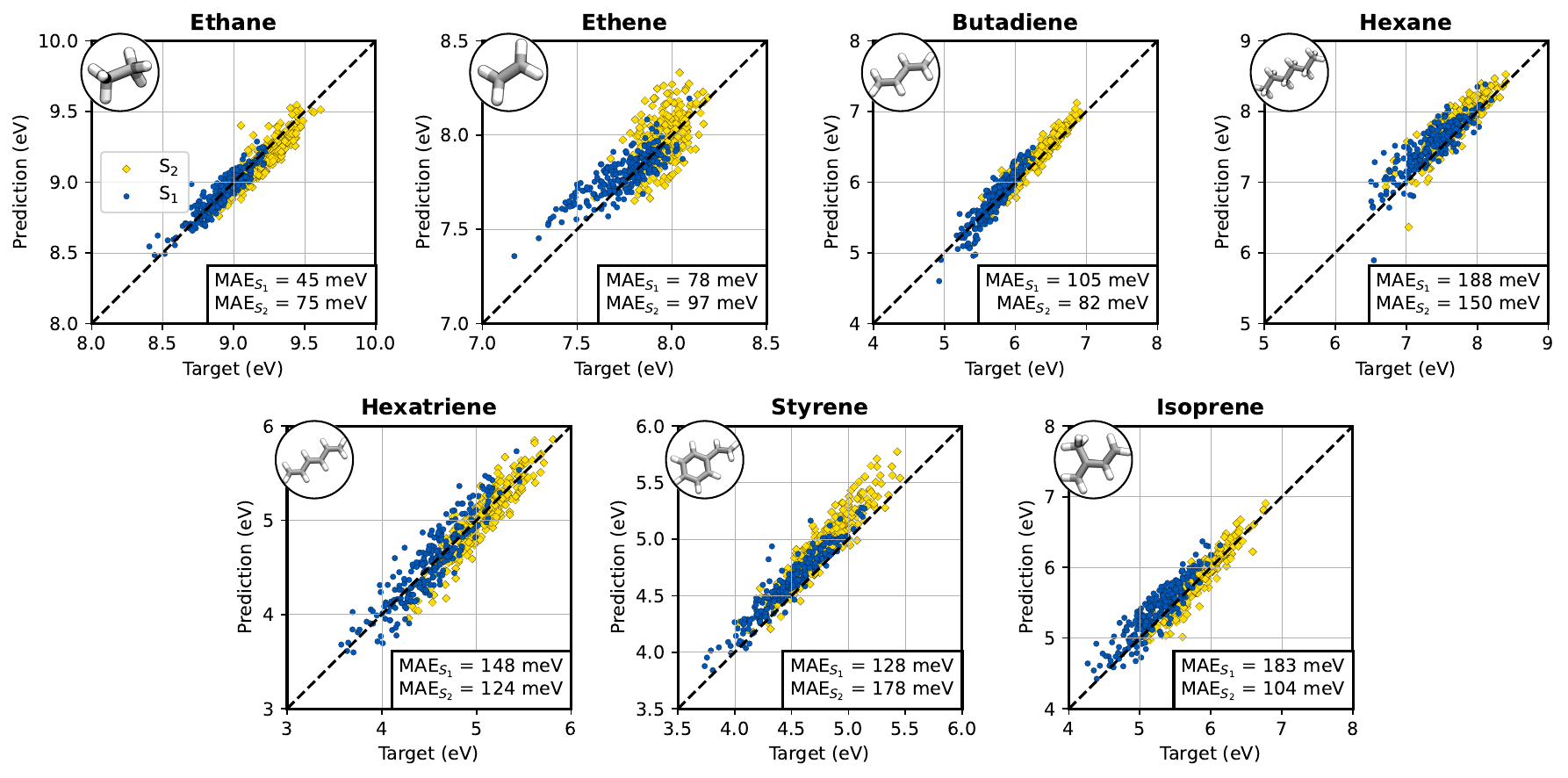}
    \caption{\textbf{Prediction of the electronic excited states.}
    Performance of the ML model on predicting the excitation energy for the first two singlet excited states.
    The target is computed with sTDA B3LYP/def2-TZVP.
    The prediction is computed by coupling the ML prediction with sTDA. The blue circles and the yellow diamonds represent the first and second excited states respectively. The MAE is reported for both.
    }
    \label{fig:excens}
\end{figure*}

\subsection{Large basis targets}

The comparison between different architectures reveals the need to balance data and physics-based considerations when optimizing the overall architecture of the ML model. 
Even though the accuracy of predictions for the indirect model (3) is remarkable, one has to keep in mind that a minimal basis description of the electronic structure is very far from converged: errors on the electronic eigenvalues, particularly for low-lying excited states, are often of the order of several eV. 
As anticipated, an indirect training strategy can also be used to predict target quantities that are computed
with larger basis sets, while keeping a model architecture consistent with a minimal basis. 
Specifically, we use as targets the valence-state eigenvalues $\varepsilon_n^{\text{LB}}$ and L\"owdin charges $q_i^{\text{LB}}$ computed from a large triple-zeta basis (that contains 1s, 2s, 2p and 3s orbitals for H and 1s, 2s, 2p, 3s, 3p, 3d, 4s, 4p, 4d, 4f and 5s for C atoms).
The performance of this model is shown in Figure~\ref{fig:mo_lowdin_allin} for both MO energies and atomic charges (see Figure~\ref{sifig:mo_lowdin} for a detailed plot).

The errors in these ``large-basis target'' (LBT) calculations are considerably larger (up to a factor of 10 for ethane) than in those targeting the properties computed with a minimal basis,
 which is unsurprising given that the electronic-structure problem is considerably underparameterized relative to the reference calculations. 
These errors, however, are at least an order of magnitude \emph{smaller} than those of an explicit minimal-basis electronic-structure calculation  (Figure~\ref{fig:mo_lowdin_allin}~(b) and (c)). The model learns an effective pseudo-Hamiltonian that reproduces to a high accuracy the desired electronic properties of a converged calculation. Such a pseudo-Hamiltonian has the symmetries of a minimal-basis $\mathbf{H}$, but is not explicitly tied to a choice of atom-centered basis functions.
Nonetheless, the MO shapes can still be inspected with a pseudo-basis with the correct angular symmetries and arbitrary radial basis function.
The molecular orbitals are shown in Figure~\ref{fig:mo_lowdin_allin}~(d) using the STO-3G atomic orbital basis.
The LUMO orbital of ethane is the only one exhibiting a mismatch with the ML prediction. The difficulty in predicting this MO lies in its Rydberg character. Rydberg orbitals are known to be present for calculations on small molecules in gas phase, and their diffuse character results particularly challenging for ML when using a minimal-basis pseudo-Hamiltonian.
All the other orbitals show that the symmetry and the nodal structure of the LBT one-electron wavefunctions are learned correctly by the model.
In the remainder of this study we will focus exclusively on this type of LBT hybrid models, that offer an excellent trade-off between accuracy and computational expense.  

\begin{figure}[htb]
    \centering
    \includegraphics[width=0.45\textwidth]{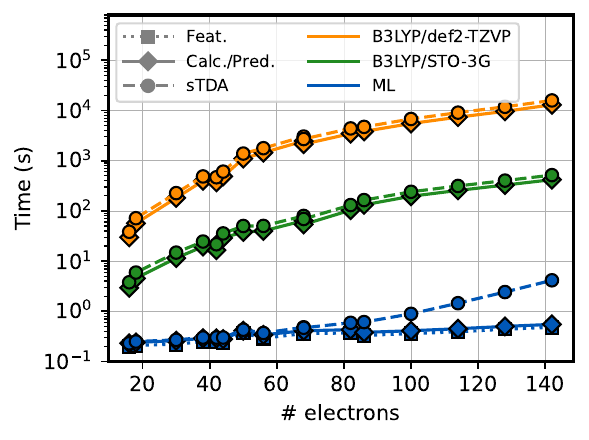}
    \caption{\textbf{Timings of the ML model and QM calculations}.
    Time to compute excited state energies for molecules with an increasing number of electrons.
    The B3LYP/def2-TZVP time (orange) and the B3LYP/STO-3G time (green) are computed as the sum of the DFT calculation and sTDA.
    The ML time (blue) is the sum of the featurization time, the prediction time, and the sTDA.
    Each time is computed on a single core of an Intel Xeon Gold 5120 Processor.
    B3LYP/def2-TZVP and B3LYP/STO-3G calculations are performed with PySCF.
    More details are in Table~\ref{sitab:timings} in the SI.
    }
    \label{fig:timings}
\end{figure}

\subsection{Electronic excitations from an ML Hamiltonian}

One of the advantages of explicit electronic-structure calculations is that, after having determined the self-consistent single-particle Hamiltonian, they allow the prediction of many molecular properties through simple and inexpensive post-processing steps. 
Our benchmarks thus far only validate the accuracy for properties that are explicitly trained on. To check whether the predicted pseudo-Hamiltonian can also be manipulated to access other properties, we consider the problem of predicting excitation energies based on time-dependent DFT (TD-DFT), which is commonly used to compute excited states for large molecules.
We use, in particular, an approximation of TD-DFT developed by Grimme~\cite{Grimme2013} called simplified Tamm-Dancoff Approximation (sTDA). 
Its appeal in our present case is that integrals in sTDA are approximated with L\"{o}wdin charges that are readily available from our ML prediction (see Materials and Methods).
As shown in Fig.~\ref{fig:excens}, the excitation energies computed from the effective ML Hamiltonian are predicted with good accuracy, with a balanced error for both the first and second excited states.
The error increases as the molecule gets larger and more flexible such as for hexane, but it always remains well below 200~meV.
As a point of comparison, excitation energies computed at the STO-3G level differ by at least 1 eV from those computed with a large basis set (see Table~\ref{sitab:errors_stda_HC}).
What makes these results particularly remarkable is that the model does not explicitly use the sTDA excitations as targets, which indicates good generalization capabilities of the ML model -- and at the same time scope for further improvement of the accuracy by including these additional targets to the model optimization step. 
Overall, our indirect learning strategy delivers predictions of electronic properties with an accuracy comparable to that of converged settings, at a cost that is much smaller than that of a minimal-basis DFT calculation, and many orders of magnitudes faster when compared to the large basis (see Fig.~\ref{fig:timings}).
The speed gain is mainly due to the ML algorithm itself, which directly predicts the blocks of the self-consistent Hamiltonian. The minimal-basis formulation reduces the cost of the diagonalization step, although - for the larger molecules - evaluating the sTDA excitations becomes the rate-limiting step for predictions that start from the ML-based pseudo-Hamiltonian. %
The local nature of the model also means that the ML Hamiltonian is highly sparse, which would be beneficial in combination with linear-scaling solvers\cite{goed99rmp}.
We stress that the current implementation is not optimized for speed, being instead focused on making it easy to test new ideas: therefore a more efficient calculation of symmetry-adapted features, as well as the use of more sophisticated model architectures, leave much room to further improve accuracy and computational requirements of the ML model.

\subsection{Extrapolative predictions}

\begin{figure*}[ht!]
    \centering
    \includegraphics[width=\textwidth]{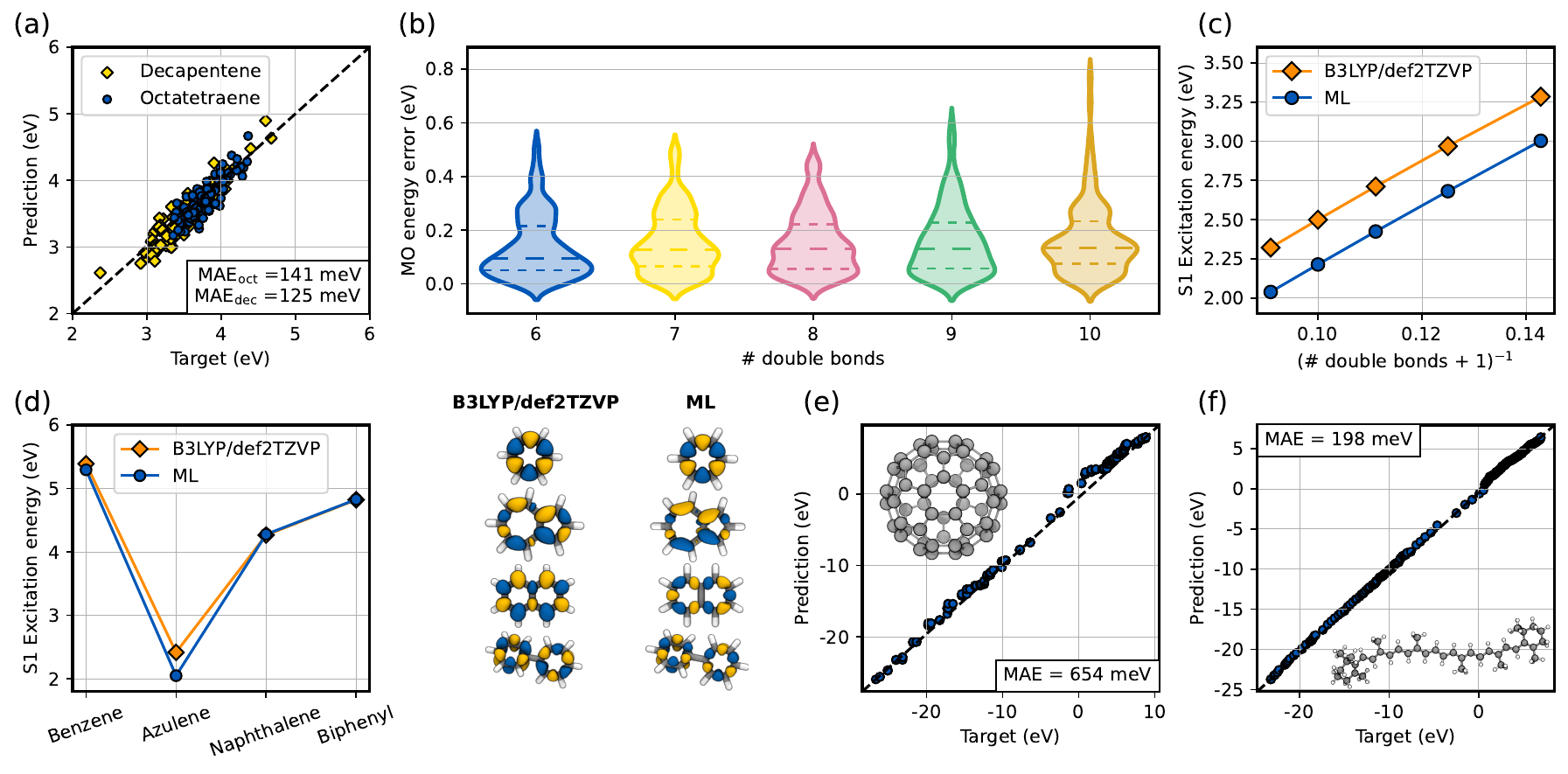}
    \caption{
    \textbf{Generalization performance of the ML model.}
    The target is B3LYP/def2-TZVP, coupled with sTDA to obtain the excitation energies.
    (a) ML prediction of the excitation energy of the first singlet excited state of octatetraene and decapentene. The MAE is reported for both molecules. The subscript ``oct" refers to octatetraene, ``dec" to decapentene.
    (b) Distribution of absolute errors for the MO energies of polyalkenes with a progressively longer conjugated chain. The dashed lines are the first, second, and third quartile of the distribution.
    (c) ML prediction of the excitation energy of the first singlet excited state for polyalkenes, from six up to ten double bonds.
    (d) ML prediction of the excitation energy of the first singlet excited state for some aromatic molecules.
    The transition density is visualized on the right. For the ML model, we have used the STO-3G basis to obtain the transition density cubes.
    (e) ML prediction of the MO energies of C60.
    (d) ML prediction of the MO energies of $\beta$-carotene.
    }
    \label{fig:extrapol}
\end{figure*}

The calculation of the the excitation energies based on the ML-derived minimal-basis Hamiltonian demonstrates the ability of our framework to generalize to new \emph{properties}, beyond those it has been trained on. 
As we shall see, it also demonstrates excellent transferability to new \emph{structures}, much larger and more complex than those included in the training set. 
As a first benchmark, we test our ML model by predicting the excitation energies for polyalkenes of different lengths.
For octatetraene and decapentene, we extract 100 conformations from a REMD trajectory spanning a broad temperature range, following a protocol analogous to that used to generate the train set.
Figure~\ref{fig:extrapol}~(a) shows the prediction of the first excited state for several conformations of octatetraene and decapentene.
Errors are comparable to those observed on the validation set, indicating excellent transferability to bigger molecules.
This is confirmed by the small errors obtained for the prediction of the MO energies (Figure~\ref{fig:extrapol}~(b) and Figure~\ref{sifig:poly_ene_parity}) of longer polyalkenes up to 10 double bonds.

In this benchmark, errors are dominated by the presence of distorted configurations. 
The excitation energy for the optimized geometries (Figure~\ref{fig:extrapol}~(c)) provides clearer insights into specific physical effects, without the noise introduced by thermal distortion.
The model perfectly captures the dependence of the S$_1$ energy on the increase in conjugation length, but there is a redshift of approximately 300~meV  relative to the target values.
This redshift illustrates a limitation of the ML framework, which uses short-ranged features to parameterize the pseudo-Hamiltonian, and all matrix elements involving atoms beyond the cutoff are predicted to be zero (Figure~\ref{sifig:poly_cutoff_fock}).
This effect results in a systematic underestimation of the HOMO-LUMO gap (Figures~\ref{sifig:poly_cutoff_fock} and \ref{sifig:homo_lumo_poly}), which translates into an underestimation of the $S_1$ excitation energy.
Increasing the cutoff of the ML features would be an obvious strategy to address this problem, which however also leads to a model that is less accurate and transferable -- as it would require training on larger molecules to thoroughly sample longer-range interactions (see Figure~\ref{sifig:large_cutoff_extrapol}).
An alternative, very effective solution is to use an explicit minimal-basis calculation as a baseline, using the ML model to learn the large-basis targets by correcting the STO-3G Hamiltonian. %
As shown in the SI, this baselined model yields much lower errors, and completely eliminates the redshift.
This is relevant in light of several recent efforts that use low-cost electronic-structure properties as molecular descriptors\cite{qiao+20jcp,fabr+22dd}, and representative of the kind of trade-offs that are necessary when designing hybrid modeling schemes. It entails, however, a substantial increase in computational cost, and we will restrict our investigation to a model that does not rely on a baseline. 

We also consider the case of four representative aromatic molecules, none of which is included in the training set (Figure~\ref{fig:extrapol}~(d)).
Even though styrene is the only aromatic molecule used during training, the prediction of $S_1$ for the test aromatic molecules  matches very well the reference value, except for azulene which contains a pentagonal motif which is completely missing in the training data.
Figure~\ref{fig:extrapol}~(d) also shows the transition density associated with the first excited state for both the ML prediction and the target.
The ML-based transition density matches well the qualitative nature of the reference, indicating that the model predicts the correct excitation despite small quantitative differences in the shape due to the use of a minimal basis set.
Naphthalene is the only exception: for this molecule, the L$_a$ and L$_b$ states are particularly challenging for computational methods~\cite{Prlj2016}: for B3LYP/def2-TZVP, in particular, they are very close in energy. As a result, the small errors in the ML model lead to an exchange in their ordering (see Figure~\ref{sifig:napht_LaLb}).

\begin{figure*}[htb!]
    \centering
    \includegraphics[width=\textwidth]{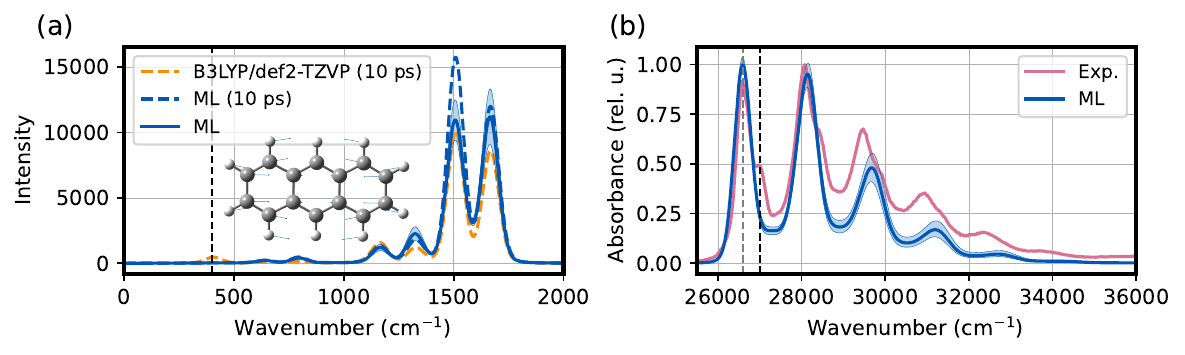}
    \caption{
    \textbf{Vibronic spectrum of anthracene.}
    (a) Spectral density predicted with the ML model (solid blue line) and corresponding 95\% confidence interval. The spectral density is an average over several windows of 100~ps of MD trajectory.
    The spectral density for the target sTDA B3LYP/def2-TZVP (orange dashed line) is reported for a single window of 10~ps, alongside the corresponding ML prediction (blue dashed line).
    (b) Absorption spectrum predicted with the ML model (blue) and compared with the experimental spectrum~\cite{Taniguchi2018} (orange).
    The ML spectrum is the average over 100~ps of anthracene MD. The blue interval denotes the 95\% confidence interval of the mean.
    The vertical dashed black line in (a) denotes a peak around 400~cm$^{-1}$ that is not captured by ML. The black dashed line in (b), shifted by 400~cm$^{-1}$ from the grey line, denotes the corresponding missing vibronic shoulder. The normal mode at around 400~cm$^{-1}$ is shown in the anthracene sketch in panel (a).    
    }
    \label{fig:vibronic}
\end{figure*}

These results, together with the ability of the model to capture the correct trend in the excitation energy of polyenes, demonstrate that using a hybrid model based on the evaluation of an effective Hamiltonian as an intermediate step allows to capture global effects such as conjugation and aromaticity, despite being modeled from local features. 
This is in stark contrast to models that target the excitation energy directly, which would not be capable of capturing changes for molecules larger than those included in the training set (see e.g. Ref~\citenum{wilk+19pnas} for an example of this effect in the case of the molecular polarizability).
An analysis of the dependence of the excitation energies of biphenyl and butadiene as a function of molecular distortions confirms that an sTDA calculation based on the ML Hamiltonian correctly captures the qualitative behavior of excited states, and is more accurate relative to the converged DFT calculations than commonly used semiempirical methods, even when used in an extrapolative regime (Figure~\ref{sifig:scandihe}).

As a final example, we test our model to predict the Hamiltonian of very large molecules, namely C$_{60}$ fullerene and $\beta$-carotene (Figure~\ref{fig:extrapol}~(e) and (f)).
Despite its size and complexity, MO energies of $\beta$-carotene are predicted with an MAE of less than 200 meV, which is largely due to the underestimation of the HOMO-LUMO gap, similar to what we observed in much simpler, linear polyalkenes (Figure~\ref{fig:extrapol}~(c)).
The peculiar structural features of C$_{60}$, with the presence of pentagonal rings, complete lack of hydrogen atoms, and high curvature, make it a complete outlier relative to the training set. 
Nevertheless, the LBT model achieves an MAE of 654~meV on the MO energies, which is much smaller than the error of a minimal-basis calculation (MAE in excess of 4.7~eV). %
For $\beta$-carotene the atomic L\"{o}wdin charges are predicted with an accuracy comparable to that observed for small molecules, while for C$_{60}$ they are exactly zero due to symmetry (Figure~\ref{sifig:c60_beta_charges}), which is captured by the ML model thanks to its equivariant structure.

\subsection{Vibronic spectra }
As a final application, we demonstrate how our hybrid model can be further combined with advanced simulation techniques to obtain accurate yet inexpensive predictions of subtle quantum-mechanical effects.
We estimate the vibronic spectrum of anthracene via second-order cumulant expansion theory (see Materials and Methods).
Within this framework, the vibronic structure of the excitation is encoded by the spectral density.
Among the different approaches to compute this quantity~\cite{dierksen2004density, dierksen2004vibronic}, we rely on %
the calculation of the autocorrelation function of the excitation energy along an MD trajectory.
This dynamical method is typically very  accurate\cite{Valleau2012}, and transparently incorporates vibrational information. 
This is an application where a fast ML model for excited states can make prohibitively-demanding quantum-chemical calculations routine: a single sTDA calculation for anthracene at the B3LYP/def2-TZVP level requires approximately 8 CPU minutes, versus half a second for the hybrid model -- a speed-up of three orders of magnitude.

The spectral density and the vibronic spectrum of anthracene are shown in Figure~\ref{fig:vibronic}~(a) and (b), respectively.
The low cost of ML calculations allows us to average the spectral density over several 10~ps windows.
The corresponding averaged spectral density for sTDA B3LYP/def2-TZVP is much more demanding, and we show its value computed for a single window.
The target spectral density is mostly within the confidence interval reported for the ML averaged spectral density (blue interval) (Figure~\ref{fig:vibronic}~(a)).
A comparison with the ML spectral density computed in the same window shows that all the peaks are slightly overestimated by the ML, with the exception of a peak at around 400~cm$^{-1}$ which is absent in the ML prediction.
This peak is responsible for the vibronic shoulder visible in the experimental absorption spectrum, which is missing in that predicted from ML simulations (Figure~\ref{fig:vibronic}~(b) and Figure~\ref{sifig:vibronic_targ_10ps}). 
Inspection of this normal mode shows that it is a global ``breathing"-like motion of the entire molecule (see inset in Figure~\ref{fig:vibronic}~(a)) arising from small alterations of the interatomic distances in anthracene, and thus hard to characterize with short-ranged features.
Besides this minor discrepancy, the absorption spectrum is in good agreement with the experimental one.
Indeed, the ML spectrum shows a competitive performance with ZINDO (Figure~\ref{sifig:vibronic_zindo}), a similarly fast semiempirical method specifically built to target singlet excited states of organic molecules, and sometimes used to compute spectral densities in complex biomolecules \cite{Chandrasekaran2015,Aghtar2017}.

\section{Discussion}

At the most fundamental level, the difference between physically-motivated and data-driven modeling approaches is that between primarily deductive and naively inductive paradigms of scientific knowledge. 
The former strives for universality, and to reduce the complexity of empirical observations to a minimal set of fundamental laws, while the latter infers patterns from data, and is usually more limited in generalization power, but can be more precise, or computationally efficient, in capturing structure-property relations.
The approach we discuss here to describe electronic excitations, combining machine-learning of an effective minimal-basis, single-particle Hamiltonian with training on a large basis set and the further application to evaluate physics-based approximations of molecular excitations, demonstrates the advantages of a middle-ground, hybrid solution.

Despite the relatively small training set composed of MD snapshots of seven small hydrocarbon molecules and the simplistic choice of ML architecture, our model shows excellent transferability, both in terms of the evaluation of derived electronic properties and in terms of making predictions for larger, more complex compounds. 
We probe the former aspect by computing excitation energies in the sTDA approximation, or the vibronic spectrum based on molecular dynamics trajectories, and the latter by making predictions on molecules that are larger and more complex than those included in the training. 
In both cases, we demonstrate excellent transferability, with similar errors observed in the interpolative and extrapolative regime -- except for cases, such as C$_{60}$, which entail dramatically different chemical motifs. 
In every case we consider, we obtain an accuracy that is much better than that afforded by an explicit minimal-basis quantum calculation, at a considerably reduced cost.
We capture qualitative physical effects, such as the dependence of molecular excitations on the extent of conjugation or on internal molecular rotations. We can also easily interpret the quantitative errors, which we trace back to the aggressive local truncation of descriptors.

We make a few observations that could guide the development of similar, hybrid models. 
(1) Reproducing the mathematical structure of the quantum mechanical approximations is more effective than explicitly targeting the value of approximate electronic-structure quantities.
(2) Using indirect properties as targets, such as single-particle eigenvalues, makes it possible to ``promote'' the model accuracy to a higher level of theory, e.g. using a larger basis set, at no additional cost. 
(3) Indirect learning should be sufficiently constrained to avoid overfitting and unphysical predictions. 
(4) The use of well-principled descriptors, that incorporate the symmetries of the problem, is beneficial in reproducing the qualitative behavior of the excitations.
(5) Locality plays a fundamental role in facilitating transferability, but there is a tradeoff with the asymptotic accuracy that the model can achieve. 
One of the main challenges is that, despite the common features\cite{musi+21cr,niga+22jcp}, the design space of ML models for chemistry is very large\cite{bata+22arxiv}. 
Incorporating physical approximations into an ML model can make the architecture easier to interpret, but also introduces more degrees of freedom that need to be explored more extensively. 
We suggest that restricting the investigation to simple, easy-to-interpret models, translating some of the insights that have become well-established in the construction of ML interatomic potentials to electronic-structure targets, and emphasizing generalization power over in-sample benchmark accuracy -- which is the main advantage of physics-based, deductive modeling -- are the next logical steps in advancing the integration between machine learning and quantum chemistry.

\section{Materials and Methods}

\paragraph{Model Architecture}
Our ML model aims to reproduce the single-particle states (MOs) from a mean-field quantum chemistry calculation such as Hartree-Fock or Kohn-Sham DFT.
These MOs are expanded on a basis set of atomic orbitals (AOs), and their coefficients are obtained from the solution of the self-consistent-field (SCF) generalized eigenvalue equation:
\begin{equation}
    \mathbf{HC} = \mathbf{SC}\mathbf{\mathcal{E}}
\end{equation}
where $\mathbf{H}$ is the Fock (or Kohn-Sham) matrix, $\mathbf{S}$ is the overlap matrix of the AO basis, and $\mathbf{\mathcal{E}}$ is a diagonal matrix containing the MO energies. As $\mathbf{H}$ depends on the orbitals themselves, the iterative solution of this equation is numerically expensive, especially for large basis sets.

We learn an effective Hamiltonian $\tilde{\mathbf{H}}$ that substitutes $\mathbf{H}$ in the eigenvalue equation and directly yields the MOs, bypassing the SCF solution. To simplify and constrain the problem, we require that $\tilde{\mathbf{H}}$ is defined on a minimal and orthonormal AO basis, akin to standard semiempirical methods.
In addition, the minimal basis over which we learn $\tilde{\mathbf{H}}$ is only implicitly defined, a feature that improves the model flexibility.
We train our model so that the solutions:
\begin{equation}
    \label{eq:ortho_roothan_ml}
    \mathbf{\tilde{H}\tilde{C}} = \mathbf{\tilde{C}}\mathbf{\mathcal{E}}
\end{equation}
generate a selected subset of MOs and MO energies as the original SCF equations.
The obtained MO coefficients and energies can be used to predict additional quantities, such as electronic excitation energies, within a physics-based model.

For model 1, which directly targets the entries of an orthogonalized Hamiltonian, we use the L\"{o}wdin-symmetrized Fock $\mathbf{H}_{\rm LSF} = \mathbf{S}^{-1/2}\mathbf{HS}^{-1/2}$ computed with B3LYP/STO-3G as our target.
When we target an SCF Fock $\mathbf{H}$ in a larger basis, which yields more accurate results, we cannot link its entries to our prediction $\tilde{\mathbf{H}}$ in an implicit minimal-basis.
Instead, we ensure that our model generates the desired subset of MOs, with energies that are as close as possible to the quantum chemical calculations. To do so, we first train our model indirectly on B3LYP/STO-3G targets, minimizing a loss on MO energies (model 2) and on MO energies and L\"{o}wdin charges (model 3).
The L\"{o}wdin charge $q_{A}$ on atom A is computed as:
\begin{equation}
    \label{eq:lowdin_charge}
    q_{A} = Z_A - 2 \sum_{\mu \in A} \sum_{i=1}^{N_{\rm occ}} \tilde{C}_{\mu i} \tilde{C}_{\mu i}
\end{equation}
where $Z_A$ is the atomic number of $A$, $\mu$ indexes an atomic orbital, $N_{\rm occ}$ is the number of occupied MOs, and the MO coefficients $\tilde{\mbf{C}}$ are obtained from Eq.~\ref{eq:ortho_roothan_ml}.
The final LBT ML model is trained on MO energies and L\"{o}wdin charges, as for model 3, but on targets computed with B3LYP/def2-TZVP.

\paragraph{Dataset Generation}

We use a dataset of 1000 different geometries of ethane, ethene, butadiene, and octatetraene that was originally presented in ref.~\citenum{gris+19acscs}. %
We extend it with configurations of hexane, hexatriene, isoprene, styrene, and decapentene following a similar protocol as in the reference. 
In summary, we carry out a replica exchange molecular dynamics (REMD) simulation with a timestep of 0.5~fs, for a total of 150~ps of sampling per replica and attempting exchanges every 2 fs. 
Molecular dynamics trajectories for each replica were integrated in the constant-temperature ensemble using a generalized Langevin equation (GLE) thermostat. 
Forces were computed at DFTB3-UFF/3OB level of theory. The simulation was run with i-PI~\cite{kapi+19cpc} in combination with DFTB+~\cite{Hourahine2020}. 
For each molecule, 1000 structures were selected from all trajectories, using farthest point sampling (FPS)~\cite{ceri+13jctc} on SOAP~\cite{bart+13prb} descriptors averaged over the structures. The SOAP descriptor was computed with \texttt{rascaline} \cite{rascaline}, using a cutoff of 4.5 $\textup{\AA}$, $n_{\rm max} =$ 6, $l_{\rm max} =$ 4, and a Gaussian width of 0.2 $\textup{\AA}$. FPS was performed with \texttt{scikit-matter} \cite{goscinski2023scikit}. 
For these datasets we then performed DFT calculations using both the STO-3G and def2-TZVP basis sets and Gaussian-like B3LYP functional (\texttt{b3lypg}) level of theory using PySCF~\cite{Sun2020}, to obtain the Fock and overlap matrices and other required electronic structure properties. All calculations were performed in the spherical atomic basis and molecular symmetry was not taken into account even when present, as for some optimized geometries.

\paragraph{Symmetry-adapted Hamiltonian regression}

The single-particle Hamiltonian is expressed in terms of AO basis functions $\phi_a(\mathbf{x} - \mathbf{r}_i)$, where $a=\tnlm$ denotes the orbital symmetry and $\mathbf{r}_i$ the atom on which the orbital is centered.
We use the shorthand $\bra{ia}\hat{H}\ket{jb} = {H}_{ia, jb}(A)$ to denote the Hamiltonian matrix element between an orbital $\phi_a$ centered on atom $i$ and $\phi_b$ centered on atom $j$ of a structure $A$. 
Due to the presence of two atomic indices, these elements must be equivariant to permutations of the orbital labels associated with each atomic center. 
The rotations of each block of the matrix (involving all the corresponding $m,m'$ indices)  %
can be decomposed into rotations of irreducible representations (irreps) of $O(3)$ %
(the transformation from the uncoupled $\ket{\tlm} \ket{\tlm*}$ basis %
to the \emph{coupled} angular basis (irrep) $\ket{\lambda \mu}$  is effected through Clebsch-Gordan coefficients).
We model each irreducible block separately, so that our targets have the form $H_{ij}^{p\tau \lambda\mu}$ where $(\lambda\mu)$ is the \SOthree{} symmetry index, $\tau$ captures additional symmetries (e.g., inversion parity) associated with this block and $p$ enumerates the other variables, i.e. the angular ($\tl, \tl*$) and radial ($\tn, \tn*$) basis, as well as the chemical nature of the atoms.
Given this symmetry-based decomposition of $\mathbf{H}$, we can employ ML models of arbitrary complexity as long as they are equivariant to the same permutation and \SOthree{} symmetries.
Here, we use linear models with descriptors $\boldsymbol{\xi}^{\tau\lambda\mu}(A_{ij})$ with the same symmetries of the target:
\begin{equation}
H_{ij}^{p\tau\lambda\mu} = 
 \mbf{w}^{p\tau\lambda} \cdot \boldsymbol{\xi}^{\tau\lambda\mu}(A_{ij}) +\delta_{\lambda 0} b^{p\tau \lambda}
\end{equation}
where $\mbf{w}^{p\tau\lambda}$ are invariant weights and $\delta_{\lambda 0} b^{p\tau \lambda}$ is the intercept for the scalar ($\lambda=0$) blocks, which in our model we take to be zero.
We stress here that this framework, while described for linear models, %
is equally compatible with any deep-learning scheme.

\paragraph{Symmetry-adapted features}
Our descriptors $\boldsymbol{\xi}^{\tau\lambda\mu}(A_{ij})$ are the two-centered features described in Ref.~\citenum{niga+22jcp}, obtained as generalizations of the atom-centered density correlation descriptors \cite{bart+13prb, drau19prb, will+19jcp} to simultaneously represent multiple atomic centers and their connectivity. 
Briefly, these features rely on \emph{pair} density coefficients $c_{nlm}(A_{ij})$ as the core ingredients that essentially specify the position of atom $j$ relative to atom $i$.
\begin{equation}
    c_{nlm}(A_{ij})
    =
    R_{nl}(r_{ij}) Y_l^m(\hat{r}_{ij})
\end{equation}
In this form, the spatial description has been discretized on a more tractable basis $R_{nl}$ (GTO-style radial functions) %
and spherical harmonics $Y_l^m(\hat{r})$ for the radial and angular degrees of freedom respectively, as is commonly done in quantum chemistry codes.

Summing over one of the indices ($j$), for all pairs within a set cutoff distance 
leads to a neighbor density $c_{nlm}(A_i)=\sum_{j}c_{nlm}(A_{ij})$, which describes the local correlations of a single center $i$ and its neighbors.
Combinations (through tensor products) of the neighbor density express higher-order correlations with multiple neighbors. 
On the other hand, the combination of the neighbor density with $c_{nlm}(A_{ij})$ yields a richer description of the specific pair between two atoms $(ij)$. In particular, the simplest such combination $c_{n'l'm'}(A_{i}) c_{nlm}(A_{ij})$ describes the correlations of three atoms - two centers $i$, $j$ and the neighbors of $i$. 
The cutoff distance enforces locality of the descriptor and usually is an optimizable hyperparameter, however, a cutoff smaller than the
interatomic separation (between $i$ and $j$) means that there will be no features corresponding to the pair, and hence a zero prediction for all the associated Hamiltonian blocks (see e.g. Figure~\ref{sifig:delta_illustr}).
The choice of spherical harmonics as the angular basis and implementation of the combinations (tensor products) through a Clebsch-Gordan coupling (similar to the one described for the Hamiltonian blocks) ensures rotational equivariance of the features and symmetry to the permutation of the atom labels can be similarly enforced by averaging over the permutation group. We direct the interested reader to Ref.~\citenum{niga+22jcp} for more details about the symmetrization of the features.

\paragraph{sTDA}
Excitation energies were computed using Grimme's simplified Tamm-Dancoff Approach (sTDA)~\cite{Grimme2013,Bannwarth2014}, solving the 
 TDA equations $\mathbf{A}\mathbf{X} = \Omega \mathbf{X}$, where $\mathbf{X}$ indicates the configuration interaction singles (CIS) amplitudes that describe the excitations. 
sTDA uses an approximated form for $\mathbf{A}$, in which exchange-correlation terms are neglected and integrals are simplified:
\begin{equation}
    \label{eq:A_sTDA}
    A_{ia,jb}^{\rm sTDA} =
    \delta_{ij} \delta_{ab} \left( \varepsilon_a - \varepsilon_i \right)
    + 2 \left( ia \vert jb \right)' - \left( ij \vert ab \right)'
\end{equation}
where $\varepsilon_i$ is the MO energy for the $i$-th orbital, $i,j$ indices refer to occupied orbitals, and $a,b$ refer to virtual orbitals.
Integrals are evaluated using a monopole approximation:
\begin{equation}
    \label{eq:int_sTDA}
    \left( pq \vert rs \right)' = \sum_{A}^{N}\sum_{B}^{N} q_{pq}^{A} q_{rs}^{B} \gamma_{AB}
\end{equation}
where $q_{pq}^{A}$ is the L\"{o}wdin transition charge between MOs $p$ and $q$ for atom $A$, the sums run over all the atoms of the molecule, and $\gamma_{AB}$ is a Matanaga-Nishimoto-Ohno-Klopman term~\cite{Grimme2013}.
The L\"{o}wdin transition charges are computed from the predicted MO coefficients $\tilde{\mbf{C}}$ as:
\begin{equation}
    \label{eq:transition_lowdin_q}
    q_{rs}^{A} = \sum_{\mu \in A} \tilde{C}_{\mu r} \tilde{C}_{\mu s}
\end{equation}
where the sum runs over atomic orbitals $\mu$ centered on atom A.
Additional approximations are present in sTDA to speed up the calculation. 
The CI space is truncated using a user-defined threshold, followed by an additional selection of the most important electronic configurations to be included. For details we refer to the original publication~\cite{Grimme2013}.

All the ingredients needed for sTDA (namely, the MO energies and the transition L\"{o}wdin charges) are available from the ML model, which makes the coupling of the ML model to sTDA straightforward.
All our calculations are performed with an in-house implementation of sTDA in a spherical basis~\cite{stdatorch} in PyTorch~\cite{PyTorch2019}.
For both polyalkenes (from five to ten double bonds), aromatic molecules (benzene, azulene, biphenyl, naphtalene), $\beta$-carotene and C$_{60}$ the geometry was optimized with DFT at the B3LYP/6-31G(d) level of theory using Gaussian~\cite{g16}.

\paragraph{Vibronic Spectra}
The anthracene trajectory was generated with a DFTB3/3OB dynamics in the NVT ensemble with the Langevin thermostat and a coupling constant of 1~ps$^{-1}$. The temperature was set to 300K. We used a timestep of 0.5~fs, for a total simulation time of 100~ps. Coordinates were saved every 1~fs. The simulation was run with AMBER~\cite{da2018amber}.
Vibronic spectra were computed using the second-order cumulant expansion formalism~\cite{mukamel1995,Loco2018}.
Starting from a correlated trajectory, the excitation energy $U(t)$ is computed for each frame, and used to evaluate its autocorrelation function $c_{UU}(t)=\langle U(t) U(0)\rangle$. The autocorrelation is used to calculate the spectral density function $J(\omega)$ encoding the vibronic coupling:
\begin{equation}
    \label{eq:sds}
    J(\omega) = \frac{\beta \omega}{\pi} \int_{-\infty}^{\infty} e^{i \omega t} c_{UU}(t)\, {\rm d}t 
\end{equation}
The autocorrelation was damped in order to fall smoothly to zero in a time window of 10~ps.
The vibronic homogeneous lineshape is obtained from the spectral density through the lineshape function $g(t)$:

\begin{align}
    \label{eq:gt}
    g(t) = - \int_{0}^{\infty} \frac{J(\omega)}{\omega^2} & \Bigg[
        \coth\left( \frac{\beta \hbar \omega}{2} \right) \left(\cos\left( \omega t \right) - 
        1\right) \nonumber \\ 
        &- i \left(\sin\left( \omega t \right) - \omega t\right) \Bigg] {\rm d}\omega
\end{align}
from which the homogeneous absorption lineshape is computed as:
\begin{equation}
    \label{eq:homo_abs}
    D(\omega - \omega_{\rm eg}) = \mathcal{R} \int_{0}^{\infty} e^{i \left(\omega - \omega_{\rm eg}\right)t - g(t)} {\rm d}t
\end{equation}
Finally, the absorption spectrum is obtained after incorporating into the homogeneous spectrum static disorder, modelled by random sampling from a Gaussian distribution with a full width at half maximum (FWHM) of 400~cm$^{-1}$.
The absorption spectrum was computed with SPECDEN~\cite{specden2020}.

\begin{acknowledgments}
\textbf{Funding}: BM, LC, and EC acknowledge funding from the European Research Council (ERC) under the Grant ERC-AdG-786714 (LIFETimeS).
MC, DS, and JN acknowledge funding from the European Research Council (ERC) under the research and innovation program (Grant Agreement No. 101001890-FIAMMA), an industrial grant from Samsung, and the NCCR  MARVEL, funded by the Swiss National Science Foundation (SNSF, grant number 182892).
\textbf{Author contributions}: 
Conceptualization: MC, BM, LC; Methodology: DS, EC, JN; Investigation: DS, EC, JN; Writing: all authors.
\textbf{Competing interests}: The authors declare that they have no competing interests.
\textbf{Data and materials availability}: All the software used in this study is freely available in the publicly accessible repositories: \url{https://github.com/ecignoni/halex} and \url{https://github.com/ecignoni/stda_torch}.
Reference datasets including molecular configurations and electronic-structure properties will be made available upon publication.

\end{acknowledgments}

\pagebreak

\setcounter{equation}{0}
\setcounter{figure}{0}
\setcounter{table}{0}
\setcounter{section}{0}
\setcounter{page}{1}
\makeatletter
\renewcommand{\theequation}{S\arabic{equation}}
\renewcommand{\thefigure}{S\arabic{figure}}
\renewcommand{\thetable}{S\arabic{table}}
\renewcommand{\thesection}{S\arabic{section}}

\onecolumngrid\clearpage
\widetext
\begin{center}
\textbf{\Large Supplementary Information for\\~\\
Electronic excited states from physically-constrained machine learning}
\\~\\
Edoardo Cignoni\textsuperscript{1}$^{\dagger}$, Divya Suman\textsuperscript{2}$^{\dagger}$, Jigyasa Nigam\textsuperscript{2}, \\
Lorenzo Cupellini\textsuperscript{1}, Benedetta Mennucci\textsuperscript{1}, and Michele Ceriotti\textsuperscript{2,3}\\~\\
{\small
\textsuperscript{1} \textit{Dipartimento di Chimica e Chimica Industriale, Università di Pisa, Pisa, Italy} \\
\textsuperscript{2} \textit{Laboratory of Computational Science and Modeling, Institut des Mat\'eriaux, \'Ecole Polytechnique F\'ed\'erale de Lausanne, 1015 Lausanne, Switzerland} \\
\textsuperscript{3} \textit{Division of Chemistry and Chemical Engineering, California Institute of Technology, Pasadena, CA, USA} \\~\\
$^{\dagger}$ \textit{These authors contributed equally to this work.}\\
}
\end{center}

\section{Model Details and Training}
We provide here the details of how we have trained the ML model presented in the main text.
We have used as features the atom-centered two neighbor correlations obtained as the product of $c_q (A_i) = c_{nlm}(A_i)  c_{n'l'm'}(A_i) $ with itself to describe the local environment around atom $i$ and the corresponding three-body correlation descriptor $c_q(A_{ij}) = c_{nlm}(A_i) c_{n'l'm'}(A_{ij})$ to describe the pair of atoms $(ij)$.
As explained in the Materials and Methods and in Ref.~\cite{niga+22jcp}, these features have been built iteratively by first expanding the atomic or pair density in a basis of radial functions and spherical harmonics, and then using combinations through Clebsch-Gordan iterations %
to build the desired equivariant features (NICE framework)\cite{niga+20jcp,niga+22jcp2}.
The spherical expansion has been computed with \texttt{librascal}\cite{musi+21jcp}.
In particular, the expansion was computed with six radial basis functions and using up to $l=4$ real spherical harmonics.
We have used a cutoff of 3.5~$\textup{\AA}$, using the default shifted cosine smoothing function with a smoothness width of 0.5~$\textup{\AA}$.
The density was built using Gaussians with a width of 0.2~$\textup{\AA}$. It is important to note here that we have used quite a small cutoff of 3.5~$\textup{\AA}$ while computing our descriptors, despite having much larger molecules in our training dataset.

The features are further post-processed, prior to their use in the ML model, with Principal Component Analysis (PCA).
In particular, we have retained, for each symmetry block of the features, up to 200 principal components.
This helps reduce some degrees of redundancy in the calculated features, and to lower the memory requirements of computing these descriptors.
In order to preserve equivariance, a separate PCA is fitted, for each symmetry block in the features, without centering them.
Indeed, feature centering is akin to fitting the intercept in a linear model, which breaks the equivariance property for all $\lambda > 0$.

Before starting the training with backpropagation on MO energies and L\"{o}wdin charges, an analytical ridge regression model is fitted to the elements of the STO-3G Fock matrix.
This provides a sensible initial guess for the ML Fock matrix, greatly facilitating the training.

Starting from this fitted ridge guess, the ML parameters are further fitted by backpropagation via gradient descent.
All our linear models are fitted without the intercept, even for the $\lambda = 0$ blocks of the Fock Hamiltonian.
We have used a composite loss formed by a mean squared error (MSE) on MO energies and L\"{o}wdin atomic charges:

\begin{equation}
    \label{eq:loss_composite}
    \mathcal{L}^{\varepsilon,{\rm q}}
    =
    \frac{\omega_{\varepsilon}}{N} \sum_{n=1}^{N} \frac{1}{O_n} \sum_{o=1}^{O_n}
    \left(
    \varepsilon_{no} - \tilde{\varepsilon}_{no}
    \right)^2
    +
    \frac{\omega_q}{N}\sum_{n=1}^{N}\frac{1}{M_n}\sum_{n=1}^{M_n}
    \left(
     q_{nm} - \tilde{q}_{nm}
    \right)^2
    +
    \omega_{r} \sum_{s=1}^{S} \frac{\lVert \boldsymbol{\omega_s} \rVert^2}{N_s}
\end{equation}
where $N$ is the number of training points, $O_n$ is the number of occupied MO orbitals in the $n$-th molecule, $\varepsilon_{no}$ is the target MO energy,
$M_n$ is the number of atoms in the $n$-th molecule, and $q_{nm}$ is the target atomic charge. The tilde in $\tilde{\varepsilon}_{no}$ and $\tilde{q}_{nm}$ denotes the model prediction.
$\omega_{\varepsilon}$ and $\omega_q$ are two hyperparameters weighting the relative importance of the two MSE losses.
They have been determined so that, at convergence, the two losses have a similar magnitude and are of the order of unity.
In this way, both terms have equal weight in the total loss.
In particular, we have used $\omega_{\varepsilon} = 1.5 \times 10^6$ and $\omega_q = 1 \times 10^6$.
The last term is an L2 regularization term. $S$ denotes the number of symmetry blocks in the target, $N_s$ is the number of samples in the $s$-th symmetry block, and $\boldsymbol{\omega}_s$ is the vector of linear weights for the $s$-th symmetry block. In this work we have used $\omega_r = 1 \times 10^{-14}$.

\begin{figure}[htb!]
    \centering
    \includegraphics[width=.5\textwidth]{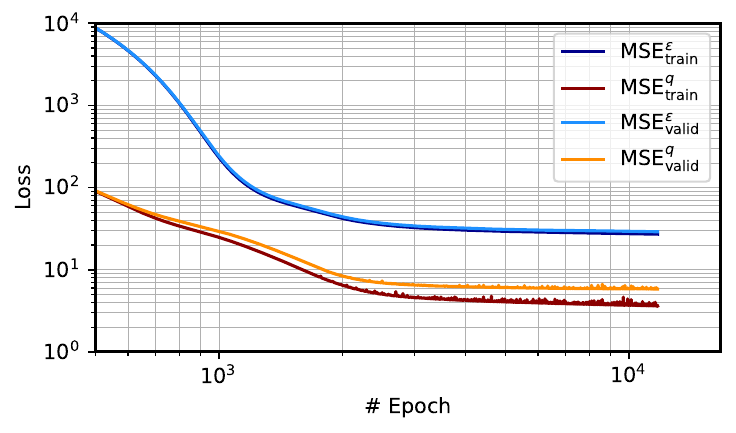}
    \caption{\textbf{Loss versus Epoch curves for training and validation losses.} The MSE on MO energies $\varepsilon$ (first term in Eq.~(\ref{eq:loss_composite})) and the MSE on L\"{o}wdin charges (second term in Eq.~(\ref{eq:loss_composite})) are shown separately.}
    \label{sifig:training_loss}
\end{figure}

The model was trained on a training dataset composed of ethane, ethene, butadiene, hexane, hexatriene, isoprene, and styrene. For each molecule, a total of 250 conformations sampled from a REMD simulation (see Materials and Methods) have been used for training, for a total of 1750 training points. A validation dataset comprising 50 conformations per molecule, for a total of 350 conformations, has been used together with the training dataset to detect possible overfitting. The training was performed in batches of 100 conformations, using the Adam optimizer implemented in PyTorch with default parameters and a learning rate of 1. The model was trained until the loss converged to a plateau, where the training was interrupted.
The loss in the train and validation set is shown in Figure~\ref{sifig:training_loss}.

\section{Training with a Larger Cutoff}
\label{sisec:large_cutoff}

We train our models with features computed for atoms within a cutoff distance of $3.5$~$\textup{\AA}$.
This translates into the impossibility of predicting matrix elements for atoms that are outside the cutoff.
While this is surely a limitation when the interaction between distant atoms becomes significant, it is also beneficial from a model regularization perspective.
In fact, a small cutoff to compute the features, reduces their complexity, and allows for better transferability and robustness of our models.

\begin{figure}[htb!]
    \centering
    \includegraphics[width=.6\textwidth]{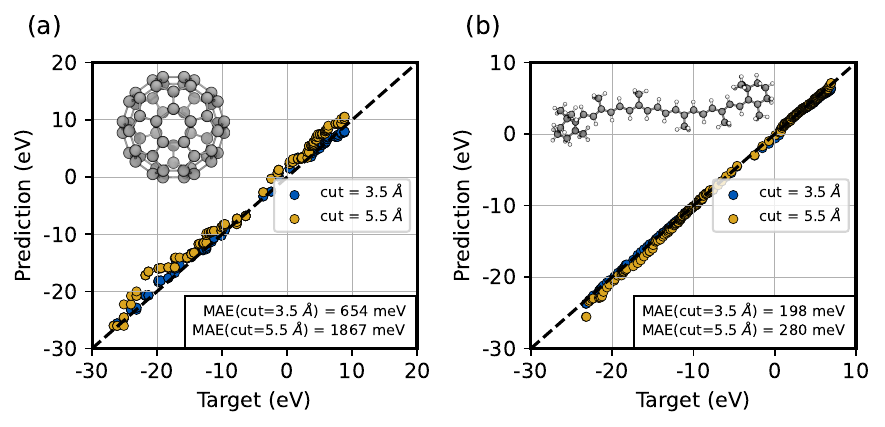}
    \caption{\textbf{Effect of using a cutoff to compute features on the model performance.} The mean absolute errors (MAE) when features are computed with a large cutoff compared to a smaller cutoff, for (a) C60 and (b) $\beta$-carotene. MAE is reported by including the valence MOs only. The MAE when core MOs are included is considerably larger.}
    \label{sifig:large_cutoff_extrapol}
\end{figure}

This can be confirmed by training a model with features computed with a larger cutoff and comparing its performance on extrapolated molecules with that of a model trained using features with smaller cutoff. 

Here, we target the B3LYP/def2-TZVP Hamiltonian, using features with a cutoff of $3.5$~$\textup{\AA}$ (the model shown in the main text) and with a larger cutoff of $5.5$~$\textup{\AA}$. This latter cutoff is only moderately larger than the first, and many elements of the resulting Hamiltonian are still strictly zero. Nonetheless, the model exhibits clear signs of overfitting when it comes to extrapolating to completely out-of-sample molecules as we can see in 
Figure~\ref{sifig:large_cutoff_extrapol}. The errors for C60 and $\beta$-carotene increase significantly with larger cutoff.
Thus, the use of a larger cutoff results in more powerful models that, if not trained extensively with a very large dataset, can lead to overfitting and loss of overall generalizability of the model.

\section{A $\Delta$ML model for the Electronic Hamiltonian}
In the main text, as well as in the preceding section, we discussed the limitations associated with using a small cutoff for features while training a model and also highlighted how using a larger cutoff instead is detrimental to the model's generalization capability. 

It is also known that the use of local features poses certain problems when long-range interactions become important~\cite{Yue2021,Anstine2023,Kabylda2023}. While this limitation can be lifted in future refinements of the model through the use of long-range features~\cite{gris-ceri19jcp}, we show here an alternative approach to mitigate the limitations arising from the use of a small cutoff by ensuring the inclusion of long-range interactions through the implementation of a $\Delta$ML model.

We use the B3LYP/STO-3G QM calculation as our baseline model from which we obtain a zeroth-order Fock matrix.
This baseline is then corrected with a $\Delta$ Fock matrix learned with ML in the very same way as the model discussed in the main text.
We are therefore able to include long-range interactions within the Fock matrix at a lower cost since they are computed with a low-level QM calculation.
A similar strategy has been reported in Ref.~\cite{unke+21nips}, although here it is used to learn an effective Fock with a minimal basis size and symmetries, targeting the properties of a Fock in a larger basis.

\begin{figure}[htb!]
    \centering
    \includegraphics[width=.65\textwidth]{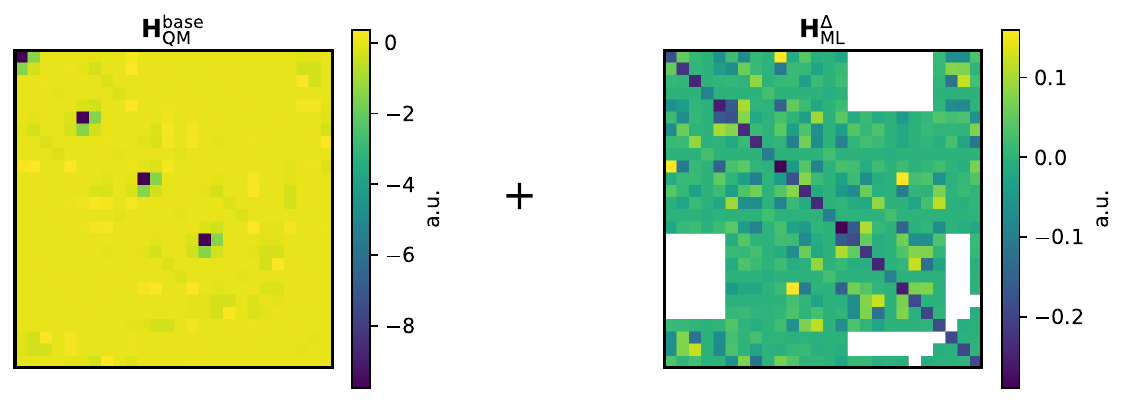}
    \caption{\textbf{Illustration of the $\Delta$ML model idea.} A baseline Fock matrix $\mathbf{H}^{\rm base}_{\rm QM}$ is computed with a low-level QM calculation, which can be DFT with a smaller basis (e.g., STO-3G as in our example) or a semiempirical method such as PM6.
    An ML model is trained to learn a $\Delta$ Fock matrix $\mathbf{H}^{\Delta}_{\rm ML}$ that corrects the baseline so as to obtain the final Fock matrix.
    Long-range interactions that are not present in the ML (empty blocks in the $\mathbf{H}^{\Delta}_{\rm ML}$ matrix shown) model are accounted for by the baseline calculation.}
    \label{sifig:delta_illustr}
\end{figure}

We denote the baseline Fock matrix obtained with e.g. STO-3G as $\mathbf{H}^{\rm base}_{\rm QM}$, and the $\Delta$ Fock matrix learned with ML as $\mathbf{H}^{\Delta}_{\rm ML}$, the Fock matrix $\mathbf{H}$ predicted by the model is given as :

\begin{equation}
    \label{eq:delta_fock}
    \mathbf{H} = \mathbf{H}^{\rm base}_{\rm QM} + \mathbf{H}^{\Delta}_{\rm ML}
\end{equation}
This matrix is diagonalized to obtain MO energies and L\"{o}wdin charges, and the $\mathbf{H}^{\Delta}_{\rm ML}$ is trained so that MO energies and charges are as close as possible to those of a higher-level, large basis QM calculation (e.g., B3LYP/def2-TZVP).

\begin{figure}[htb!]
    \centering
    \includegraphics[width=.6\textwidth]{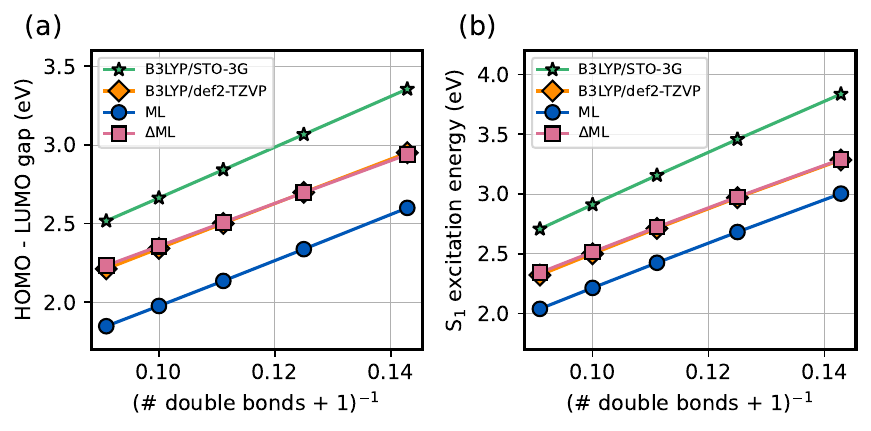}
    \caption{\textbf{Comparison of the ML and $\Delta$ML model.} Predictions on polyalkenes from 6 up to 10 double bonds.
    (a) Prediction of the HOMO-LUMO gap.
    (b) Prediction of the S$_1$ excitation energy.}
    \label{sifig:delta_hlexc}
\end{figure}

\begin{figure}[htb!]
    \centering
    \includegraphics[width=.6\textwidth]{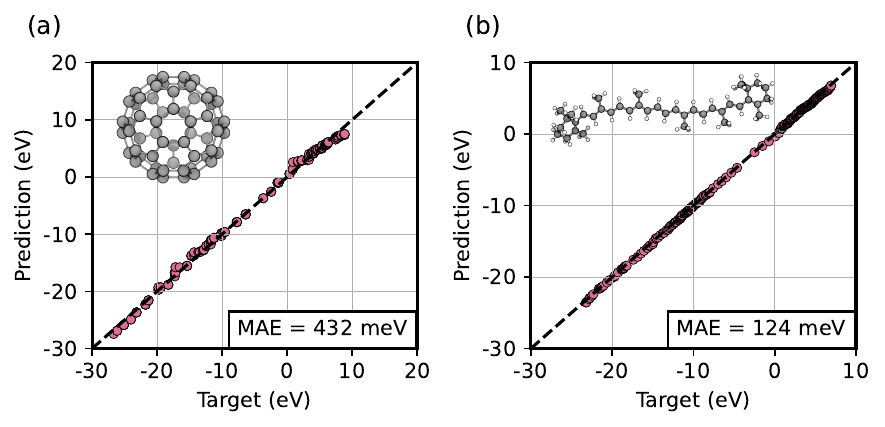}
    \caption{\textbf{$\Delta$ML model prediction for larger molecules.} The MO energies of (a) C$_{60}$ and (b) $\beta$-carotene as predicted by the $\Delta$ML model. The MAE is reported in both cases, and does not include core MOs.}
    \label{sifig:delta_c60_betacaro}
\end{figure}

\begin{figure}[htb!]
    \centering
    \includegraphics[width=.4\textwidth]{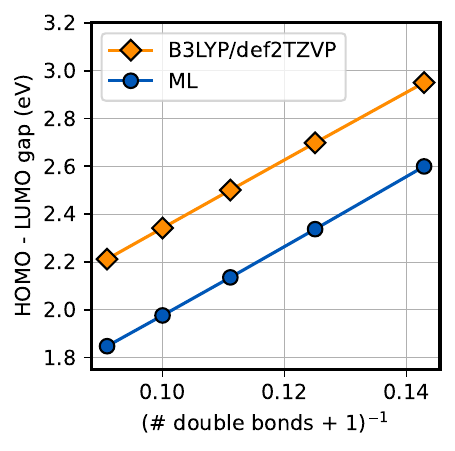}
    \caption{\textbf{Target versus predicted HOMO-LUMO gaps for polyalkenes.} HOMO-LUMO gap For polyalkenes, from 6 up to 10 double bonds we compare the HOMO-LUMO gap values for the target computed at B3LYP/def2-TZVP level from QM and the ones predicted by the model.}
    \label{sifig:homo_lumo_poly}
\end{figure}

With the $\Delta$ML model we achieve comparable errors on the test dataset compared to the ML model (presented in the main text). The real advantage of the $\Delta$ML can be seen when trying to extrapolate to much larger molecules. Figure~\ref{sifig:delta_hlexc} shows the predictions from the $\Delta$ML model as well as ML model on longer polyalkenes, for both the HOMO-LUMO gap and the excitation energy of S$_1$ (see also Figure~\ref{fig:extrapol}~c and Figure~\ref{sifig:homo_lumo_poly}).
As we can see, we are able to get rid of the redshift that arises from the locality of the features using the $\Delta$ML model.
Figure~\ref{sifig:delta_c60_betacaro} further shows that the $\Delta$ML model generalizes well to very large molecules, albeit in this case, the improvement in generalization is less striking (see Figure~\ref{fig:extrapol}~(e) and (f)).

We stress here that other choices of the baseline Fock matrix are possible as well.
For example, using a baseline obtained from a semiempirical method (e.g. PM6 or DFTB3) instead of a QM one could result in a faster $\Delta$ML model.
This improvement in generalization comes of course at the expense of a baseline calculation, making the $\Delta$ML model between two and three orders of magnitude slower than the ML model (Figure~\ref{fig:timings}). Nonetheless, it surely is a good starting point for the development of a $\Delta$ML model when a cheaper baseline is adopted.

\newpage
\section{Supplementary Figures}

\begin{figure*}[htb!]
    \centering
    \includegraphics[width=.9\textwidth]{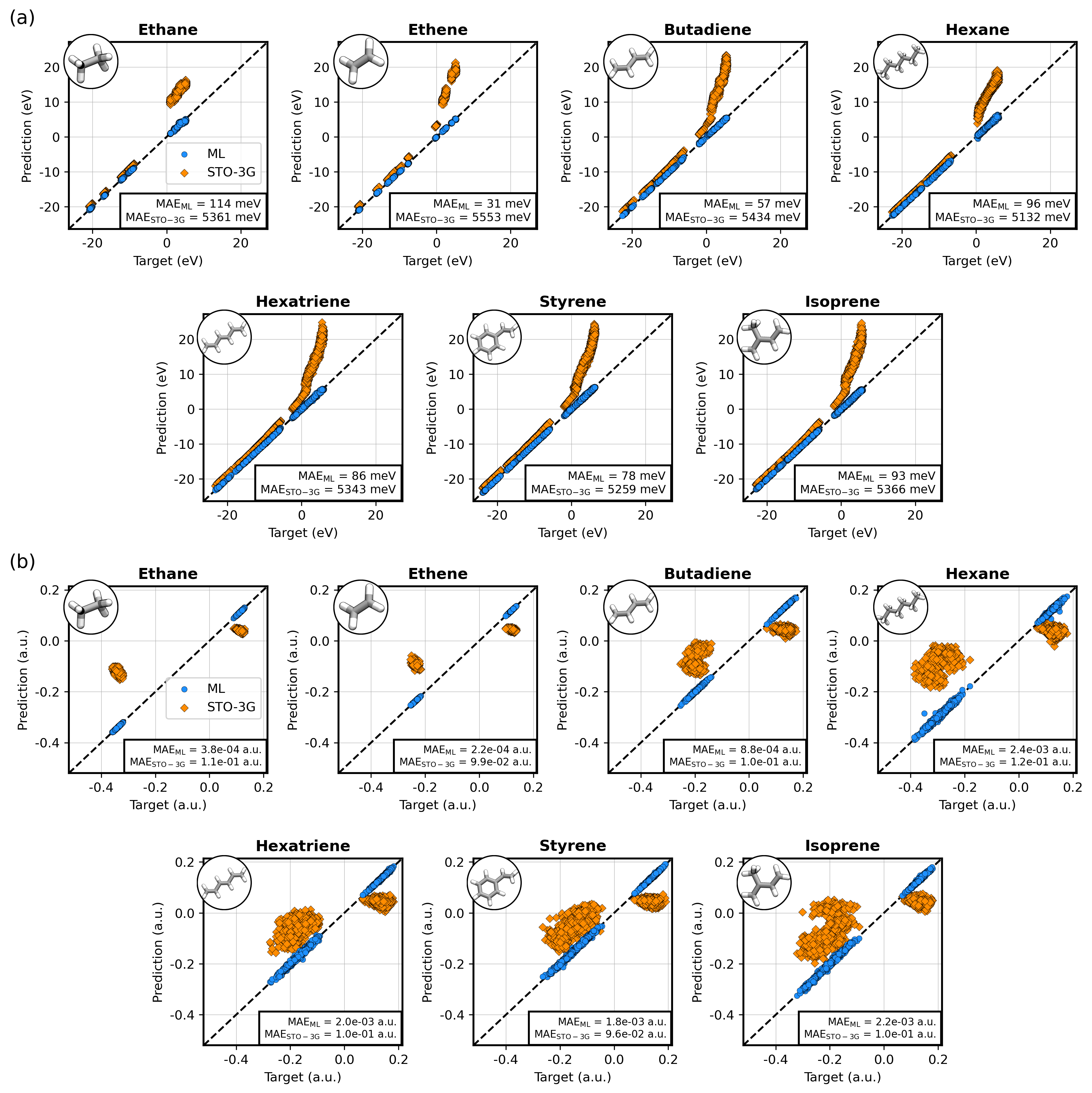}
    \caption{
    \textbf{Performance of the LBT model on validation geometries.}
    The target is computed with B3LYP/def2-TZVP.
    (a) Prediction of MO energies. Core orbitals are not shown.
    (b) Prediction of L\"{o}wdin atomic charges.
    The ML prediction is shown in blue circles.
    The B3LYP/STO-3G baseline is shown in orange diamonds.
    }
    \label{sifig:mo_lowdin}
\end{figure*}

\begin{figure}[htb!]
    \centering
    \includegraphics[width=.2\textwidth]{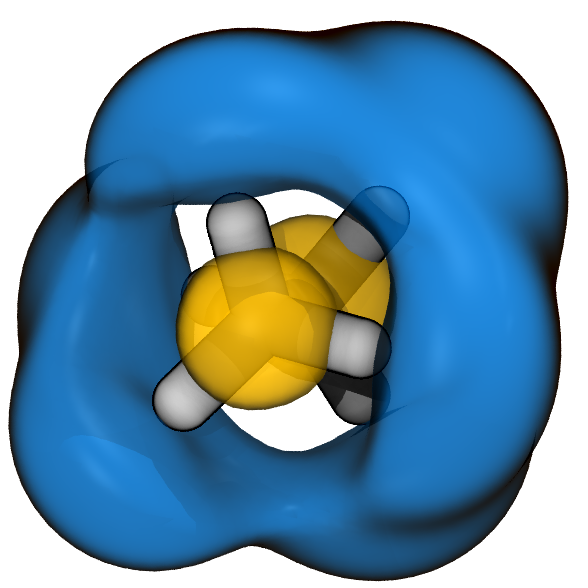}
    \caption{\textbf{Visualisation of ethane's LUMO orbital.} The LUMO orbital of ethane computed with B3LYP/def2-TZVP, as displayed in Figure~\ref{fig:mo_lowdin_allin}~(d) presented with a reduced isodensity value instead, to emphasize its Rydberg character more effectively.}
    \label{sifig:ethane_lumo_rydberg}
\end{figure}

\begin{figure}[htb!]
    \centering
    \includegraphics[width=.9\textwidth]{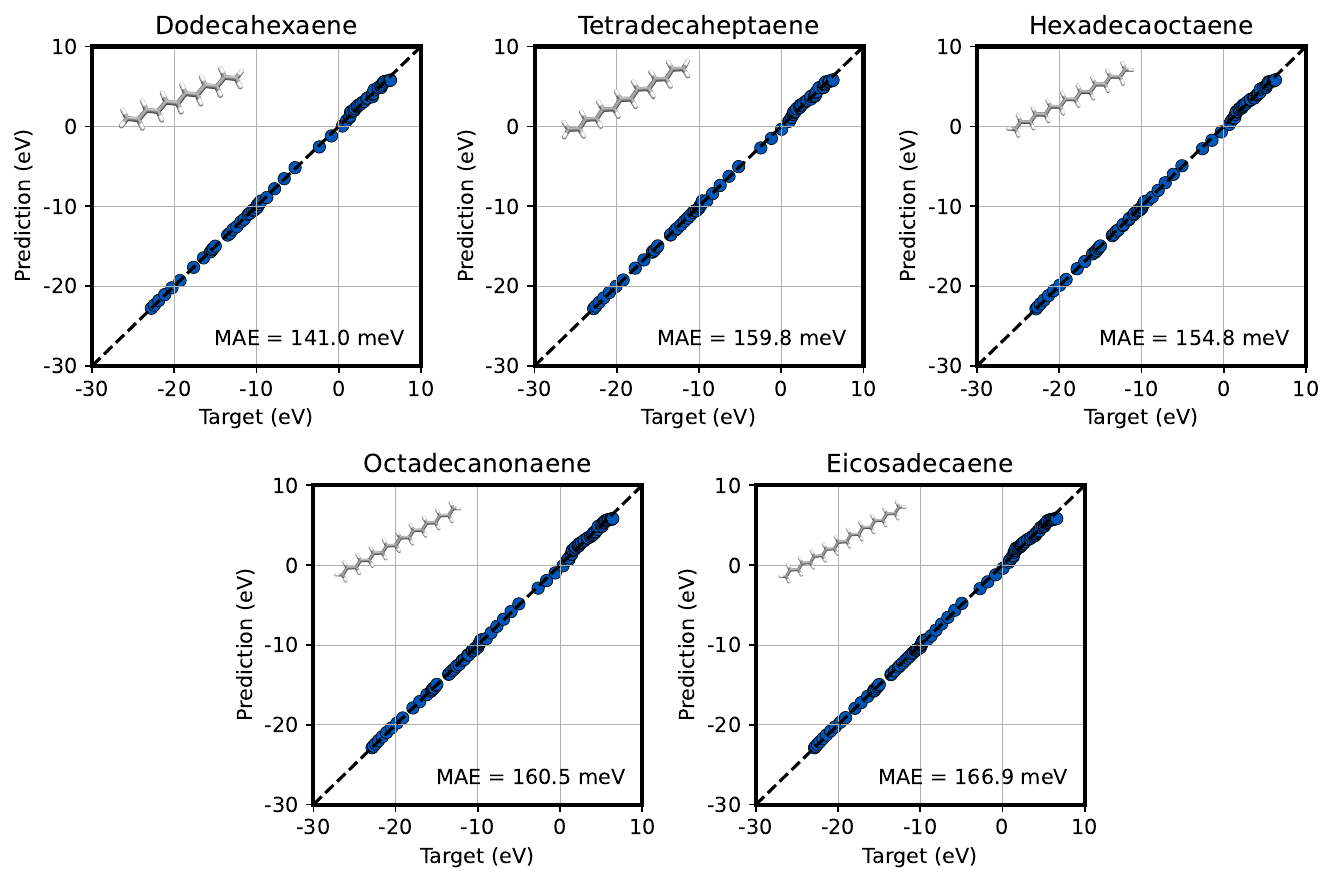}
    \caption{
    \textbf{Prediction of the MO energy spectrum for several polyalkenes.} The target is computed at B3LYP/def2-TZVP level at the optimized geometry. The energy of the core MOs is not shown. The mean absolute error (MAE) is shown for each molecule. The energy of the core MOs is not included in the MAE. Their inclusion slightly reduces the MAE.
    }
    \label{sifig:poly_ene_parity}
\end{figure}

\begin{figure}[htb!]
    \centering
    \includegraphics[width=.9\textwidth]{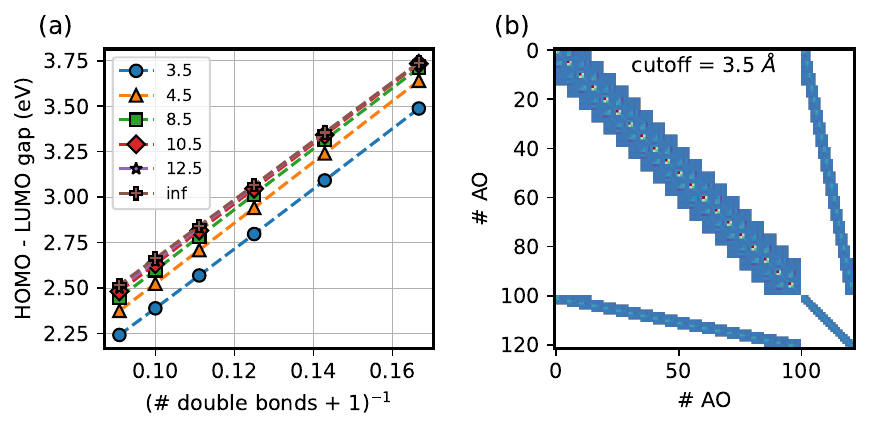}
    \caption{
    \textbf{Effect of applying a cutoff on the elements of the Fock matrix.}
    (a) HOMO - LUMO gap for polyalkenes from 5 up to 10 double bonds, for various cutoff values (see legend). The calculation is run at B3LYP/STO-3G level.
    The Fock matrix elements between two AOs whose center-to-center distance is larger than the cutoff are zeroed, the MO energies are computed, and the HOMO - LUMO gap is derived.
    (b) Example of a Fock matrix for eicosadecaene (10 double bonds) at its optimized geometry.
    A cutoff of 3.5 $\textup{\AA}$ is applied. The white regions in the matrix correspond to elements that are zeroed by the application of the cutoff.
    }
    \label{sifig:poly_cutoff_fock}
\end{figure}

\begin{figure}[htb!]
    \centering
    \includegraphics[width=.5\textwidth]{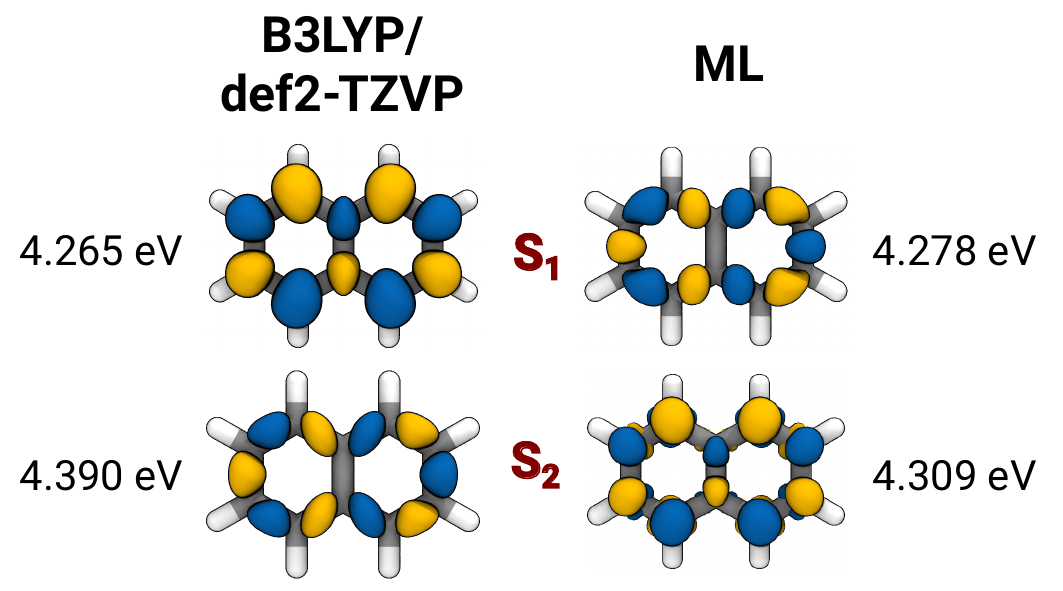}
    \caption{\textbf{Excited States of Naphthalene: sTDA B3LYP/def2-TZVP vs. sTDA ML Predictions.}
    Transition density of the first and second excited states of naphthalene, as computed with sTDA B3LYP/def2-TZVP (left) and with sTDA ML (right).
    The L$_a$ state (predominant HOMO - LUMO character) is the S$_1$ state according to sTDA B3LYP/def2-TZVP, with a close-lying L$_b$ state (nearly equal contributions of HOMO-1 - LUMO and HOMO - LUMO+1) about 130~meV above in energy.
    The order is inverted with sTDA ML, which predicts the L$_b$ state as about 30~meV below the L$_a$ state.
    }
    \label{sifig:napht_LaLb}
\end{figure}

\begin{figure}[htb!]
    \centering
    \includegraphics[width=.9\textwidth]{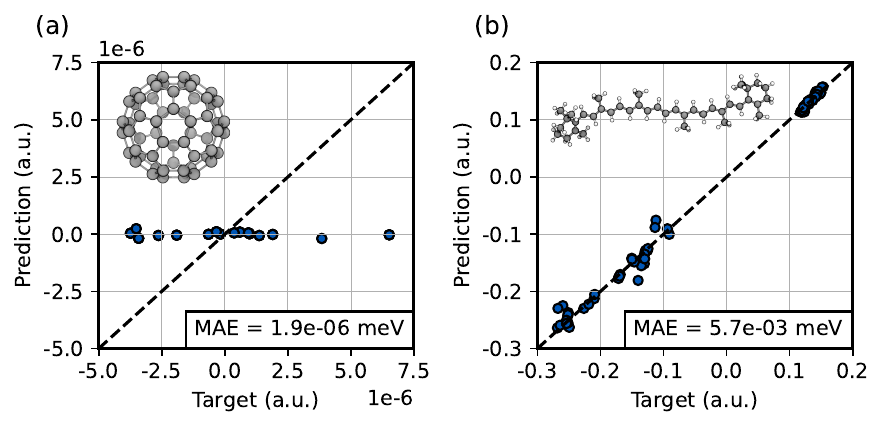}
    \caption{\textbf{Predicted L\"{o}wdin charges for extrapolated molecules.} Prediction of atomic L\"{o}wdin charges for (a) C$_{60}$ and (b) $\beta$-carotene.
    The target charges are computed with B3LYP/def2-TZVP.
    The MAE is reported in both cases. Note that the scale of panel (a) is multiplied by 10$^{-6}$: the charges should be zero because of symmetry, and the data shows simply that the ML model has less numerical noise than the reference calculations.
    }
    \label{sifig:c60_beta_charges}
\end{figure}

\begin{figure}[htb!]
    \centering
    \includegraphics[width=.9\textwidth]{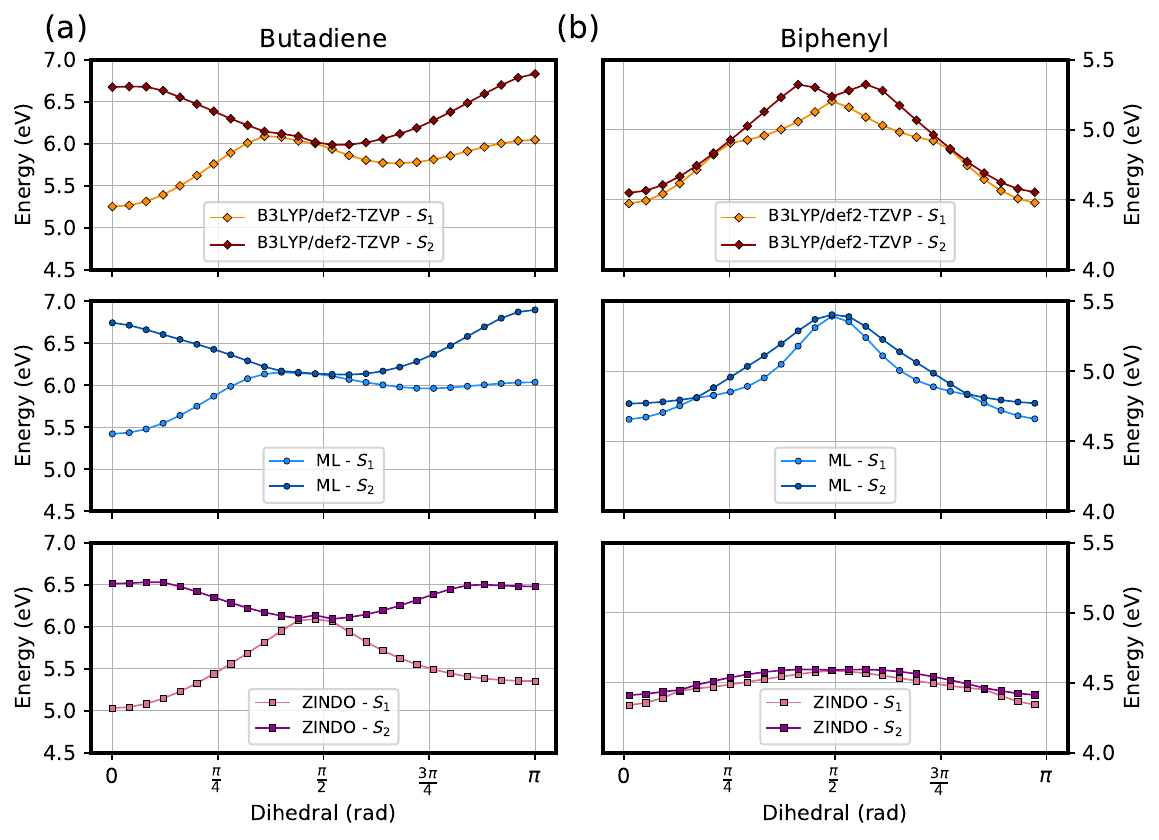}
    \caption{\textbf{Comparison of the excitation energies from sTDA B3LYP/def2-TZVP, ML Model, and ZINDO.}
    Rigid scan around the central single bond in (a) butadiene and (b) biphenyl.
    The excitation energies for the first two singlet excited states are reported for three different methods sTDA B3LYP/def2-TZVP (top row), the ML model (middle row), and ZINDO (bottom row).
    }
    \label{sifig:scandihe}
\end{figure}

\begin{figure}[htb!]
    \centering
    \includegraphics[width=.9\textwidth]{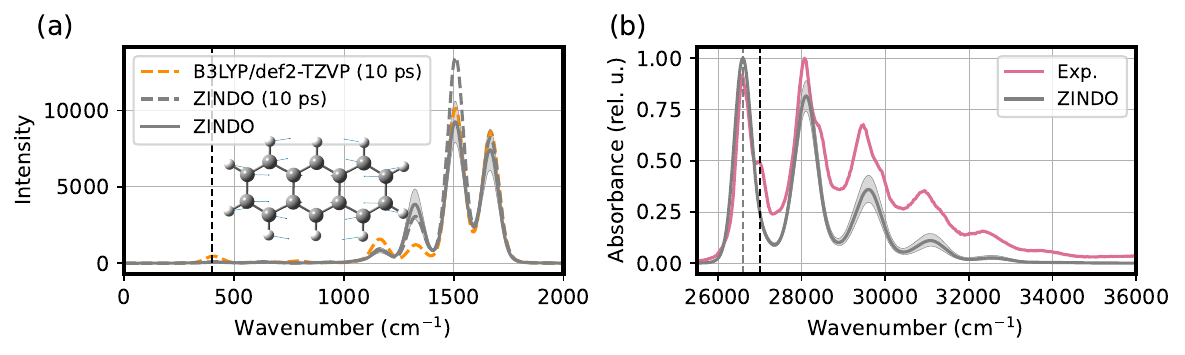}
    \caption{\textbf{Spectral density and vibron spectrum.}
    (a) Spectral density and (b) vibronic spectrum as computed with the semi-empirical method ZINDO (gray) and compared against sTDA B3LYP/def2-TZVP (orange, panel (a)) and the experimental spectrum (magenta, panel (b)).
    The solid line corresponds to the average over 10~ps windows along the anthracene MD.
    Confidence interval of 95\% around the mean is reported.
    }
    \label{sifig:vibronic_zindo}
\end{figure}

\begin{figure}[htb!]
    \centering
    \includegraphics[width=.5\textwidth]{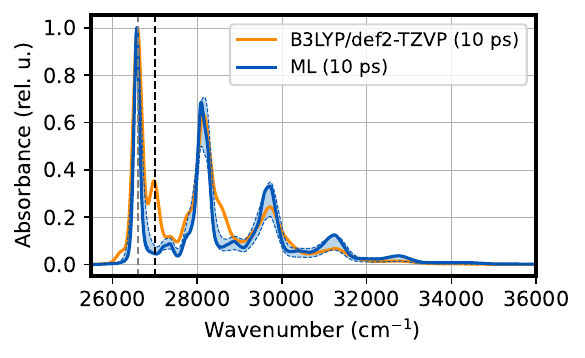}
    \caption{\textbf{Vibronic spectrum of anthracene.}
    Vibronic spectrum in a 10~ps window of the anthracene trajectory, computed with ML (blue) and with sTDA B3LYP/def2-TZVP (orange).
    The static disorder is not added in this spectrum, to better compare the vibronic peaks of the two spectra.
    The vertical dashed line denotes the peak missing from the ML prediction.
    The 95\% confidence interval of the mean is reported for the ML model.
    }
    \label{sifig:vibronic_targ_10ps}
\end{figure}

\clearpage
\section{Supplementary Tables}

\begin{table}[htb!]
    \centering
    \label{sitab:errors_stda_HC}
    \begin{tabular}{ccccc}
    \hline
    \multirow{2}{*}{
    \textbf{Molecule}
    } & \multicolumn{2}{c}{\textbf{ML}} & \multicolumn{2}{c}{\textbf{B3LYP/STO-3G}} \\
      & MAE$_{S_1}$ & MAE$_{S_2}$ & MAE$_{S_1}$ & MAE$_{S_2}$ \\
    \hline
    \textbf{Ethane} & 45 & 75 & 8585 & 8513 \\
    \textbf{Ethene} & 78 & 97 & 2484 & 2864 \\
    \textbf{Butadiene} & 105 & 82 & 1455 & 1529 \\
    \textbf{Hexane} & 188 & 150 & 5514 & 5538 \\
    \textbf{Hexatriene} & 148 & 124 & 1123 & 1163 \\
    \textbf{Isoprene} & 183 & 104 & 1451 & 1476 \\
    \textbf{Styrene} & 128 & 178 & 1110 & 1249 \\
    \hline
    \end{tabular}
    \caption{\textbf{Mean absolute error (MAE) computed for the first and second singlet excited states.} The MAE is computed as $N^{-1}\sum_n=1^N \vert \mathcal{E}_n - \hat{\mathcal{E}}_n \vert$, where $N$ is the number of samples and $\mathcal{E}_i$ is the excitation energy of the $i$-th data point.}
\end{table}

\begin{table}[htb!]
    \centering
    
    \label{sitab:timings}
    \begin{tabular}{cccccc}
    \hline
    \textbf{Molecule} & \textbf{Method} & \textbf{Featurization (s)} & \textbf{Prediction (s)} & \textbf{sTDA (s)} & \textbf{Total (s)} \\
    \hline
    \multirow{2}*{\textbf{Azulene}}
     & B3LYP/def2-TZVP & - & 2466.89 & 563.93 & 3030.82 \\
     & B3LYP/STO-3G & - & 64.04 & 15.46 & 79.50 \\
     & ML & 0.36 & 0.05 & 0.07 & 0.48 \\
    \multirow{2}*{\textbf{Benzene}}
     & B3LYP/def2-TZVP & - & 370.86 & 97.22 & 468.08 \\
     & B3LYP/STO-3G & - & 16.48 & 5.18 & 21.66 \\
     & ML & 0.25 & 0.04 & 0.01 & 0.30 \\
    \multirow{2}*{\textbf{Biphenyl}}
     & B3LYP/def2-TZVP & - & 3466.63 & 932.91 & 4399.54 \\
     & B3LYP/STO-3G & - & 103.53 & 27.31 & 130.84 \\
     & ML & 0.37 & 0.06 & 0.16 & 0.59 \\
    \multirow{2}*{\textbf{Butadiene}}
     & B3LYP/def2-TZVP & - & 181.38 & 44.83 & 226.21 \\
     & B3LYP/STO-3G & - & 11.51 & 3.27 & 14.78 \\
     & ML & 0.22 & 0.04 & 0.01 & 0.27 \\
    \multirow{2}*{\textbf{Dodecahexaene}}
     & B3LYP/def2-TZVP & - & 3837.90 & 883.43 & 4721.33 \\
     & B3LYP/STO-3G & - & 132.11 & 32.29 & 164.40 \\
     & ML & 0.33 & 0.05 & 0.23 & 0.61 \\
    \multirow{2}*{\textbf{Eicosadecaene}}
     & B3LYP/def2-TZVP & - & 13023.30 & 2861.36 & 15884.66 \\
     & B3LYP/STO-3G & - & 418.76 & 97.83 & 516.59 \\
     & ML & 0.48 & 0.07 & 3.60 & 4.15 \\
    \multirow{2}*{\textbf{Hexane}}
     & B3LYP/def2-TZVP & - & 1100.52 & 288.16 & 1388.68 \\
     & B3LYP/STO-3G & - & 38.13 & 11.96 & 50.09 \\
     & ML & 0.37 & 0.04 & 0.02 & 0.43 \\
    \multirow{2}*{\textbf{Hexatriene}}
     & B3LYP/def2-TZVP & - & 487.39 & 119.38 & 606.77 \\
     & B3LYP/STO-3G & - & 28.90 & 6.75 & 35.65 \\
     & ML & 0.24 & 0.04 & 0.02 & 0.30 \\
    \multirow{2}*{\textbf{Ethane}}
     & B3LYP/def2-TZVP & - & 56.57 & 15.03 & 71.60 \\
     & B3LYP/STO-3G & - & 4.55 & 1.37 & 5.92 \\
     & ML & 0.21 & 0.03 & 0.01 & 0.25 \\
    \multirow{2}*{\textbf{Ethene}}
     & B3LYP/def2-TZVP & - & 29.73 & 8.70 & 38.43 \\
     & B3LYP/STO-3G & - & 2.94 & 0.85 & 3.79 \\
     & ML & 0.20 & 0.03 & 0.00 & 0.23 \\
    \multirow{2}*{\textbf{Hexadecaoctaene}}
     & B3LYP/def2-TZVP & - & 7374.33 & 1707.01 & 9081.34 \\
     & B3LYP/STO-3G & - & 256.04 & 58.11 & 314.15 \\
     & ML & 0.39 & 0.06 & 1.00 & 1.45 \\
    \multirow{2}*{\textbf{Isoprene}}
     & B3LYP/def2-TZVP & - & 396.94 & 89.22 & 486.16 \\
     & B3LYP/STO-3G & - & 19.49 & 5.07 & 24.56 \\
     & ML & 0.25 & 0.04 & 0.01 & 0.30 \\
    \multirow{2}*{\textbf{Naphtalene}}
     & B3LYP/def2-TZVP & - & 2136.56 & 568.42 & 2704.98 \\
     & B3LYP/STO-3G & - & 53.69 & 15.39 & 69.08 \\
     & ML & 0.36 & 0.05 & 0.06 & 0.47 \\
    \multirow{2}*{\textbf{Octadecanonaene}}
     & B3LYP/def2-TZVP & - & 9621.86 & 2238.88 & 11860.74 \\
     & B3LYP/STO-3G & - & 327.18 & 74.85 & 402.03 \\
     & ML & 0.44 & 0.06 & 1.92 & 2.42 \\
    \multirow{2}*{\textbf{Octatetraene}}
     & B3LYP/def2-TZVP & - & 1345.15 & 326.00 & 1671.15 \\
     & B3LYP/STO-3G & - & 50.00 & 13.18 & 63.18 \\
     & ML & 0.27 & 0.04 & 0.04 & 0.35 \\
    \multirow{2}*{\textbf{Styrene}}
     & B3LYP/def2-TZVP & - & 1447.99 & 329.12 & 1777.11 \\
     & B3LYP/STO-3G & - & 39.86 & 10.40 & 50.26 \\
     & ML & 0.29 & 0.05 & 0.03 & 0.37 \\
    \multirow{2}*{\textbf{Tetradecaheptaene}}
     & B3LYP/def2-TZVP & - & 5507.33 & 1269.84 & 6777.17 \\
     & B3LYP/STO-3G & - & 195.91 & 44.48 & 240.39 \\
     & ML & 0.36 & 0.05 & 0.48 & 0.89 \\
    \hline
    \end{tabular}
    \caption{\textbf{Time required to compute an excited state for the molecules from different methods.}
    The time required to featurize a molecule is reported under ``Featurization", and is absent for non-ML methods.
    The time required to predict (or calculate, in the case of non-ML methods) the Fock matrix is reported under ``Prediction".
    The time required to run the sTDA is reported under ``sTDA".
    Each time is computed on a single core of an Intel Xeon Gold 5120 Processor.
    }
\end{table}


\begin{thebibliography}{74}%
\makeatletter
\providecommand \@ifxundefined [1]{%
 \@ifx{#1\undefined}
}%
\providecommand \@ifnum [1]{%
 \ifnum #1\expandafter \@firstoftwo
 \else \expandafter \@secondoftwo
 \fi
}%
\providecommand \@ifx [1]{%
 \ifx #1\expandafter \@firstoftwo
 \else \expandafter \@secondoftwo
 \fi
}%
\providecommand \natexlab [1]{#1}%
\providecommand \enquote  [1]{``#1''}%
\providecommand \bibnamefont  [1]{#1}%
\providecommand \bibfnamefont [1]{#1}%
\providecommand \citenamefont [1]{#1}%
\providecommand \href@noop [0]{\@secondoftwo}%
\providecommand \href [0]{\begingroup \@sanitize@url \@href}%
\providecommand \@href[1]{\@@startlink{#1}\@@href}%
\providecommand \@@href[1]{\endgroup#1\@@endlink}%
\providecommand \@sanitize@url [0]{\catcode `\\12\catcode `\$12\catcode
  `\&12\catcode `\#12\catcode `\^12\catcode `\_12\catcode `\%12\relax}%
\providecommand \@@startlink[1]{}%
\providecommand \@@endlink[0]{}%
\providecommand \url  [0]{\begingroup\@sanitize@url \@url }%
\providecommand \@url [1]{\endgroup\@href {#1}{\urlprefix }}%
\providecommand \urlprefix  [0]{URL }%
\providecommand \Eprint [0]{\href }%
\providecommand \doibase [0]{http://dx.doi.org/}%
\providecommand \selectlanguage [0]{\@gobble}%
\providecommand \bibinfo  [0]{\@secondoftwo}%
\providecommand \bibfield  [0]{\@secondoftwo}%
\providecommand \translation [1]{[#1]}%
\providecommand \BibitemOpen [0]{}%
\providecommand \bibitemStop [0]{}%
\providecommand \bibitemNoStop [0]{.\EOS\space}%
\providecommand \EOS [0]{\spacefactor3000\relax}%
\providecommand \BibitemShut  [1]{\csname bibitem#1\endcsname}%
\let\auto@bib@innerbib\@empty
\bibitem [{\citenamefont {Carleo}\ \emph {et~al.}(2019)\citenamefont {Carleo},
  \citenamefont {Cirac}, \citenamefont {Cranmer}, \citenamefont {Daudet},
  \citenamefont {Schuld}, \citenamefont {Tishby}, \citenamefont
  {{Vogt-Maranto}},\ and\ \citenamefont {Zdeborov{\'a}}}]{carl+19rmp}%
  \BibitemOpen
  \bibfield  {author} {\bibinfo {author} {\bibfnamefont {Giuseppe}\
  \bibnamefont {Carleo}}, \bibinfo {author} {\bibfnamefont {Ignacio}\
  \bibnamefont {Cirac}}, \bibinfo {author} {\bibfnamefont {Kyle}\ \bibnamefont
  {Cranmer}}, \bibinfo {author} {\bibfnamefont {Laurent}\ \bibnamefont
  {Daudet}}, \bibinfo {author} {\bibfnamefont {Maria}\ \bibnamefont {Schuld}},
  \bibinfo {author} {\bibfnamefont {Naftali}\ \bibnamefont {Tishby}}, \bibinfo
  {author} {\bibfnamefont {Leslie}\ \bibnamefont {{Vogt-Maranto}}}, \ and\
  \bibinfo {author} {\bibfnamefont {Lenka}\ \bibnamefont {Zdeborov{\'a}}},\
  }\bibfield  {title} {\enquote {\bibinfo {title} {Machine learning and the
  physical sciences},}\ }\href {\doibase 10.1103/RevModPhys.91.045002}
  {\bibfield  {journal} {\bibinfo  {journal} {Reviews of Modern Physics}\
  }\textbf {\bibinfo {volume} {91}},\ \bibinfo {pages} {045002} (\bibinfo
  {year} {2019})}\BibitemShut {NoStop}%
\bibitem [{\citenamefont {Ceriotti}\ \emph {et~al.}(2021)\citenamefont
  {Ceriotti}, \citenamefont {Clementi},\ and\ \citenamefont {{Anatole von
  Lilienfeld}}}]{ceri+21cr}%
  \BibitemOpen
  \bibfield  {author} {\bibinfo {author} {\bibfnamefont {Michele}\ \bibnamefont
  {Ceriotti}}, \bibinfo {author} {\bibfnamefont {Cecilia}\ \bibnamefont
  {Clementi}}, \ and\ \bibinfo {author} {\bibfnamefont {O.}~\bibnamefont
  {{Anatole von Lilienfeld}}},\ }\bibfield  {title} {\enquote {\bibinfo {title}
  {Introduction: {{Machine Learning}} at the {{Atomic Scale}}},}\ }\href
  {\doibase 10.1021/acs.chemrev.1c00598} {\bibfield  {journal} {\bibinfo
  {journal} {Chemical Reviews}\ }\textbf {\bibinfo {volume} {121}},\ \bibinfo
  {pages} {9719--9721} (\bibinfo {year} {2021})}\BibitemShut {NoStop}%
\bibitem [{\citenamefont {Cheng}\ \emph {et~al.}(2019)\citenamefont {Cheng},
  \citenamefont {Engel}, \citenamefont {Behler}, \citenamefont {Dellago},\ and\
  \citenamefont {Ceriotti}}]{chen+19pnas}%
  \BibitemOpen
  \bibfield  {author} {\bibinfo {author} {\bibfnamefont {Bingqing}\
  \bibnamefont {Cheng}}, \bibinfo {author} {\bibfnamefont {Edgar~A.}\
  \bibnamefont {Engel}}, \bibinfo {author} {\bibfnamefont {J{\"o}rg}\
  \bibnamefont {Behler}}, \bibinfo {author} {\bibfnamefont {Christoph}\
  \bibnamefont {Dellago}}, \ and\ \bibinfo {author} {\bibfnamefont {Michele}\
  \bibnamefont {Ceriotti}},\ }\bibfield  {title} {\enquote {\bibinfo {title}
  {Ab initio thermodynamics of liquid and solid water},}\ }\href {\doibase
  10.1073/pnas.1815117116} {\bibfield  {journal} {\bibinfo  {journal}
  {Proceedings of the National Academy of Sciences of the United States of
  America}\ }\textbf {\bibinfo {volume} {116}},\ \bibinfo {pages} {1110--1115}
  (\bibinfo {year} {2019})}\BibitemShut {NoStop}%
\bibitem [{\citenamefont {Deringer}\ \emph {et~al.}(2021)\citenamefont
  {Deringer}, \citenamefont {Bernstein}, \citenamefont {Cs{\'a}nyi},
  \citenamefont {Ben~Mahmoud}, \citenamefont {Ceriotti}, \citenamefont
  {Wilson}, \citenamefont {Drabold},\ and\ \citenamefont
  {Elliott}}]{deri+21nature}%
  \BibitemOpen
  \bibfield  {author} {\bibinfo {author} {\bibfnamefont {Volker~L.}\
  \bibnamefont {Deringer}}, \bibinfo {author} {\bibfnamefont {Noam}\
  \bibnamefont {Bernstein}}, \bibinfo {author} {\bibfnamefont {G{\'a}bor}\
  \bibnamefont {Cs{\'a}nyi}}, \bibinfo {author} {\bibfnamefont {Chiheb}\
  \bibnamefont {Ben~Mahmoud}}, \bibinfo {author} {\bibfnamefont {Michele}\
  \bibnamefont {Ceriotti}}, \bibinfo {author} {\bibfnamefont {Mark}\
  \bibnamefont {Wilson}}, \bibinfo {author} {\bibfnamefont {David~A.}\
  \bibnamefont {Drabold}}, \ and\ \bibinfo {author} {\bibfnamefont
  {Stephen~R.}\ \bibnamefont {Elliott}},\ }\bibfield  {title} {\enquote
  {\bibinfo {title} {Origins of structural and electronic transitions in
  disordered silicon},}\ }\href {\doibase 10.1038/s41586-020-03072-z}
  {\bibfield  {journal} {\bibinfo  {journal} {Nature}\ }\textbf {\bibinfo
  {volume} {589}},\ \bibinfo {pages} {59--64} (\bibinfo {year}
  {2021})}\BibitemShut {NoStop}%
\bibitem [{\citenamefont {Zhou}\ \emph {et~al.}(2023)\citenamefont {Zhou},
  \citenamefont {Zhang}, \citenamefont {Ma},\ and\ \citenamefont
  {Deringer}}]{zhou+23ne}%
  \BibitemOpen
  \bibfield  {author} {\bibinfo {author} {\bibfnamefont {Yuxing}\ \bibnamefont
  {Zhou}}, \bibinfo {author} {\bibfnamefont {Wei}\ \bibnamefont {Zhang}},
  \bibinfo {author} {\bibfnamefont {En}~\bibnamefont {Ma}}, \ and\ \bibinfo
  {author} {\bibfnamefont {Volker~L.}\ \bibnamefont {Deringer}},\ }\bibfield
  {title} {\enquote {\bibinfo {title} {Device-scale atomistic modelling of
  phase-change memory materials},}\ }\href {\doibase
  10.1038/s41928-023-01030-x} {\bibfield  {journal} {\bibinfo  {journal}
  {Nature Electronics}\ } (\bibinfo {year} {2023}),\
  10.1038/s41928-023-01030-x}\BibitemShut {NoStop}%
\bibitem [{\citenamefont {Ceriotti}(2022)}]{ceri22mrsb}%
  \BibitemOpen
  \bibfield  {author} {\bibinfo {author} {\bibfnamefont {Michele}\ \bibnamefont
  {Ceriotti}},\ }\bibfield  {title} {\enquote {\bibinfo {title} {Beyond
  potentials: {{Integrated}} machine learning models for materials},}\ }\href
  {\doibase 10.1557/s43577-022-00440-0} {\bibfield  {journal} {\bibinfo
  {journal} {MRS Bulletin}\ }\textbf {\bibinfo {volume} {47}},\ \bibinfo
  {pages} {1045--1053} (\bibinfo {year} {2022})}\BibitemShut {NoStop}%
\bibitem [{\citenamefont {Brockherde}\ \emph {et~al.}(2017)\citenamefont
  {Brockherde}, \citenamefont {Vogt}, \citenamefont {Li}, \citenamefont
  {Tuckerman}, \citenamefont {Burke},\ and\ \citenamefont
  {M{\"u}ller}}]{broc+17nc}%
  \BibitemOpen
  \bibfield  {author} {\bibinfo {author} {\bibfnamefont {Felix}\ \bibnamefont
  {Brockherde}}, \bibinfo {author} {\bibfnamefont {Leslie}\ \bibnamefont
  {Vogt}}, \bibinfo {author} {\bibfnamefont {Li}~\bibnamefont {Li}}, \bibinfo
  {author} {\bibfnamefont {Mark~E.}\ \bibnamefont {Tuckerman}}, \bibinfo
  {author} {\bibfnamefont {Kieron}\ \bibnamefont {Burke}}, \ and\ \bibinfo
  {author} {\bibfnamefont {Klaus~Robert}\ \bibnamefont {M{\"u}ller}},\
  }\bibfield  {title} {\enquote {\bibinfo {title} {Bypassing the {{Kohn-Sham}}
  equations with machine learning},}\ }\href {\doibase
  10.1038/s41467-017-00839-3} {\bibfield  {journal} {\bibinfo  {journal}
  {Nature Communications}\ }\textbf {\bibinfo {volume} {8}},\ \bibinfo {pages}
  {872} (\bibinfo {year} {2017})}\BibitemShut {NoStop}%
\bibitem [{\citenamefont {Grisafi}\ \emph {et~al.}(2019)\citenamefont
  {Grisafi}, \citenamefont {Fabrizio}, \citenamefont {Meyer}, \citenamefont
  {Wilkins}, \citenamefont {Corminboeuf},\ and\ \citenamefont
  {Ceriotti}}]{gris+19acscs}%
  \BibitemOpen
  \bibfield  {author} {\bibinfo {author} {\bibfnamefont {Andrea}\ \bibnamefont
  {Grisafi}}, \bibinfo {author} {\bibfnamefont {Alberto}\ \bibnamefont
  {Fabrizio}}, \bibinfo {author} {\bibfnamefont {Benjamin}\ \bibnamefont
  {Meyer}}, \bibinfo {author} {\bibfnamefont {David~M.}\ \bibnamefont
  {Wilkins}}, \bibinfo {author} {\bibfnamefont {Clemence}\ \bibnamefont
  {Corminboeuf}}, \ and\ \bibinfo {author} {\bibfnamefont {Michele}\
  \bibnamefont {Ceriotti}},\ }\bibfield  {title} {\enquote {\bibinfo {title}
  {Transferable {{Machine-Learning Model}} of the {{Electron Density}}},}\
  }\href {\doibase 10.1021/acscentsci.8b00551} {\bibfield  {journal} {\bibinfo
  {journal} {ACS Central Science}\ }\textbf {\bibinfo {volume} {5}},\ \bibinfo
  {pages} {57--64} (\bibinfo {year} {2019})}\BibitemShut {NoStop}%
\bibitem [{\citenamefont {Shao}\ \emph {et~al.}(2023)\citenamefont {Shao},
  \citenamefont {Paetow}, \citenamefont {Tuckerman},\ and\ \citenamefont
  {Pavanello}}]{Shao2023}%
  \BibitemOpen
  \bibfield  {author} {\bibinfo {author} {\bibfnamefont {Xuecheng}\
  \bibnamefont {Shao}}, \bibinfo {author} {\bibfnamefont {Lukas}\ \bibnamefont
  {Paetow}}, \bibinfo {author} {\bibfnamefont {Mark~E.}\ \bibnamefont
  {Tuckerman}}, \ and\ \bibinfo {author} {\bibfnamefont {Michele}\ \bibnamefont
  {Pavanello}},\ }\bibfield  {title} {\enquote {\bibinfo {title} {Machine
  learning electronic structure methods based on the one-electron reduced
  density matrix},}\ }\href {\doibase 10.1038/s41467-023-41953-9} {\bibfield
  {journal} {\bibinfo  {journal} {Nature Communications}\ }\textbf {\bibinfo
  {volume} {14}} (\bibinfo {year} {2023}),\
  10.1038/s41467-023-41953-9}\BibitemShut {NoStop}%
\bibitem [{\citenamefont {Dral}\ and\ \citenamefont
  {Barbatti}(2021)}]{Dral2021}%
  \BibitemOpen
  \bibfield  {author} {\bibinfo {author} {\bibfnamefont {Pavlo~O.}\
  \bibnamefont {Dral}}\ and\ \bibinfo {author} {\bibfnamefont {Mario}\
  \bibnamefont {Barbatti}},\ }\bibfield  {title} {\enquote {\bibinfo {title}
  {Molecular excited states through a machine learning lens},}\ }\href
  {\doibase 10.1038/s41570-021-00278-1} {\bibfield  {journal} {\bibinfo
  {journal} {Nature Reviews Chemistry}\ }\textbf {\bibinfo {volume} {5}},\
  \bibinfo {pages} {388--405} (\bibinfo {year} {2021})}\BibitemShut {NoStop}%
\bibitem [{\citenamefont {Cignoni}\ \emph {et~al.}(2023)\citenamefont
  {Cignoni}, \citenamefont {Cupellini},\ and\ \citenamefont
  {Mennucci}}]{Cignoni2023}%
  \BibitemOpen
  \bibfield  {author} {\bibinfo {author} {\bibfnamefont {Edoardo}\ \bibnamefont
  {Cignoni}}, \bibinfo {author} {\bibfnamefont {Lorenzo}\ \bibnamefont
  {Cupellini}}, \ and\ \bibinfo {author} {\bibfnamefont {Benedetta}\
  \bibnamefont {Mennucci}},\ }\bibfield  {title} {\enquote {\bibinfo {title}
  {Machine learning exciton hamiltonians in light-harvesting complexes},}\
  }\href {\doibase 10.1021/acs.jctc.2c01044} {\bibfield  {journal} {\bibinfo
  {journal} {Journal of Chemical Theory and Computation}\ }\textbf {\bibinfo
  {volume} {19}},\ \bibinfo {pages} {965--977} (\bibinfo {year}
  {2023})}\BibitemShut {NoStop}%
\bibitem [{\citenamefont {Westermayr}\ and\ \citenamefont
  {Marquetand}(2020)}]{Westermayr2020ChemRev}%
  \BibitemOpen
  \bibfield  {author} {\bibinfo {author} {\bibfnamefont {Julia}\ \bibnamefont
  {Westermayr}}\ and\ \bibinfo {author} {\bibfnamefont {Philipp}\ \bibnamefont
  {Marquetand}},\ }\bibfield  {title} {\enquote {\bibinfo {title} {Machine
  learning for electronically excited states of molecules},}\ }\href {\doibase
  10.1021/acs.chemrev.0c00749} {\bibfield  {journal} {\bibinfo  {journal}
  {Chem. Rev.}\ }\textbf {\bibinfo {volume} {121}},\ \bibinfo {pages}
  {9873--9926} (\bibinfo {year} {2020})}\BibitemShut {NoStop}%
\bibitem [{\citenamefont {Dral}\ \emph {et~al.}(2018)\citenamefont {Dral},
  \citenamefont {Barbatti},\ and\ \citenamefont {Thiel}}]{dral+18jpcl}%
  \BibitemOpen
  \bibfield  {author} {\bibinfo {author} {\bibfnamefont {Pavlo~O.}\
  \bibnamefont {Dral}}, \bibinfo {author} {\bibfnamefont {Mario}\ \bibnamefont
  {Barbatti}}, \ and\ \bibinfo {author} {\bibfnamefont {Walter}\ \bibnamefont
  {Thiel}},\ }\bibfield  {title} {\enquote {\bibinfo {title} {Nonadiabatic
  {{Excited-State Dynamics}} with {{Machine Learning}}},}\ }\href {\doibase
  10.1021/acs.jpclett.8b02469} {\bibfield  {journal} {\bibinfo  {journal} {The
  Journal of Physical Chemistry Letters}\ }\textbf {\bibinfo {volume} {9}},\
  \bibinfo {pages} {5660--5663} (\bibinfo {year} {2018})}\BibitemShut {NoStop}%
\bibitem [{\citenamefont {Chen}\ \emph {et~al.}(2020)\citenamefont {Chen},
  \citenamefont {Zuehlsdorff}, \citenamefont {Morawietz}, \citenamefont
  {Isborn},\ and\ \citenamefont {Markland}}]{chen_exploiting_2020}%
  \BibitemOpen
  \bibfield  {author} {\bibinfo {author} {\bibfnamefont {Michael~S.}\
  \bibnamefont {Chen}}, \bibinfo {author} {\bibfnamefont {Tim~J.}\ \bibnamefont
  {Zuehlsdorff}}, \bibinfo {author} {\bibfnamefont {Tobias}\ \bibnamefont
  {Morawietz}}, \bibinfo {author} {\bibfnamefont {Christine~M.}\ \bibnamefont
  {Isborn}}, \ and\ \bibinfo {author} {\bibfnamefont {Thomas~E.}\ \bibnamefont
  {Markland}},\ }\bibfield  {title} {\enquote {\bibinfo {title} {Exploiting
  {Machine} {Learning} to {Efficiently} {Predict} {Multidimensional} {Optical}
  {Spectra} in {Complex} {Environments}},}\ }\href {\doibase
  10.1021/acs.jpclett.0c02168} {\bibfield  {journal} {\bibinfo  {journal} {J.
  Phys. Chem. Lett.}\ }\textbf {\bibinfo {volume} {11}},\ \bibinfo {pages}
  {7559--7568} (\bibinfo {year} {2020})}\BibitemShut {NoStop}%
\bibitem [{\citenamefont {Szabo}\ and\ \citenamefont
  {Ostlund}(1996)}]{szabo_MQC}%
  \BibitemOpen
  \bibfield  {author} {\bibinfo {author} {\bibfnamefont {Attila}\ \bibnamefont
  {Szabo}}\ and\ \bibinfo {author} {\bibfnamefont {Neil~S.}\ \bibnamefont
  {Ostlund}},\ }\href@noop {} {\emph {\bibinfo {title} {Modern Quantum
  Chemistry: Introduction to Advanced Electronic Structure Theory}}}\ (\bibinfo
   {publisher} {Dover Publications, Inc.},\ \bibinfo {year} {1996})\BibitemShut
  {NoStop}%
\bibitem [{\citenamefont {Becke}(1993)}]{Becke1993}%
  \BibitemOpen
  \bibfield  {author} {\bibinfo {author} {\bibfnamefont {Axel~D.}\ \bibnamefont
  {Becke}},\ }\bibfield  {title} {\enquote {\bibinfo {title} {A new mixing of
  hartree{\textendash}fock and local density-functional theories},}\ }\href
  {\doibase 10.1063/1.464304} {\bibfield  {journal} {\bibinfo  {journal} {The
  Journal of Chemical Physics}\ }\textbf {\bibinfo {volume} {98}},\ \bibinfo
  {pages} {1372--1377} (\bibinfo {year} {1993})}\BibitemShut {NoStop}%
\bibitem [{\citenamefont {Baird}\ and\ \citenamefont
  {Dewar}(1969)}]{Baird1969}%
  \BibitemOpen
  \bibfield  {author} {\bibinfo {author} {\bibfnamefont {N.~Colin}\
  \bibnamefont {Baird}}\ and\ \bibinfo {author} {\bibfnamefont {Michael J.~S.}\
  \bibnamefont {Dewar}},\ }\bibfield  {title} {\enquote {\bibinfo {title}
  {Ground states of $\sigma$-bonded molecules. {IV}. the {MINDO} method and its
  application to hydrocarbons},}\ }\href {\doibase 10.1063/1.1671186}
  {\bibfield  {journal} {\bibinfo  {journal} {The Journal of Chemical Physics}\
  }\textbf {\bibinfo {volume} {50}},\ \bibinfo {pages} {1262--1274} (\bibinfo
  {year} {1969})}\BibitemShut {NoStop}%
\bibitem [{\citenamefont {Dewar}\ and\ \citenamefont
  {Thiel}(1977)}]{Dewar1977}%
  \BibitemOpen
  \bibfield  {author} {\bibinfo {author} {\bibfnamefont {Michael~JS}\
  \bibnamefont {Dewar}}\ and\ \bibinfo {author} {\bibfnamefont {Walter}\
  \bibnamefont {Thiel}},\ }\bibfield  {title} {\enquote {\bibinfo {title}
  {Ground states of molecules. 38. the mndo method. approximations and
  parameters},}\ }\href@noop {} {\bibfield  {journal} {\bibinfo  {journal}
  {Journal of the American Chemical Society}\ }\textbf {\bibinfo {volume}
  {99}},\ \bibinfo {pages} {4899--4907} (\bibinfo {year} {1977})}\BibitemShut
  {NoStop}%
\bibitem [{\citenamefont {Stewart}(2012)}]{Stewart2012}%
  \BibitemOpen
  \bibfield  {author} {\bibinfo {author} {\bibfnamefont {James J.~P.}\
  \bibnamefont {Stewart}},\ }\bibfield  {title} {\enquote {\bibinfo {title}
  {Optimization of parameters for semiempirical methods {VI}: more
  modifications to the {NDDO} approximations and re-optimization of
  parameters},}\ }\href {\doibase 10.1007/s00894-012-1667-x} {\bibfield
  {journal} {\bibinfo  {journal} {Journal of Molecular Modeling}\ }\textbf
  {\bibinfo {volume} {19}},\ \bibinfo {pages} {1--32} (\bibinfo {year}
  {2012})}\BibitemShut {NoStop}%
\bibitem [{\citenamefont {Watson}\ and\ \citenamefont
  {Chan}(2016)}]{wats-chan16jctc}%
  \BibitemOpen
  \bibfield  {author} {\bibinfo {author} {\bibfnamefont {Thomas~J.}\
  \bibnamefont {Watson}}\ and\ \bibinfo {author} {\bibfnamefont {Garnet
  Kin-Lic}\ \bibnamefont {Chan}},\ }\bibfield  {title} {\enquote {\bibinfo
  {title} {Correct {{Quantum Chemistry}} in a {{Minimal Basis}} from
  {{Effective Hamiltonians}}},}\ }\href {\doibase 10.1021/acs.jctc.5b00138}
  {\bibfield  {journal} {\bibinfo  {journal} {Journal of Chemical Theory and
  Computation}\ }\textbf {\bibinfo {volume} {12}},\ \bibinfo {pages} {512--522}
  (\bibinfo {year} {2016})}\BibitemShut {NoStop}%
\bibitem [{\citenamefont {Sch{\"u}tt}\ and\ \citenamefont
  {VandeVondele}(2018)}]{schu-vand18jctc}%
  \BibitemOpen
  \bibfield  {author} {\bibinfo {author} {\bibfnamefont {Ole}\ \bibnamefont
  {Sch{\"u}tt}}\ and\ \bibinfo {author} {\bibfnamefont {Joost}\ \bibnamefont
  {VandeVondele}},\ }\bibfield  {title} {\enquote {\bibinfo {title} {Machine
  {{Learning Adaptive Basis Sets}} for {{Efficient Large Scale Density
  Functional Theory Simulation}}},}\ }\href {\doibase 10.1021/acs.jctc.8b00378}
  {\bibfield  {journal} {\bibinfo  {journal} {Journal of Chemical Theory and
  Computation}\ }\textbf {\bibinfo {volume} {14}},\ \bibinfo {pages}
  {4168--4175} (\bibinfo {year} {2018})}\BibitemShut {NoStop}%
\bibitem [{\citenamefont {Changlani}\ \emph {et~al.}(2015)\citenamefont
  {Changlani}, \citenamefont {Zheng},\ and\ \citenamefont
  {Wagner}}]{chan+15jcp}%
  \BibitemOpen
  \bibfield  {author} {\bibinfo {author} {\bibfnamefont {Hitesh~J.}\
  \bibnamefont {Changlani}}, \bibinfo {author} {\bibfnamefont {Huihuo}\
  \bibnamefont {Zheng}}, \ and\ \bibinfo {author} {\bibfnamefont {Lucas~K.}\
  \bibnamefont {Wagner}},\ }\bibfield  {title} {\enquote {\bibinfo {title}
  {Density-matrix based determination of low-energy model {{Hamiltonians}} from
  {\emph{ab initio}} wavefunctions},}\ }\href {\doibase 10.1063/1.4927664}
  {\bibfield  {journal} {\bibinfo  {journal} {The Journal of Chemical Physics}\
  }\textbf {\bibinfo {volume} {143}},\ \bibinfo {pages} {102814} (\bibinfo
  {year} {2015})}\BibitemShut {NoStop}%
\bibitem [{\citenamefont {Fedik}\ \emph {et~al.}(2023)\citenamefont {Fedik},
  \citenamefont {Nebgen}, \citenamefont {Lubbers}, \citenamefont {Barros},
  \citenamefont {Kulichenko}, \citenamefont {Li}, \citenamefont {Zubatyuk},
  \citenamefont {Messerly}, \citenamefont {Isayev},\ and\ \citenamefont
  {Tretiak}}]{Fedik2023}%
  \BibitemOpen
  \bibfield  {author} {\bibinfo {author} {\bibfnamefont {Nikita}\ \bibnamefont
  {Fedik}}, \bibinfo {author} {\bibfnamefont {Benjamin}\ \bibnamefont
  {Nebgen}}, \bibinfo {author} {\bibfnamefont {Nicholas}\ \bibnamefont
  {Lubbers}}, \bibinfo {author} {\bibfnamefont {Kipton}\ \bibnamefont
  {Barros}}, \bibinfo {author} {\bibfnamefont {Maksim}\ \bibnamefont
  {Kulichenko}}, \bibinfo {author} {\bibfnamefont {Ying~Wai}\ \bibnamefont
  {Li}}, \bibinfo {author} {\bibfnamefont {Roman}\ \bibnamefont {Zubatyuk}},
  \bibinfo {author} {\bibfnamefont {Richard}\ \bibnamefont {Messerly}},
  \bibinfo {author} {\bibfnamefont {Olexandr}\ \bibnamefont {Isayev}}, \ and\
  \bibinfo {author} {\bibfnamefont {Sergei}\ \bibnamefont {Tretiak}},\
  }\bibfield  {title} {\enquote {\bibinfo {title} {Synergy of semiempirical
  models and machine learning in computational chemistry},}\ }\href {\doibase
  10.1063/5.0151833} {\bibfield  {journal} {\bibinfo  {journal} {The Journal of
  Chemical Physics}\ }\textbf {\bibinfo {volume} {159}} (\bibinfo {year}
  {2023}),\ 10.1063/5.0151833}\BibitemShut {NoStop}%
\bibitem [{\citenamefont {Nigam}\ \emph
  {et~al.}(2022{\natexlab{a}})\citenamefont {Nigam}, \citenamefont {Willatt},\
  and\ \citenamefont {Ceriotti}}]{niga+22jcp}%
  \BibitemOpen
  \bibfield  {author} {\bibinfo {author} {\bibfnamefont {Jigyasa}\ \bibnamefont
  {Nigam}}, \bibinfo {author} {\bibfnamefont {Michael~J.}\ \bibnamefont
  {Willatt}}, \ and\ \bibinfo {author} {\bibfnamefont {Michele}\ \bibnamefont
  {Ceriotti}},\ }\bibfield  {title} {\enquote {\bibinfo {title} {Equivariant
  representations for molecular {{Hamiltonians}} and {{{\emph{N}}}} -center
  atomic-scale properties},}\ }\href {\doibase 10.1063/5.0072784} {\bibfield
  {journal} {\bibinfo  {journal} {The Journal of Chemical Physics}\ }\textbf
  {\bibinfo {volume} {156}},\ \bibinfo {pages} {014115} (\bibinfo {year}
  {2022}{\natexlab{a}})}\BibitemShut {NoStop}%
\bibitem [{\citenamefont {Sch{\"u}tt}\ \emph {et~al.}(2019)\citenamefont
  {Sch{\"u}tt}, \citenamefont {Gastegger}, \citenamefont {Tkatchenko},
  \citenamefont {M{\"u}ller},\ and\ \citenamefont {Maurer}}]{schu+19nc}%
  \BibitemOpen
  \bibfield  {author} {\bibinfo {author} {\bibfnamefont {K.~T.}\ \bibnamefont
  {Sch{\"u}tt}}, \bibinfo {author} {\bibfnamefont {M.}~\bibnamefont
  {Gastegger}}, \bibinfo {author} {\bibfnamefont {A.}~\bibnamefont
  {Tkatchenko}}, \bibinfo {author} {\bibfnamefont {K.-R.}\ \bibnamefont
  {M{\"u}ller}}, \ and\ \bibinfo {author} {\bibfnamefont {R.~J.}\ \bibnamefont
  {Maurer}},\ }\bibfield  {title} {\enquote {\bibinfo {title} {Unifying machine
  learning and quantum chemistry with a deep neural network for molecular
  wavefunctions},}\ }\href {\doibase 10.1038/s41467-019-12875-2} {\bibfield
  {journal} {\bibinfo  {journal} {Nature Communications}\ }\textbf {\bibinfo
  {volume} {10}},\ \bibinfo {pages} {5024} (\bibinfo {year}
  {2019})}\BibitemShut {NoStop}%
\bibitem [{\citenamefont {Hegde}\ and\ \citenamefont
  {Bowen}(2017)}]{hegd-bowe17sr}%
  \BibitemOpen
  \bibfield  {author} {\bibinfo {author} {\bibfnamefont {Ganesh}\ \bibnamefont
  {Hegde}}\ and\ \bibinfo {author} {\bibfnamefont {R.~Chris}\ \bibnamefont
  {Bowen}},\ }\bibfield  {title} {\enquote {\bibinfo {title} {Machine-learned
  approximations to {{Density Functional Theory Hamiltonians}}},}\ }\href
  {\doibase 10.1038/srep42669} {\bibfield  {journal} {\bibinfo  {journal}
  {Scientific Reports}\ }\textbf {\bibinfo {volume} {7}},\ \bibinfo {pages}
  {42669} (\bibinfo {year} {2017})}\BibitemShut {NoStop}%
\bibitem [{\citenamefont {Li}\ \emph {et~al.}(2022)\citenamefont {Li},
  \citenamefont {Wang}, \citenamefont {Zou}, \citenamefont {Ye}, \citenamefont
  {Xu}, \citenamefont {Gong}, \citenamefont {Duan},\ and\ \citenamefont
  {Xu}}]{li2022deep}%
  \BibitemOpen
  \bibfield  {author} {\bibinfo {author} {\bibfnamefont {He}~\bibnamefont
  {Li}}, \bibinfo {author} {\bibfnamefont {Zun}\ \bibnamefont {Wang}}, \bibinfo
  {author} {\bibfnamefont {Nianlong}\ \bibnamefont {Zou}}, \bibinfo {author}
  {\bibfnamefont {Meng}\ \bibnamefont {Ye}}, \bibinfo {author} {\bibfnamefont
  {Runzhang}\ \bibnamefont {Xu}}, \bibinfo {author} {\bibfnamefont {Xiaoxun}\
  \bibnamefont {Gong}}, \bibinfo {author} {\bibfnamefont {Wenhui}\ \bibnamefont
  {Duan}}, \ and\ \bibinfo {author} {\bibfnamefont {Yong}\ \bibnamefont {Xu}},\
  }\bibfield  {title} {\enquote {\bibinfo {title} {Deep-learning density
  functional theory hamiltonian for efficient ab initio electronic-structure
  calculation},}\ }\href@noop {} {\bibfield  {journal} {\bibinfo  {journal}
  {Nature Computational Science}\ }\textbf {\bibinfo {volume} {2}},\ \bibinfo
  {pages} {367--377} (\bibinfo {year} {2022})}\BibitemShut {NoStop}%
\bibitem [{\citenamefont {Gong}\ \emph {et~al.}(2023)\citenamefont {Gong},
  \citenamefont {Li}, \citenamefont {Zou}, \citenamefont {Xu}, \citenamefont
  {Duan},\ and\ \citenamefont {Xu}}]{deephe3}%
  \BibitemOpen
  \bibfield  {author} {\bibinfo {author} {\bibfnamefont {Xiaoxun}\ \bibnamefont
  {Gong}}, \bibinfo {author} {\bibfnamefont {He}~\bibnamefont {Li}}, \bibinfo
  {author} {\bibfnamefont {Nianlong}\ \bibnamefont {Zou}}, \bibinfo {author}
  {\bibfnamefont {Runzhang}\ \bibnamefont {Xu}}, \bibinfo {author}
  {\bibfnamefont {Wenhui}\ \bibnamefont {Duan}}, \ and\ \bibinfo {author}
  {\bibfnamefont {Yong}\ \bibnamefont {Xu}},\ }\bibfield  {title} {\enquote
  {\bibinfo {title} {General framework for e(3)-equivariant neural network
  representation of density functional theory hamiltonian},}\ }\href {\doibase
  10.1038/s41467-023-38468-8} {\bibfield  {journal} {\bibinfo  {journal}
  {Nature Communications}\ }\textbf {\bibinfo {volume} {14}},\ \bibinfo {pages}
  {2848} (\bibinfo {year} {2023})}\BibitemShut {NoStop}%
\bibitem [{\citenamefont {Yu}\ \emph {et~al.}(2023)\citenamefont {Yu},
  \citenamefont {Xu}, \citenamefont {Qian}, \citenamefont {Qian},\ and\
  \citenamefont {Ji}}]{yu2023efficient}%
  \BibitemOpen
  \bibfield  {author} {\bibinfo {author} {\bibfnamefont {Haiyang}\ \bibnamefont
  {Yu}}, \bibinfo {author} {\bibfnamefont {Zhao}\ \bibnamefont {Xu}}, \bibinfo
  {author} {\bibfnamefont {Xiaofeng}\ \bibnamefont {Qian}}, \bibinfo {author}
  {\bibfnamefont {Xiaoning}\ \bibnamefont {Qian}}, \ and\ \bibinfo {author}
  {\bibfnamefont {Shuiwang}\ \bibnamefont {Ji}},\ }\bibfield  {title} {\enquote
  {\bibinfo {title} {Efficient and equivariant graph networks for predicting
  quantum hamiltonian},}\ }\href@noop {} {\bibfield  {journal} {\bibinfo
  {journal} {arXiv preprint arXiv:2306.04922}\ } (\bibinfo {year}
  {2023})}\BibitemShut {NoStop}%
\bibitem [{\citenamefont {Zhang}\ \emph
  {et~al.}(2022{\natexlab{a}})\citenamefont {Zhang}, \citenamefont {Onat},
  \citenamefont {Dusson}, \citenamefont {McSloy}, \citenamefont {Anand},
  \citenamefont {Maurer}, \citenamefont {Ortner},\ and\ \citenamefont
  {Kermode}}]{zhang2022equivariant}%
  \BibitemOpen
  \bibfield  {author} {\bibinfo {author} {\bibfnamefont {Liwei}\ \bibnamefont
  {Zhang}}, \bibinfo {author} {\bibfnamefont {Berk}\ \bibnamefont {Onat}},
  \bibinfo {author} {\bibfnamefont {Genevi{\`e}ve}\ \bibnamefont {Dusson}},
  \bibinfo {author} {\bibfnamefont {Adam}\ \bibnamefont {McSloy}}, \bibinfo
  {author} {\bibfnamefont {Gautam}\ \bibnamefont {Anand}}, \bibinfo {author}
  {\bibfnamefont {Reinhard~J}\ \bibnamefont {Maurer}}, \bibinfo {author}
  {\bibfnamefont {Christoph}\ \bibnamefont {Ortner}}, \ and\ \bibinfo {author}
  {\bibfnamefont {James~R}\ \bibnamefont {Kermode}},\ }\bibfield  {title}
  {\enquote {\bibinfo {title} {Equivariant analytical mapping of first
  principles hamiltonians to accurate and transferable materials models},}\
  }\href@noop {} {\bibfield  {journal} {\bibinfo  {journal} {Npj Computational
  Materials}\ }\textbf {\bibinfo {volume} {8}},\ \bibinfo {pages} {158}
  (\bibinfo {year} {2022}{\natexlab{a}})}\BibitemShut {NoStop}%
\bibitem [{\citenamefont {Unke}\ \emph {et~al.}(2021)\citenamefont {Unke},
  \citenamefont {Bogojeski}, \citenamefont {Gastegger}, \citenamefont {Geiger},
  \citenamefont {Smidt},\ and\ \citenamefont {M{\"u}ller}}]{unke+21nips}%
  \BibitemOpen
  \bibfield  {author} {\bibinfo {author} {\bibfnamefont {Oliver}\ \bibnamefont
  {Unke}}, \bibinfo {author} {\bibfnamefont {Mihail}\ \bibnamefont
  {Bogojeski}}, \bibinfo {author} {\bibfnamefont {Michael}\ \bibnamefont
  {Gastegger}}, \bibinfo {author} {\bibfnamefont {Mario}\ \bibnamefont
  {Geiger}}, \bibinfo {author} {\bibfnamefont {Tess}\ \bibnamefont {Smidt}}, \
  and\ \bibinfo {author} {\bibfnamefont {Klaus-Robert}\ \bibnamefont
  {M{\"u}ller}},\ }\bibfield  {title} {\enquote {\bibinfo {title} {{{SE}}
  (3)-equivariant prediction of molecular wavefunctions and electronic
  densities},}\ }\href@noop {} {\bibfield  {journal} {\bibinfo  {journal}
  {Advances in Neural Information Processing Systems}\ }\textbf {\bibinfo
  {volume} {34}} (\bibinfo {year} {2021})}\BibitemShut {NoStop}%
\bibitem [{\citenamefont {Zhang}\ \emph
  {et~al.}(2022{\natexlab{b}})\citenamefont {Zhang}, \citenamefont {Onat},
  \citenamefont {Dusson}, \citenamefont {McSloy}, \citenamefont {Anand},
  \citenamefont {Maurer}, \citenamefont {Ortner},\ and\ \citenamefont
  {Kermode}}]{zhan+22npjcm}%
  \BibitemOpen
  \bibfield  {author} {\bibinfo {author} {\bibfnamefont {Liwei}\ \bibnamefont
  {Zhang}}, \bibinfo {author} {\bibfnamefont {Berk}\ \bibnamefont {Onat}},
  \bibinfo {author} {\bibfnamefont {Genevi{\`e}ve}\ \bibnamefont {Dusson}},
  \bibinfo {author} {\bibfnamefont {Adam}\ \bibnamefont {McSloy}}, \bibinfo
  {author} {\bibfnamefont {G.}~\bibnamefont {Anand}}, \bibinfo {author}
  {\bibfnamefont {Reinhard~J.}\ \bibnamefont {Maurer}}, \bibinfo {author}
  {\bibfnamefont {Christoph}\ \bibnamefont {Ortner}}, \ and\ \bibinfo {author}
  {\bibfnamefont {James~R.}\ \bibnamefont {Kermode}},\ }\bibfield  {title}
  {\enquote {\bibinfo {title} {Equivariant analytical mapping of first
  principles {{Hamiltonians}} to accurate and transferable materials models},}\
  }\href {\doibase 10.1038/s41524-022-00843-2} {\bibfield  {journal} {\bibinfo
  {journal} {npj Computational Materials}\ }\textbf {\bibinfo {volume} {8}},\
  \bibinfo {pages} {158} (\bibinfo {year} {2022}{\natexlab{b}})}\BibitemShut
  {NoStop}%
\bibitem [{\citenamefont {Ramakrishnan}\ \emph {et~al.}(2015)\citenamefont
  {Ramakrishnan}, \citenamefont {Hartmann}, \citenamefont {Tapavicza},\ and\
  \citenamefont {Von~Lilienfeld}}]{ramakrishnan2015electronic}%
  \BibitemOpen
  \bibfield  {author} {\bibinfo {author} {\bibfnamefont {Raghunathan}\
  \bibnamefont {Ramakrishnan}}, \bibinfo {author} {\bibfnamefont {Mia}\
  \bibnamefont {Hartmann}}, \bibinfo {author} {\bibfnamefont {Enrico}\
  \bibnamefont {Tapavicza}}, \ and\ \bibinfo {author} {\bibfnamefont
  {O~Anatole}\ \bibnamefont {Von~Lilienfeld}},\ }\bibfield  {title} {\enquote
  {\bibinfo {title} {Electronic spectra from tddft and machine learning in
  chemical space},}\ }\href@noop {} {\bibfield  {journal} {\bibinfo  {journal}
  {The Journal of chemical physics}\ }\textbf {\bibinfo {volume} {143}}
  (\bibinfo {year} {2015})}\BibitemShut {NoStop}%
\bibitem [{\citenamefont {Westermayr}\ \emph {et~al.}(2020)\citenamefont
  {Westermayr}, \citenamefont {Faber}, \citenamefont {Christensen},
  \citenamefont {von Lilienfeld},\ and\ \citenamefont
  {Marquetand}}]{westermayr2020neural}%
  \BibitemOpen
  \bibfield  {author} {\bibinfo {author} {\bibfnamefont {Julia}\ \bibnamefont
  {Westermayr}}, \bibinfo {author} {\bibfnamefont {Felix~A}\ \bibnamefont
  {Faber}}, \bibinfo {author} {\bibfnamefont {Anders~S}\ \bibnamefont
  {Christensen}}, \bibinfo {author} {\bibfnamefont {O~Anatole}\ \bibnamefont
  {von Lilienfeld}}, \ and\ \bibinfo {author} {\bibfnamefont {Philipp}\
  \bibnamefont {Marquetand}},\ }\bibfield  {title} {\enquote {\bibinfo {title}
  {Neural networks and kernel ridge regression for excited states dynamics of
  ch2nh: From single-state to multi-state representations and multi-property
  machine learning models},}\ }\href@noop {} {\bibfield  {journal} {\bibinfo
  {journal} {Machine Learning: Science and Technology}\ }\textbf {\bibinfo
  {volume} {1}},\ \bibinfo {pages} {025009} (\bibinfo {year}
  {2020})}\BibitemShut {NoStop}%
\bibitem [{\citenamefont {Mazouin}\ \emph {et~al.}(2022)\citenamefont
  {Mazouin}, \citenamefont {Sch{\"o}pfer},\ and\ \citenamefont {von
  Lilienfeld}}]{mazouin2022selected}%
  \BibitemOpen
  \bibfield  {author} {\bibinfo {author} {\bibfnamefont {Bernard}\ \bibnamefont
  {Mazouin}}, \bibinfo {author} {\bibfnamefont {Alexandre~Alain}\ \bibnamefont
  {Sch{\"o}pfer}}, \ and\ \bibinfo {author} {\bibfnamefont {O~Anatole}\
  \bibnamefont {von Lilienfeld}},\ }\bibfield  {title} {\enquote {\bibinfo
  {title} {Selected machine learning of homo--lumo gaps with improved
  data-efficiency},}\ }\href@noop {} {\bibfield  {journal} {\bibinfo  {journal}
  {Materials Advances}\ }\textbf {\bibinfo {volume} {3}},\ \bibinfo {pages}
  {8306--8316} (\bibinfo {year} {2022})}\BibitemShut {NoStop}%
\bibitem [{\citenamefont {Grimme}(2013)}]{Grimme2013}%
  \BibitemOpen
  \bibfield  {author} {\bibinfo {author} {\bibfnamefont {Stefan}\ \bibnamefont
  {Grimme}},\ }\bibfield  {title} {\enquote {\bibinfo {title} {A simplified
  tamm-dancoff density functional approach for the electronic excitation
  spectra of very large molecules},}\ }\href {\doibase 10.1063/1.4811331}
  {\bibfield  {journal} {\bibinfo  {journal} {The Journal of Chemical Physics}\
  }\textbf {\bibinfo {volume} {138}} (\bibinfo {year} {2013}),\
  10.1063/1.4811331}\BibitemShut {NoStop}%
\bibitem [{\citenamefont {Bannwarth}\ and\ \citenamefont
  {Grimme}(2014)}]{Bannwarth2014}%
  \BibitemOpen
  \bibfield  {author} {\bibinfo {author} {\bibfnamefont {Christoph}\
  \bibnamefont {Bannwarth}}\ and\ \bibinfo {author} {\bibfnamefont {Stefan}\
  \bibnamefont {Grimme}},\ }\bibfield  {title} {\enquote {\bibinfo {title} {A
  simplified time-dependent density functional theory approach for electronic
  ultraviolet and circular dichroism spectra of very large molecules},}\ }\href
  {\doibase 10.1016/j.comptc.2014.02.023} {\bibfield  {journal} {\bibinfo
  {journal} {Computational and Theoretical Chemistry}\ }\textbf {\bibinfo
  {volume} {1040-1041}},\ \bibinfo {pages} {45--53} (\bibinfo {year}
  {2014})}\BibitemShut {NoStop}%
\bibitem [{\citenamefont {Westermayr}\ and\ \citenamefont
  {Maurer}(2021)}]{west-maur21cs}%
  \BibitemOpen
  \bibfield  {author} {\bibinfo {author} {\bibfnamefont {Julia}\ \bibnamefont
  {Westermayr}}\ and\ \bibinfo {author} {\bibfnamefont {Reinhard~J.}\
  \bibnamefont {Maurer}},\ }\bibfield  {title} {\enquote {\bibinfo {title}
  {Physically inspired deep learning of molecular excitations and photoemission
  spectra},}\ }\href {\doibase 10.1039/D1SC01542G} {\bibfield  {journal}
  {\bibinfo  {journal} {Chemical Science}\ }\textbf {\bibinfo {volume} {12}},\
  \bibinfo {pages} {10755--10764} (\bibinfo {year} {2021})}\BibitemShut
  {NoStop}%
\bibitem [{\citenamefont {Goedecker}(1999)}]{goed99rmp}%
  \BibitemOpen
  \bibfield  {author} {\bibinfo {author} {\bibfnamefont {S}~\bibnamefont
  {Goedecker}},\ }\bibfield  {title} {\enquote {\bibinfo {title} {Linear
  scaling electronic structure methods},}\ }\href@noop {} {\bibfield  {journal}
  {\bibinfo  {journal} {Reviews of Modern Physics}\ }\textbf {\bibinfo {volume}
  {71}},\ \bibinfo {pages} {1085--1123} (\bibinfo {year} {1999})}\BibitemShut
  {NoStop}%
\bibitem [{\citenamefont {Qiao}\ \emph {et~al.}(2020)\citenamefont {Qiao},
  \citenamefont {Welborn}, \citenamefont {Anandkumar}, \citenamefont {Manby},\
  and\ \citenamefont {Miller}}]{qiao+20jcp}%
  \BibitemOpen
  \bibfield  {author} {\bibinfo {author} {\bibfnamefont {Zhuoran}\ \bibnamefont
  {Qiao}}, \bibinfo {author} {\bibfnamefont {Matthew}\ \bibnamefont {Welborn}},
  \bibinfo {author} {\bibfnamefont {Animashree}\ \bibnamefont {Anandkumar}},
  \bibinfo {author} {\bibfnamefont {Frederick~R.}\ \bibnamefont {Manby}}, \
  and\ \bibinfo {author} {\bibfnamefont {Thomas~F.}\ \bibnamefont {Miller}},\
  }\bibfield  {title} {\enquote {\bibinfo {title} {{{OrbNet}}: {{Deep}}
  learning for quantum chemistry using symmetry-adapted atomic-orbital
  features},}\ }\href {\doibase 10.1063/5.0021955} {\bibfield  {journal}
  {\bibinfo  {journal} {The Journal of Chemical Physics}\ }\textbf {\bibinfo
  {volume} {153}},\ \bibinfo {pages} {124111} (\bibinfo {year}
  {2020})}\BibitemShut {NoStop}%
\bibitem [{\citenamefont {Fabrizio}\ \emph {et~al.}(2022)\citenamefont
  {Fabrizio}, \citenamefont {Briling},\ and\ \citenamefont
  {Corminboeuf}}]{fabr+22dd}%
  \BibitemOpen
  \bibfield  {author} {\bibinfo {author} {\bibfnamefont {Alberto}\ \bibnamefont
  {Fabrizio}}, \bibinfo {author} {\bibfnamefont {Ksenia~R.}\ \bibnamefont
  {Briling}}, \ and\ \bibinfo {author} {\bibfnamefont {Clemence}\ \bibnamefont
  {Corminboeuf}},\ }\bibfield  {title} {\enquote {\bibinfo {title} {{{SPAHM}}:
  The spectrum of approximated {{Hamiltonian}} matrices representations},}\
  }\href {\doibase 10.1039/D1DD00050K} {\bibfield  {journal} {\bibinfo
  {journal} {Digital Discovery}\ ,\ \bibinfo {pages} {10.1039.D1DD00050K}}
  (\bibinfo {year} {2022})}\BibitemShut {NoStop}%
\bibitem [{\citenamefont {Prlj}\ \emph {et~al.}(2016)\citenamefont {Prlj},
  \citenamefont {Sandoval-Salinas}, \citenamefont {Casanova}, \citenamefont
  {Jacquemin},\ and\ \citenamefont {Corminboeuf}}]{Prlj2016}%
  \BibitemOpen
  \bibfield  {author} {\bibinfo {author} {\bibfnamefont {Antonio}\ \bibnamefont
  {Prlj}}, \bibinfo {author} {\bibfnamefont {Mar{\'{\i}}a~Eugenia}\
  \bibnamefont {Sandoval-Salinas}}, \bibinfo {author} {\bibfnamefont {David}\
  \bibnamefont {Casanova}}, \bibinfo {author} {\bibfnamefont {Denis}\
  \bibnamefont {Jacquemin}}, \ and\ \bibinfo {author} {\bibfnamefont
  {Cl{\'{e}}mence}\ \bibnamefont {Corminboeuf}},\ }\bibfield  {title} {\enquote
  {\bibinfo {title} {Low-lying $\pi\pi^{\ast}$ states of heteroaromatic
  molecules: A challenge for excited state methods},}\ }\href {\doibase
  10.1021/acs.jctc.6b00245} {\bibfield  {journal} {\bibinfo  {journal} {Journal
  of Chemical Theory and Computation}\ }\textbf {\bibinfo {volume} {12}},\
  \bibinfo {pages} {2652--2660} (\bibinfo {year} {2016})}\BibitemShut {NoStop}%
\bibitem [{\citenamefont {Taniguchi}\ and\ \citenamefont
  {Lindsey}(2018)}]{Taniguchi2018}%
  \BibitemOpen
  \bibfield  {author} {\bibinfo {author} {\bibfnamefont {Masahiko}\
  \bibnamefont {Taniguchi}}\ and\ \bibinfo {author} {\bibfnamefont
  {Jonathan~S.}\ \bibnamefont {Lindsey}},\ }\bibfield  {title} {\enquote
  {\bibinfo {title} {Database of absorption and fluorescence spectra of >300
  common compounds for use in photochemcad},}\ }\href {\doibase
  10.1111/php.12860} {\bibfield  {journal} {\bibinfo  {journal} {Photochemistry
  and Photobiology}\ }\textbf {\bibinfo {volume} {94}},\ \bibinfo {pages}
  {290--327} (\bibinfo {year} {2018})}\BibitemShut {NoStop}%
\bibitem [{\citenamefont {Wilkins}\ \emph {et~al.}(2019)\citenamefont
  {Wilkins}, \citenamefont {Grisafi}, \citenamefont {Yang}, \citenamefont
  {Lao}, \citenamefont {DiStasio},\ and\ \citenamefont
  {Ceriotti}}]{wilk+19pnas}%
  \BibitemOpen
  \bibfield  {author} {\bibinfo {author} {\bibfnamefont {David~M.}\
  \bibnamefont {Wilkins}}, \bibinfo {author} {\bibfnamefont {Andrea}\
  \bibnamefont {Grisafi}}, \bibinfo {author} {\bibfnamefont {Yang}\
  \bibnamefont {Yang}}, \bibinfo {author} {\bibfnamefont {Ka~Un}\ \bibnamefont
  {Lao}}, \bibinfo {author} {\bibfnamefont {Robert~A.}\ \bibnamefont
  {DiStasio}}, \ and\ \bibinfo {author} {\bibfnamefont {Michele}\ \bibnamefont
  {Ceriotti}},\ }\bibfield  {title} {\enquote {\bibinfo {title} {Accurate
  molecular polarizabilities with coupled cluster theory and machine
  learning},}\ }\href {\doibase 10.1073/pnas.1816132116} {\bibfield  {journal}
  {\bibinfo  {journal} {Proceedings of the National Academy of Sciences of the
  United States of America}\ }\textbf {\bibinfo {volume} {116}},\ \bibinfo
  {pages} {3401--3406} (\bibinfo {year} {2019})}\BibitemShut {NoStop}%
\bibitem [{\citenamefont {Dierksen}\ and\ \citenamefont
  {Grimme}(2004{\natexlab{a}})}]{dierksen2004density}%
  \BibitemOpen
  \bibfield  {author} {\bibinfo {author} {\bibfnamefont {Marc}\ \bibnamefont
  {Dierksen}}\ and\ \bibinfo {author} {\bibfnamefont {Stefan}\ \bibnamefont
  {Grimme}},\ }\bibfield  {title} {\enquote {\bibinfo {title} {Density
  functional calculations of the vibronic structure of electronic absorption
  spectra},}\ }\href@noop {} {\bibfield  {journal} {\bibinfo  {journal} {The
  Journal of chemical physics}\ }\textbf {\bibinfo {volume} {120}},\ \bibinfo
  {pages} {3544--3554} (\bibinfo {year} {2004}{\natexlab{a}})}\BibitemShut
  {NoStop}%
\bibitem [{\citenamefont {Dierksen}\ and\ \citenamefont
  {Grimme}(2004{\natexlab{b}})}]{dierksen2004vibronic}%
  \BibitemOpen
  \bibfield  {author} {\bibinfo {author} {\bibfnamefont {Marc}\ \bibnamefont
  {Dierksen}}\ and\ \bibinfo {author} {\bibfnamefont {Stefan}\ \bibnamefont
  {Grimme}},\ }\bibfield  {title} {\enquote {\bibinfo {title} {The vibronic
  structure of electronic absorption spectra of large molecules: a
  time-dependent density functional study on the influence of “exact”
  hartree- fock exchange},}\ }\href@noop {} {\bibfield  {journal} {\bibinfo
  {journal} {The Journal of Physical Chemistry A}\ }\textbf {\bibinfo {volume}
  {108}},\ \bibinfo {pages} {10225--10237} (\bibinfo {year}
  {2004}{\natexlab{b}})}\BibitemShut {NoStop}%
\bibitem [{\citenamefont {Valleau}\ \emph {et~al.}(2012)\citenamefont
  {Valleau}, \citenamefont {Eisfeld},\ and\ \citenamefont
  {Aspuru-Guzik}}]{Valleau2012}%
  \BibitemOpen
  \bibfield  {author} {\bibinfo {author} {\bibfnamefont {St{\'{e}}phanie}\
  \bibnamefont {Valleau}}, \bibinfo {author} {\bibfnamefont {Alexander}\
  \bibnamefont {Eisfeld}}, \ and\ \bibinfo {author} {\bibfnamefont
  {Al{\'{a}}n}\ \bibnamefont {Aspuru-Guzik}},\ }\bibfield  {title} {\enquote
  {\bibinfo {title} {On the alternatives for bath correlators and spectral
  densities from mixed quantum-classical simulations},}\ }\href {\doibase
  10.1063/1.4769079} {\bibfield  {journal} {\bibinfo  {journal} {The Journal of
  Chemical Physics}\ }\textbf {\bibinfo {volume} {137}} (\bibinfo {year}
  {2012}),\ 10.1063/1.4769079}\BibitemShut {NoStop}%
\bibitem [{\citenamefont {Chandrasekaran}\ \emph {et~al.}(2015)\citenamefont
  {Chandrasekaran}, \citenamefont {Aghtar}, \citenamefont {Valleau},
  \citenamefont {Aspuru-Guzik},\ and\ \citenamefont
  {Kleinekath\"{o}fer}}]{Chandrasekaran2015}%
  \BibitemOpen
  \bibfield  {author} {\bibinfo {author} {\bibfnamefont {Suryanarayanan}\
  \bibnamefont {Chandrasekaran}}, \bibinfo {author} {\bibfnamefont {Mortaza}\
  \bibnamefont {Aghtar}}, \bibinfo {author} {\bibfnamefont {St{\'{e}}phanie}\
  \bibnamefont {Valleau}}, \bibinfo {author} {\bibfnamefont {Al{\'{a}}n}\
  \bibnamefont {Aspuru-Guzik}}, \ and\ \bibinfo {author} {\bibfnamefont
  {Ulrich}\ \bibnamefont {Kleinekath\"{o}fer}},\ }\bibfield  {title} {\enquote
  {\bibinfo {title} {Influence of force fields and quantum chemistry approach
  on spectral densities of {BChl} \textit{a} in solution and in {FMO}
  proteins},}\ }\href {\doibase 10.1021/acs.jpcb.5b03654} {\bibfield  {journal}
  {\bibinfo  {journal} {The Journal of Physical Chemistry B}\ }\textbf
  {\bibinfo {volume} {119}},\ \bibinfo {pages} {9995--10004} (\bibinfo {year}
  {2015})}\BibitemShut {NoStop}%
\bibitem [{\citenamefont {Aghtar}\ \emph {et~al.}(2017)\citenamefont {Aghtar},
  \citenamefont {Kleinekath\"{o}fer}, \citenamefont {Curutchet},\ and\
  \citenamefont {Mennucci}}]{Aghtar2017}%
  \BibitemOpen
  \bibfield  {author} {\bibinfo {author} {\bibfnamefont {Mortaza}\ \bibnamefont
  {Aghtar}}, \bibinfo {author} {\bibfnamefont {Ulrich}\ \bibnamefont
  {Kleinekath\"{o}fer}}, \bibinfo {author} {\bibfnamefont {Carles}\
  \bibnamefont {Curutchet}}, \ and\ \bibinfo {author} {\bibfnamefont
  {Benedetta}\ \bibnamefont {Mennucci}},\ }\bibfield  {title} {\enquote
  {\bibinfo {title} {Impact of electronic fluctuations and their description on
  the exciton dynamics in the light-harvesting complex {PE}545},}\ }\href
  {\doibase 10.1021/acs.jpcb.6b10772} {\bibfield  {journal} {\bibinfo
  {journal} {The Journal of Physical Chemistry B}\ }\textbf {\bibinfo {volume}
  {121}},\ \bibinfo {pages} {1330--1339} (\bibinfo {year} {2017})}\BibitemShut
  {NoStop}%
\bibitem [{\citenamefont {Musil}\ \emph
  {et~al.}(2021{\natexlab{a}})\citenamefont {Musil}, \citenamefont {Grisafi},
  \citenamefont {Bart{\'o}k}, \citenamefont {Ortner}, \citenamefont
  {Cs{\'a}nyi},\ and\ \citenamefont {Ceriotti}}]{musi+21cr}%
  \BibitemOpen
  \bibfield  {author} {\bibinfo {author} {\bibfnamefont {Felix}\ \bibnamefont
  {Musil}}, \bibinfo {author} {\bibfnamefont {Andrea}\ \bibnamefont {Grisafi}},
  \bibinfo {author} {\bibfnamefont {Albert~P.}\ \bibnamefont {Bart{\'o}k}},
  \bibinfo {author} {\bibfnamefont {Christoph}\ \bibnamefont {Ortner}},
  \bibinfo {author} {\bibfnamefont {G{\'a}bor}\ \bibnamefont {Cs{\'a}nyi}}, \
  and\ \bibinfo {author} {\bibfnamefont {Michele}\ \bibnamefont {Ceriotti}},\
  }\bibfield  {title} {\enquote {\bibinfo {title} {Physics-{{Inspired
  Structural Representations}} for {{Molecules}} and {{Materials}}},}\ }\href
  {\doibase 10.1021/acs.chemrev.1c00021} {\bibfield  {journal} {\bibinfo
  {journal} {Chemical Reviews}\ }\textbf {\bibinfo {volume} {121}},\ \bibinfo
  {pages} {9759--9815} (\bibinfo {year} {2021}{\natexlab{a}})}\BibitemShut
  {NoStop}%
\bibitem [{\citenamefont {Batatia}\ \emph {et~al.}(2022)\citenamefont
  {Batatia}, \citenamefont {Batzner}, \citenamefont {Kov{\'a}cs}, \citenamefont
  {Musaelian}, \citenamefont {Simm}, \citenamefont {Drautz}, \citenamefont
  {Ortner}, \citenamefont {Kozinsky},\ and\ \citenamefont
  {Cs{\'a}nyi}}]{bata+22arxiv}%
  \BibitemOpen
  \bibfield  {author} {\bibinfo {author} {\bibfnamefont {Ilyes}\ \bibnamefont
  {Batatia}}, \bibinfo {author} {\bibfnamefont {Simon}\ \bibnamefont
  {Batzner}}, \bibinfo {author} {\bibfnamefont {D{\'a}vid~P{\'e}ter}\
  \bibnamefont {Kov{\'a}cs}}, \bibinfo {author} {\bibfnamefont {Albert}\
  \bibnamefont {Musaelian}}, \bibinfo {author} {\bibfnamefont {Gregor N.~C.}\
  \bibnamefont {Simm}}, \bibinfo {author} {\bibfnamefont {Ralf}\ \bibnamefont
  {Drautz}}, \bibinfo {author} {\bibfnamefont {Christoph}\ \bibnamefont
  {Ortner}}, \bibinfo {author} {\bibfnamefont {Boris}\ \bibnamefont
  {Kozinsky}}, \ and\ \bibinfo {author} {\bibfnamefont {G{\'a}bor}\
  \bibnamefont {Cs{\'a}nyi}},\ }\bibfield  {title} {\enquote {\bibinfo {title}
  {The design space of {{E}}(3)-equivariant atom-centered interatomic
  potentials},}\ }\href@noop {} {\bibfield  {journal} {\bibinfo  {journal}
  {arxiv:2205.06643}\ } (\bibinfo {year} {2022})}\BibitemShut {NoStop}%
\bibitem [{\citenamefont {Kapil}\ \emph {et~al.}(2019)\citenamefont {Kapil},
  \citenamefont {Rossi}, \citenamefont {Marsalek}, \citenamefont {Petraglia},
  \citenamefont {Litman}, \citenamefont {Spura}, \citenamefont {Cheng},
  \citenamefont {Cuzzocrea}, \citenamefont {Mei{\ss}ner}, \citenamefont
  {Wilkins}, \citenamefont {Helfrecht}, \citenamefont {Juda}, \citenamefont
  {Bienvenue}, \citenamefont {Fang}, \citenamefont {Kessler}, \citenamefont
  {Poltavsky}, \citenamefont {Vandenbrande}, \citenamefont {Wieme},
  \citenamefont {Corminboeuf}, \citenamefont {K{\"u}hne}, \citenamefont
  {Manolopoulos}, \citenamefont {Markland}, \citenamefont {Richardson},
  \citenamefont {Tkatchenko}, \citenamefont {Tribello}, \citenamefont
  {Van~Speybroeck},\ and\ \citenamefont {Ceriotti}}]{kapi+19cpc}%
  \BibitemOpen
  \bibfield  {author} {\bibinfo {author} {\bibfnamefont {Venkat}\ \bibnamefont
  {Kapil}}, \bibinfo {author} {\bibfnamefont {Mariana}\ \bibnamefont {Rossi}},
  \bibinfo {author} {\bibfnamefont {Ondrej}\ \bibnamefont {Marsalek}}, \bibinfo
  {author} {\bibfnamefont {Riccardo}\ \bibnamefont {Petraglia}}, \bibinfo
  {author} {\bibfnamefont {Yair}\ \bibnamefont {Litman}}, \bibinfo {author}
  {\bibfnamefont {Thomas}\ \bibnamefont {Spura}}, \bibinfo {author}
  {\bibfnamefont {Bingqing}\ \bibnamefont {Cheng}}, \bibinfo {author}
  {\bibfnamefont {Alice}\ \bibnamefont {Cuzzocrea}}, \bibinfo {author}
  {\bibfnamefont {Robert~H.}\ \bibnamefont {Mei{\ss}ner}}, \bibinfo {author}
  {\bibfnamefont {David~M.}\ \bibnamefont {Wilkins}}, \bibinfo {author}
  {\bibfnamefont {Benjamin~A.}\ \bibnamefont {Helfrecht}}, \bibinfo {author}
  {\bibfnamefont {Przemys{\l}aw}\ \bibnamefont {Juda}}, \bibinfo {author}
  {\bibfnamefont {S{\'e}bastien~P.}\ \bibnamefont {Bienvenue}}, \bibinfo
  {author} {\bibfnamefont {Wei}\ \bibnamefont {Fang}}, \bibinfo {author}
  {\bibfnamefont {Jan}\ \bibnamefont {Kessler}}, \bibinfo {author}
  {\bibfnamefont {Igor}\ \bibnamefont {Poltavsky}}, \bibinfo {author}
  {\bibfnamefont {Steven}\ \bibnamefont {Vandenbrande}}, \bibinfo {author}
  {\bibfnamefont {Jelle}\ \bibnamefont {Wieme}}, \bibinfo {author}
  {\bibfnamefont {Clemence}\ \bibnamefont {Corminboeuf}}, \bibinfo {author}
  {\bibfnamefont {Thomas~D.}\ \bibnamefont {K{\"u}hne}}, \bibinfo {author}
  {\bibfnamefont {David~E.}\ \bibnamefont {Manolopoulos}}, \bibinfo {author}
  {\bibfnamefont {Thomas~E.}\ \bibnamefont {Markland}}, \bibinfo {author}
  {\bibfnamefont {Jeremy~O.}\ \bibnamefont {Richardson}}, \bibinfo {author}
  {\bibfnamefont {Alexandre}\ \bibnamefont {Tkatchenko}}, \bibinfo {author}
  {\bibfnamefont {Gareth~A.}\ \bibnamefont {Tribello}}, \bibinfo {author}
  {\bibfnamefont {Veronique}\ \bibnamefont {Van~Speybroeck}}, \ and\ \bibinfo
  {author} {\bibfnamefont {Michele}\ \bibnamefont {Ceriotti}},\ }\bibfield
  {title} {\enquote {\bibinfo {title} {I-{{PI}} 2.0: {{A}} universal force
  engine for advanced molecular simulations},}\ }\href {\doibase
  10.1016/j.cpc.2018.09.020} {\bibfield  {journal} {\bibinfo  {journal}
  {Computer Physics Communications}\ }\textbf {\bibinfo {volume} {236}},\
  \bibinfo {pages} {214--223} (\bibinfo {year} {2019})}\BibitemShut {NoStop}%
\bibitem [{\citenamefont {Hourahine}\ \emph {et~al.}(2020)\citenamefont
  {Hourahine}, \citenamefont {Aradi}, \citenamefont {Blum}, \citenamefont
  {Bonaf{\'{e}}}, \citenamefont {Buccheri}, \citenamefont {Camacho},
  \citenamefont {Cevallos}, \citenamefont {Deshaye}, \citenamefont
  {Dumitric{\u{a}}}, \citenamefont {Dominguez}, \citenamefont {Ehlert},
  \citenamefont {Elstner}, \citenamefont {van~der Heide}, \citenamefont
  {Hermann}, \citenamefont {Irle}, \citenamefont {Kranz}, \citenamefont
  {K\"{o}hler}, \citenamefont {Kowalczyk}, \citenamefont {Kuba{\v{r}}},
  \citenamefont {Lee}, \citenamefont {Lutsker}, \citenamefont {Maurer},
  \citenamefont {Min}, \citenamefont {Mitchell}, \citenamefont {Negre},
  \citenamefont {Niehaus}, \citenamefont {Niklasson}, \citenamefont {Page},
  \citenamefont {Pecchia}, \citenamefont {Penazzi}, \citenamefont {Persson},
  \citenamefont {{\v{R}}ez{\'{a}}{\v{c}}}, \citenamefont {S{\'{a}}nchez},
  \citenamefont {Sternberg}, \citenamefont {St\"{o}hr}, \citenamefont
  {Stuckenberg}, \citenamefont {Tkatchenko}, \citenamefont {z.~Yu},\ and\
  \citenamefont {Frauenheim}}]{Hourahine2020}%
  \BibitemOpen
  \bibfield  {author} {\bibinfo {author} {\bibfnamefont {B.}~\bibnamefont
  {Hourahine}}, \bibinfo {author} {\bibfnamefont {B.}~\bibnamefont {Aradi}},
  \bibinfo {author} {\bibfnamefont {V.}~\bibnamefont {Blum}}, \bibinfo {author}
  {\bibfnamefont {F.}~\bibnamefont {Bonaf{\'{e}}}}, \bibinfo {author}
  {\bibfnamefont {A.}~\bibnamefont {Buccheri}}, \bibinfo {author}
  {\bibfnamefont {C.}~\bibnamefont {Camacho}}, \bibinfo {author} {\bibfnamefont
  {C.}~\bibnamefont {Cevallos}}, \bibinfo {author} {\bibfnamefont {M.~Y.}\
  \bibnamefont {Deshaye}}, \bibinfo {author} {\bibfnamefont {T.}~\bibnamefont
  {Dumitric{\u{a}}}}, \bibinfo {author} {\bibfnamefont {A.}~\bibnamefont
  {Dominguez}}, \bibinfo {author} {\bibfnamefont {S.}~\bibnamefont {Ehlert}},
  \bibinfo {author} {\bibfnamefont {M.}~\bibnamefont {Elstner}}, \bibinfo
  {author} {\bibfnamefont {T.}~\bibnamefont {van~der Heide}}, \bibinfo {author}
  {\bibfnamefont {J.}~\bibnamefont {Hermann}}, \bibinfo {author} {\bibfnamefont
  {S.}~\bibnamefont {Irle}}, \bibinfo {author} {\bibfnamefont {J.~J.}\
  \bibnamefont {Kranz}}, \bibinfo {author} {\bibfnamefont {C.}~\bibnamefont
  {K\"{o}hler}}, \bibinfo {author} {\bibfnamefont {T.}~\bibnamefont
  {Kowalczyk}}, \bibinfo {author} {\bibfnamefont {T.}~\bibnamefont
  {Kuba{\v{r}}}}, \bibinfo {author} {\bibfnamefont {I.~S.}\ \bibnamefont
  {Lee}}, \bibinfo {author} {\bibfnamefont {V.}~\bibnamefont {Lutsker}},
  \bibinfo {author} {\bibfnamefont {R.~J.}\ \bibnamefont {Maurer}}, \bibinfo
  {author} {\bibfnamefont {S.~K.}\ \bibnamefont {Min}}, \bibinfo {author}
  {\bibfnamefont {I.}~\bibnamefont {Mitchell}}, \bibinfo {author}
  {\bibfnamefont {C.}~\bibnamefont {Negre}}, \bibinfo {author} {\bibfnamefont
  {T.~A.}\ \bibnamefont {Niehaus}}, \bibinfo {author} {\bibfnamefont
  {A.~M.~N.}\ \bibnamefont {Niklasson}}, \bibinfo {author} {\bibfnamefont
  {A.~J.}\ \bibnamefont {Page}}, \bibinfo {author} {\bibfnamefont
  {A.}~\bibnamefont {Pecchia}}, \bibinfo {author} {\bibfnamefont
  {G.}~\bibnamefont {Penazzi}}, \bibinfo {author} {\bibfnamefont {M.~P.}\
  \bibnamefont {Persson}}, \bibinfo {author} {\bibfnamefont {J.}~\bibnamefont
  {{\v{R}}ez{\'{a}}{\v{c}}}}, \bibinfo {author} {\bibfnamefont {C.~G.}\
  \bibnamefont {S{\'{a}}nchez}}, \bibinfo {author} {\bibfnamefont
  {M.}~\bibnamefont {Sternberg}}, \bibinfo {author} {\bibfnamefont
  {M.}~\bibnamefont {St\"{o}hr}}, \bibinfo {author} {\bibfnamefont
  {F.}~\bibnamefont {Stuckenberg}}, \bibinfo {author} {\bibfnamefont
  {A.}~\bibnamefont {Tkatchenko}}, \bibinfo {author} {\bibfnamefont {V.~W.}\
  \bibnamefont {z.~Yu}}, \ and\ \bibinfo {author} {\bibfnamefont
  {T.}~\bibnamefont {Frauenheim}},\ }\bibfield  {title} {\enquote {\bibinfo
  {title} {{DFTB}+, a software package for efficient approximate density
  functional theory based atomistic simulations},}\ }\href {\doibase
  10.1063/1.5143190} {\bibfield  {journal} {\bibinfo  {journal} {The Journal of
  Chemical Physics}\ }\textbf {\bibinfo {volume} {152}} (\bibinfo {year}
  {2020}),\ 10.1063/1.5143190}\BibitemShut {NoStop}%
\bibitem [{\citenamefont {Ceriotti}\ \emph {et~al.}(2013)\citenamefont
  {Ceriotti}, \citenamefont {Tribello},\ and\ \citenamefont
  {Parrinello}}]{ceri+13jctc}%
  \BibitemOpen
  \bibfield  {author} {\bibinfo {author} {\bibfnamefont {Michele}\ \bibnamefont
  {Ceriotti}}, \bibinfo {author} {\bibfnamefont {Gareth~A.}\ \bibnamefont
  {Tribello}}, \ and\ \bibinfo {author} {\bibfnamefont {Michele}\ \bibnamefont
  {Parrinello}},\ }\bibfield  {title} {\enquote {\bibinfo {title}
  {Demonstrating the transferability and the descriptive power of
  sketch-map},}\ }\href {\doibase 10.1021/ct3010563} {\bibfield  {journal}
  {\bibinfo  {journal} {Journal of Chemical Theory and Computation}\ }\textbf
  {\bibinfo {volume} {9}},\ \bibinfo {pages} {1521--1532} (\bibinfo {year}
  {2013})}\BibitemShut {NoStop}%
\bibitem [{\citenamefont {Bart{\'o}k}\ \emph {et~al.}(2013)\citenamefont
  {Bart{\'o}k}, \citenamefont {Kondor},\ and\ \citenamefont
  {Cs{\'a}nyi}}]{bart+13prb}%
  \BibitemOpen
  \bibfield  {author} {\bibinfo {author} {\bibfnamefont {Albert~P.}\
  \bibnamefont {Bart{\'o}k}}, \bibinfo {author} {\bibfnamefont {Risi}\
  \bibnamefont {Kondor}}, \ and\ \bibinfo {author} {\bibfnamefont {G{\'a}bor}\
  \bibnamefont {Cs{\'a}nyi}},\ }\bibfield  {title} {\enquote {\bibinfo {title}
  {On representing chemical environments},}\ }\href {\doibase
  10.1103/PhysRevB.87.184115} {\bibfield  {journal} {\bibinfo  {journal}
  {Physical Review B}\ }\textbf {\bibinfo {volume} {87}},\ \bibinfo {pages}
  {184115} (\bibinfo {year} {2013})}\BibitemShut {NoStop}%
\bibitem [{ras(2023)}]{rascaline}%
  \BibitemOpen
  \href {https://github.com/Luthaf/rascaline} {\enquote {\bibinfo {title}
  {{Rascaline}},}\ }\bibinfo {howpublished}
  {\url{https://github.com/Luthaf/rascaline}} (\bibinfo {year}
  {2023})\BibitemShut {NoStop}%
\bibitem [{\citenamefont {Goscinski}\ \emph {et~al.}(2023)\citenamefont
  {Goscinski}, \citenamefont {Principe}, \citenamefont {Fraux}, \citenamefont
  {Kliavinek}, \citenamefont {Helfrecht}, \citenamefont {Loche}, \citenamefont
  {Ceriotti},\ and\ \citenamefont {Cersonsky}}]{goscinski2023scikit}%
  \BibitemOpen
  \bibfield  {author} {\bibinfo {author} {\bibfnamefont {Alexander}\
  \bibnamefont {Goscinski}}, \bibinfo {author} {\bibfnamefont {Victor~Paul}\
  \bibnamefont {Principe}}, \bibinfo {author} {\bibfnamefont {Guillaume}\
  \bibnamefont {Fraux}}, \bibinfo {author} {\bibfnamefont {Sergei}\
  \bibnamefont {Kliavinek}}, \bibinfo {author} {\bibfnamefont {Benjamin~Aaron}\
  \bibnamefont {Helfrecht}}, \bibinfo {author} {\bibfnamefont {Philip}\
  \bibnamefont {Loche}}, \bibinfo {author} {\bibfnamefont {Michele}\
  \bibnamefont {Ceriotti}}, \ and\ \bibinfo {author} {\bibfnamefont
  {Rose~Kathleen}\ \bibnamefont {Cersonsky}},\ }\bibfield  {title} {\enquote
  {\bibinfo {title} {scikit-matter: A suite of generalisable machine learning
  methods born out of chemistry and materials science},}\ }\href@noop {}
  {\bibfield  {journal} {\bibinfo  {journal} {Open Research Europe}\ }\textbf
  {\bibinfo {volume} {3}},\ \bibinfo {pages} {81} (\bibinfo {year}
  {2023})}\BibitemShut {NoStop}%
\bibitem [{\citenamefont {Sun}\ \emph {et~al.}(2020)\citenamefont {Sun},
  \citenamefont {Zhang}, \citenamefont {Banerjee}, \citenamefont {Bao},
  \citenamefont {Barbry}, \citenamefont {Blunt}, \citenamefont {Bogdanov},
  \citenamefont {Booth}, \citenamefont {Chen}, \citenamefont {Cui},
  \citenamefont {Eriksen}, \citenamefont {Gao}, \citenamefont {Guo},
  \citenamefont {Hermann}, \citenamefont {Hermes}, \citenamefont {Koh},
  \citenamefont {Koval}, \citenamefont {Lehtola}, \citenamefont {Li},
  \citenamefont {Liu}, \citenamefont {Mardirossian}, \citenamefont {McClain},
  \citenamefont {Motta}, \citenamefont {Mussard}, \citenamefont {Pham},
  \citenamefont {Pulkin}, \citenamefont {Purwanto}, \citenamefont {Robinson},
  \citenamefont {Ronca}, \citenamefont {Sayfutyarova}, \citenamefont
  {Scheurer}, \citenamefont {Schurkus}, \citenamefont {Smith}, \citenamefont
  {Sun}, \citenamefont {Sun}, \citenamefont {Upadhyay}, \citenamefont {Wagner},
  \citenamefont {Wang}, \citenamefont {White}, \citenamefont {Whitfield},
  \citenamefont {Williamson}, \citenamefont {Wouters}, \citenamefont {Yang},
  \citenamefont {Yu}, \citenamefont {Zhu}, \citenamefont {Berkelbach},
  \citenamefont {Sharma}, \citenamefont {Sokolov},\ and\ \citenamefont
  {Chan}}]{Sun2020}%
  \BibitemOpen
  \bibfield  {author} {\bibinfo {author} {\bibfnamefont {Qiming}\ \bibnamefont
  {Sun}}, \bibinfo {author} {\bibfnamefont {Xing}\ \bibnamefont {Zhang}},
  \bibinfo {author} {\bibfnamefont {Samragni}\ \bibnamefont {Banerjee}},
  \bibinfo {author} {\bibfnamefont {Peng}\ \bibnamefont {Bao}}, \bibinfo
  {author} {\bibfnamefont {Marc}\ \bibnamefont {Barbry}}, \bibinfo {author}
  {\bibfnamefont {Nick~S.}\ \bibnamefont {Blunt}}, \bibinfo {author}
  {\bibfnamefont {Nikolay~A.}\ \bibnamefont {Bogdanov}}, \bibinfo {author}
  {\bibfnamefont {George~H.}\ \bibnamefont {Booth}}, \bibinfo {author}
  {\bibfnamefont {Jia}\ \bibnamefont {Chen}}, \bibinfo {author} {\bibfnamefont
  {Zhi-Hao}\ \bibnamefont {Cui}}, \bibinfo {author} {\bibfnamefont {Janus~J.}\
  \bibnamefont {Eriksen}}, \bibinfo {author} {\bibfnamefont {Yang}\
  \bibnamefont {Gao}}, \bibinfo {author} {\bibfnamefont {Sheng}\ \bibnamefont
  {Guo}}, \bibinfo {author} {\bibfnamefont {Jan}\ \bibnamefont {Hermann}},
  \bibinfo {author} {\bibfnamefont {Matthew~R.}\ \bibnamefont {Hermes}},
  \bibinfo {author} {\bibfnamefont {Kevin}\ \bibnamefont {Koh}}, \bibinfo
  {author} {\bibfnamefont {Peter}\ \bibnamefont {Koval}}, \bibinfo {author}
  {\bibfnamefont {Susi}\ \bibnamefont {Lehtola}}, \bibinfo {author}
  {\bibfnamefont {Zhendong}\ \bibnamefont {Li}}, \bibinfo {author}
  {\bibfnamefont {Junzi}\ \bibnamefont {Liu}}, \bibinfo {author} {\bibfnamefont
  {Narbe}\ \bibnamefont {Mardirossian}}, \bibinfo {author} {\bibfnamefont
  {James~D.}\ \bibnamefont {McClain}}, \bibinfo {author} {\bibfnamefont
  {Mario}\ \bibnamefont {Motta}}, \bibinfo {author} {\bibfnamefont {Bastien}\
  \bibnamefont {Mussard}}, \bibinfo {author} {\bibfnamefont {Hung~Q.}\
  \bibnamefont {Pham}}, \bibinfo {author} {\bibfnamefont {Artem}\ \bibnamefont
  {Pulkin}}, \bibinfo {author} {\bibfnamefont {Wirawan}\ \bibnamefont
  {Purwanto}}, \bibinfo {author} {\bibfnamefont {Paul~J.}\ \bibnamefont
  {Robinson}}, \bibinfo {author} {\bibfnamefont {Enrico}\ \bibnamefont
  {Ronca}}, \bibinfo {author} {\bibfnamefont {Elvira~R.}\ \bibnamefont
  {Sayfutyarova}}, \bibinfo {author} {\bibfnamefont {Maximilian}\ \bibnamefont
  {Scheurer}}, \bibinfo {author} {\bibfnamefont {Henry~F.}\ \bibnamefont
  {Schurkus}}, \bibinfo {author} {\bibfnamefont {James E.~T.}\ \bibnamefont
  {Smith}}, \bibinfo {author} {\bibfnamefont {Chong}\ \bibnamefont {Sun}},
  \bibinfo {author} {\bibfnamefont {Shi-Ning}\ \bibnamefont {Sun}}, \bibinfo
  {author} {\bibfnamefont {Shiv}\ \bibnamefont {Upadhyay}}, \bibinfo {author}
  {\bibfnamefont {Lucas~K.}\ \bibnamefont {Wagner}}, \bibinfo {author}
  {\bibfnamefont {Xiao}\ \bibnamefont {Wang}}, \bibinfo {author} {\bibfnamefont
  {Alec}\ \bibnamefont {White}}, \bibinfo {author} {\bibfnamefont
  {James~Daniel}\ \bibnamefont {Whitfield}}, \bibinfo {author} {\bibfnamefont
  {Mark~J.}\ \bibnamefont {Williamson}}, \bibinfo {author} {\bibfnamefont
  {Sebastian}\ \bibnamefont {Wouters}}, \bibinfo {author} {\bibfnamefont {Jun}\
  \bibnamefont {Yang}}, \bibinfo {author} {\bibfnamefont {Jason~M.}\
  \bibnamefont {Yu}}, \bibinfo {author} {\bibfnamefont {Tianyu}\ \bibnamefont
  {Zhu}}, \bibinfo {author} {\bibfnamefont {Timothy~C.}\ \bibnamefont
  {Berkelbach}}, \bibinfo {author} {\bibfnamefont {Sandeep}\ \bibnamefont
  {Sharma}}, \bibinfo {author} {\bibfnamefont {Alexander~Yu.}\ \bibnamefont
  {Sokolov}}, \ and\ \bibinfo {author} {\bibfnamefont {Garnet Kin-Lic}\
  \bibnamefont {Chan}},\ }\bibfield  {title} {\enquote {\bibinfo {title}
  {Recent developments in the pyscf program package},}\ }\href {\doibase
  10.1063/5.0006074} {\bibfield  {journal} {\bibinfo  {journal} {The Journal of
  Chemical Physics}\ }\textbf {\bibinfo {volume} {153}} (\bibinfo {year}
  {2020}),\ 10.1063/5.0006074}\BibitemShut {NoStop}%
\bibitem [{\citenamefont {Drautz}(2019)}]{drau19prb}%
  \BibitemOpen
  \bibfield  {author} {\bibinfo {author} {\bibfnamefont {Ralf}\ \bibnamefont
  {Drautz}},\ }\bibfield  {title} {\enquote {\bibinfo {title} {Atomic cluster
  expansion for accurate and transferable interatomic potentials},}\ }\href
  {\doibase 10.1103/PhysRevB.99.014104} {\bibfield  {journal} {\bibinfo
  {journal} {Physical Review B}\ }\textbf {\bibinfo {volume} {99}},\ \bibinfo
  {pages} {014104} (\bibinfo {year} {2019})}\BibitemShut {NoStop}%
\bibitem [{\citenamefont {Willatt}\ \emph {et~al.}(2019)\citenamefont
  {Willatt}, \citenamefont {Musil},\ and\ \citenamefont
  {Ceriotti}}]{will+19jcp}%
  \BibitemOpen
  \bibfield  {author} {\bibinfo {author} {\bibfnamefont {Michael~J.}\
  \bibnamefont {Willatt}}, \bibinfo {author} {\bibfnamefont {F{\'e}lix}\
  \bibnamefont {Musil}}, \ and\ \bibinfo {author} {\bibfnamefont {Michele}\
  \bibnamefont {Ceriotti}},\ }\bibfield  {title} {\enquote {\bibinfo {title}
  {Atom-density representations for machine learning},}\ }\href {\doibase
  10.1063/1.5090481} {\bibfield  {journal} {\bibinfo  {journal} {Journal of
  Chemical Physics}\ }\textbf {\bibinfo {volume} {150}},\ \bibinfo {pages}
  {154110} (\bibinfo {year} {2019})}\BibitemShut {NoStop}%
\bibitem [{std(2023)}]{stdatorch}%
  \BibitemOpen
  \href {{https://github.com/ecignoni/stda\_torch}} {\enquote {\bibinfo {title}
  {{stda\_torch}},}\ }\bibinfo {howpublished}
  {\url{https://github.com/ecignoni/stda\_torch}} (\bibinfo {year}
  {2023})\BibitemShut {NoStop}%
\bibitem [{\citenamefont {Paszke}\ \emph {et~al.}(2019)\citenamefont {Paszke},
  \citenamefont {Gross}, \citenamefont {Massa}, \citenamefont {Lerer},
  \citenamefont {Bradbury}, \citenamefont {Chanan}, \citenamefont {Killeen},
  \citenamefont {Lin}, \citenamefont {Gimelshein}, \citenamefont {Antiga},
  \citenamefont {Desmaison}, \citenamefont {Kopf}, \citenamefont {Yang},
  \citenamefont {DeVito}, \citenamefont {Raison}, \citenamefont {Tejani},
  \citenamefont {Chilamkurthy}, \citenamefont {Steiner}, \citenamefont {Fang},
  \citenamefont {Bai},\ and\ \citenamefont {Chintala}}]{PyTorch2019}%
  \BibitemOpen
  \bibfield  {author} {\bibinfo {author} {\bibfnamefont {Adam}\ \bibnamefont
  {Paszke}}, \bibinfo {author} {\bibfnamefont {Sam}\ \bibnamefont {Gross}},
  \bibinfo {author} {\bibfnamefont {Francisco}\ \bibnamefont {Massa}}, \bibinfo
  {author} {\bibfnamefont {Adam}\ \bibnamefont {Lerer}}, \bibinfo {author}
  {\bibfnamefont {James}\ \bibnamefont {Bradbury}}, \bibinfo {author}
  {\bibfnamefont {Gregory}\ \bibnamefont {Chanan}}, \bibinfo {author}
  {\bibfnamefont {Trevor}\ \bibnamefont {Killeen}}, \bibinfo {author}
  {\bibfnamefont {Zeming}\ \bibnamefont {Lin}}, \bibinfo {author}
  {\bibfnamefont {Natalia}\ \bibnamefont {Gimelshein}}, \bibinfo {author}
  {\bibfnamefont {Luca}\ \bibnamefont {Antiga}}, \bibinfo {author}
  {\bibfnamefont {Alban}\ \bibnamefont {Desmaison}}, \bibinfo {author}
  {\bibfnamefont {Andreas}\ \bibnamefont {Kopf}}, \bibinfo {author}
  {\bibfnamefont {Edward}\ \bibnamefont {Yang}}, \bibinfo {author}
  {\bibfnamefont {Zachary}\ \bibnamefont {DeVito}}, \bibinfo {author}
  {\bibfnamefont {Martin}\ \bibnamefont {Raison}}, \bibinfo {author}
  {\bibfnamefont {Alykhan}\ \bibnamefont {Tejani}}, \bibinfo {author}
  {\bibfnamefont {Sasank}\ \bibnamefont {Chilamkurthy}}, \bibinfo {author}
  {\bibfnamefont {Benoit}\ \bibnamefont {Steiner}}, \bibinfo {author}
  {\bibfnamefont {Lu}~\bibnamefont {Fang}}, \bibinfo {author} {\bibfnamefont
  {Junjie}\ \bibnamefont {Bai}}, \ and\ \bibinfo {author} {\bibfnamefont
  {Soumith}\ \bibnamefont {Chintala}},\ }\bibfield  {title} {\enquote {\bibinfo
  {title} {Pytorch: An imperative style, high-performance deep learning
  library},}\ }in\ \href
  {http://papers.neurips.cc/paper/9015-pytorch-an-imperative-style-high-performance-deep-learning-library.pdf}
  {\emph {\bibinfo {booktitle} {Advances in Neural Information Processing
  Systems 32}}}\ (\bibinfo  {publisher} {Curran Associates, Inc.},\ \bibinfo
  {year} {2019})\ pp.\ \bibinfo {pages} {8024--8035}\BibitemShut {NoStop}%
\bibitem [{\citenamefont {Frisch}\ \emph {et~al.}(2016)\citenamefont {Frisch},
  \citenamefont {Trucks}, \citenamefont {Schlegel}, \citenamefont {Scuseria},
  \citenamefont {Robb}, \citenamefont {Cheeseman}, \citenamefont {Scalmani},
  \citenamefont {Barone}, \citenamefont {Petersson}, \citenamefont {Nakatsuji},
  \citenamefont {Li}, \citenamefont {Caricato}, \citenamefont {Marenich},
  \citenamefont {Bloino}, \citenamefont {Janesko}, \citenamefont {Gomperts},
  \citenamefont {Mennucci}, \citenamefont {Hratchian}, \citenamefont {Ortiz},
  \citenamefont {Izmaylov}, \citenamefont {Sonnenberg}, \citenamefont
  {Williams-Young}, \citenamefont {Ding}, \citenamefont {Lipparini},
  \citenamefont {Egidi}, \citenamefont {Goings}, \citenamefont {Peng},
  \citenamefont {Petrone}, \citenamefont {Henderson}, \citenamefont
  {Ranasinghe}, \citenamefont {Zakrzewski}, \citenamefont {Gao}, \citenamefont
  {Rega}, \citenamefont {Zheng}, \citenamefont {Liang}, \citenamefont {Hada},
  \citenamefont {Ehara}, \citenamefont {Toyota}, \citenamefont {Fukuda},
  \citenamefont {Hasegawa}, \citenamefont {Ishida}, \citenamefont {Nakajima},
  \citenamefont {Honda}, \citenamefont {Kitao}, \citenamefont {Nakai},
  \citenamefont {Vreven}, \citenamefont {Throssell}, \citenamefont
  {Montgomery}, \citenamefont {Peralta}, \citenamefont {Ogliaro}, \citenamefont
  {Bearpark}, \citenamefont {Heyd}, \citenamefont {Brothers}, \citenamefont
  {Kudin}, \citenamefont {Staroverov}, \citenamefont {Keith}, \citenamefont
  {Kobayashi}, \citenamefont {Normand}, \citenamefont {Raghavachari},
  \citenamefont {Rendell}, \citenamefont {Burant}, \citenamefont {Iyengar},
  \citenamefont {Tomasi}, \citenamefont {Cossi}, \citenamefont {Millam},
  \citenamefont {Klene}, \citenamefont {Adamo}, \citenamefont {Cammi},
  \citenamefont {Ochterski}, \citenamefont {Martin}, \citenamefont {Morokuma},
  \citenamefont {Farkas}, \citenamefont {Foresman},\ and\ \citenamefont
  {Fox}}]{g16}%
  \BibitemOpen
  \bibfield  {author} {\bibinfo {author} {\bibfnamefont {M.~J.}\ \bibnamefont
  {Frisch}}, \bibinfo {author} {\bibfnamefont {G.~W.}\ \bibnamefont {Trucks}},
  \bibinfo {author} {\bibfnamefont {H.~B.}\ \bibnamefont {Schlegel}}, \bibinfo
  {author} {\bibfnamefont {G.~E.}\ \bibnamefont {Scuseria}}, \bibinfo {author}
  {\bibfnamefont {M.~A.}\ \bibnamefont {Robb}}, \bibinfo {author}
  {\bibfnamefont {J.~R.}\ \bibnamefont {Cheeseman}}, \bibinfo {author}
  {\bibfnamefont {G.}~\bibnamefont {Scalmani}}, \bibinfo {author}
  {\bibfnamefont {V.}~\bibnamefont {Barone}}, \bibinfo {author} {\bibfnamefont
  {G.~A.}\ \bibnamefont {Petersson}}, \bibinfo {author} {\bibfnamefont
  {H.}~\bibnamefont {Nakatsuji}}, \bibinfo {author} {\bibfnamefont
  {X.}~\bibnamefont {Li}}, \bibinfo {author} {\bibfnamefont {M.}~\bibnamefont
  {Caricato}}, \bibinfo {author} {\bibfnamefont {A.~V.}\ \bibnamefont
  {Marenich}}, \bibinfo {author} {\bibfnamefont {J.}~\bibnamefont {Bloino}},
  \bibinfo {author} {\bibfnamefont {B.~G.}\ \bibnamefont {Janesko}}, \bibinfo
  {author} {\bibfnamefont {R.}~\bibnamefont {Gomperts}}, \bibinfo {author}
  {\bibfnamefont {B.}~\bibnamefont {Mennucci}}, \bibinfo {author}
  {\bibfnamefont {H.~P.}\ \bibnamefont {Hratchian}}, \bibinfo {author}
  {\bibfnamefont {J.~V.}\ \bibnamefont {Ortiz}}, \bibinfo {author}
  {\bibfnamefont {A.~F.}\ \bibnamefont {Izmaylov}}, \bibinfo {author}
  {\bibfnamefont {J.~L.}\ \bibnamefont {Sonnenberg}}, \bibinfo {author}
  {\bibfnamefont {D.}~\bibnamefont {Williams-Young}}, \bibinfo {author}
  {\bibfnamefont {F.}~\bibnamefont {Ding}}, \bibinfo {author} {\bibfnamefont
  {F.}~\bibnamefont {Lipparini}}, \bibinfo {author} {\bibfnamefont
  {F.}~\bibnamefont {Egidi}}, \bibinfo {author} {\bibfnamefont
  {J.}~\bibnamefont {Goings}}, \bibinfo {author} {\bibfnamefont
  {B.}~\bibnamefont {Peng}}, \bibinfo {author} {\bibfnamefont {A.}~\bibnamefont
  {Petrone}}, \bibinfo {author} {\bibfnamefont {T.}~\bibnamefont {Henderson}},
  \bibinfo {author} {\bibfnamefont {D.}~\bibnamefont {Ranasinghe}}, \bibinfo
  {author} {\bibfnamefont {V.~G.}\ \bibnamefont {Zakrzewski}}, \bibinfo
  {author} {\bibfnamefont {J.}~\bibnamefont {Gao}}, \bibinfo {author}
  {\bibfnamefont {N.}~\bibnamefont {Rega}}, \bibinfo {author} {\bibfnamefont
  {G.}~\bibnamefont {Zheng}}, \bibinfo {author} {\bibfnamefont
  {W.}~\bibnamefont {Liang}}, \bibinfo {author} {\bibfnamefont
  {M.}~\bibnamefont {Hada}}, \bibinfo {author} {\bibfnamefont {M.}~\bibnamefont
  {Ehara}}, \bibinfo {author} {\bibfnamefont {K.}~\bibnamefont {Toyota}},
  \bibinfo {author} {\bibfnamefont {R.}~\bibnamefont {Fukuda}}, \bibinfo
  {author} {\bibfnamefont {J.}~\bibnamefont {Hasegawa}}, \bibinfo {author}
  {\bibfnamefont {M.}~\bibnamefont {Ishida}}, \bibinfo {author} {\bibfnamefont
  {T.}~\bibnamefont {Nakajima}}, \bibinfo {author} {\bibfnamefont
  {Y.}~\bibnamefont {Honda}}, \bibinfo {author} {\bibfnamefont
  {O.}~\bibnamefont {Kitao}}, \bibinfo {author} {\bibfnamefont
  {H.}~\bibnamefont {Nakai}}, \bibinfo {author} {\bibfnamefont
  {T.}~\bibnamefont {Vreven}}, \bibinfo {author} {\bibfnamefont
  {K.}~\bibnamefont {Throssell}}, \bibinfo {author} {\bibfnamefont {J.~A.}\
  \bibnamefont {Montgomery}, \bibfnamefont {{Jr.}}}, \bibinfo {author}
  {\bibfnamefont {J.~E.}\ \bibnamefont {Peralta}}, \bibinfo {author}
  {\bibfnamefont {F.}~\bibnamefont {Ogliaro}}, \bibinfo {author} {\bibfnamefont
  {M.~J.}\ \bibnamefont {Bearpark}}, \bibinfo {author} {\bibfnamefont {J.~J.}\
  \bibnamefont {Heyd}}, \bibinfo {author} {\bibfnamefont {E.~N.}\ \bibnamefont
  {Brothers}}, \bibinfo {author} {\bibfnamefont {K.~N.}\ \bibnamefont {Kudin}},
  \bibinfo {author} {\bibfnamefont {V.~N.}\ \bibnamefont {Staroverov}},
  \bibinfo {author} {\bibfnamefont {T.~A.}\ \bibnamefont {Keith}}, \bibinfo
  {author} {\bibfnamefont {R.}~\bibnamefont {Kobayashi}}, \bibinfo {author}
  {\bibfnamefont {J.}~\bibnamefont {Normand}}, \bibinfo {author} {\bibfnamefont
  {K.}~\bibnamefont {Raghavachari}}, \bibinfo {author} {\bibfnamefont {A.~P.}\
  \bibnamefont {Rendell}}, \bibinfo {author} {\bibfnamefont {J.~C.}\
  \bibnamefont {Burant}}, \bibinfo {author} {\bibfnamefont {S.~S.}\
  \bibnamefont {Iyengar}}, \bibinfo {author} {\bibfnamefont {J.}~\bibnamefont
  {Tomasi}}, \bibinfo {author} {\bibfnamefont {M.}~\bibnamefont {Cossi}},
  \bibinfo {author} {\bibfnamefont {J.~M.}\ \bibnamefont {Millam}}, \bibinfo
  {author} {\bibfnamefont {M.}~\bibnamefont {Klene}}, \bibinfo {author}
  {\bibfnamefont {C.}~\bibnamefont {Adamo}}, \bibinfo {author} {\bibfnamefont
  {R.}~\bibnamefont {Cammi}}, \bibinfo {author} {\bibfnamefont {J.~W.}\
  \bibnamefont {Ochterski}}, \bibinfo {author} {\bibfnamefont {R.~L.}\
  \bibnamefont {Martin}}, \bibinfo {author} {\bibfnamefont {K.}~\bibnamefont
  {Morokuma}}, \bibinfo {author} {\bibfnamefont {O.}~\bibnamefont {Farkas}},
  \bibinfo {author} {\bibfnamefont {J.~B.}\ \bibnamefont {Foresman}}, \ and\
  \bibinfo {author} {\bibfnamefont {D.~J.}\ \bibnamefont {Fox}},\ }\href@noop
  {} {\enquote {\bibinfo {title} {Gaussian˜16 {R}evision {A}.03},}\ }
  (\bibinfo {year} {2016}),\ \bibinfo {note} {gaussian Inc. Wallingford
  CT}\BibitemShut {NoStop}%
\bibitem [{\citenamefont {Case}\ \emph {et~al.}(2018)\citenamefont {Case},
  \citenamefont {Ben-Shalom}, \citenamefont {Brozell}, \citenamefont {Cerutti},
  \citenamefont {Cheatham}, \citenamefont {III}, \citenamefont {Cruzeiro},
  \citenamefont {Darden}, \citenamefont {Duke}, \citenamefont {Ghoreishi},
  \citenamefont {Gilson}, \citenamefont {Gohlke}, \citenamefont {Goetz},
  \citenamefont {Greene}, \citenamefont {Harris}, \citenamefont {Homeyer},
  \citenamefont {Izadi}, \citenamefont {Kovalenko}, \citenamefont {Kurtzman},
  \citenamefont {Lee}, \citenamefont {LeGrand}, \citenamefont {Li},
  \citenamefont {Lin}, \citenamefont {Liu}, \citenamefont {Luchko},
  \citenamefont {Luo}, \citenamefont {Mermelstein}, \citenamefont {Merz},
  \citenamefont {Miao}, \citenamefont {Monard}, \citenamefont {Nguyen},
  \citenamefont {Nguyen}, \citenamefont {Omelyan}, \citenamefont {Onufriev},
  \citenamefont {Pan}, \citenamefont {Qi}, \citenamefont {Roe}, \citenamefont
  {Roitberg}, \citenamefont {Sagui}, \citenamefont {Schott-Verdugo},
  \citenamefont {Shen}, \citenamefont {Simmerling}, \citenamefont {Smith},
  \citenamefont {Salomon-Ferrer}, \citenamefont {Swails}, \citenamefont
  {Walker}, \citenamefont {Wang}, \citenamefont {Wei}, \citenamefont {Wolf},
  \citenamefont {Wu}, \citenamefont {Xiao}, \citenamefont {York},\ and\
  \citenamefont {Kollman}}]{da2018amber}%
  \BibitemOpen
  \bibfield  {author} {\bibinfo {author} {\bibfnamefont {D.~A.}\ \bibnamefont
  {Case}}, \bibinfo {author} {\bibfnamefont {I.~Y.}\ \bibnamefont
  {Ben-Shalom}}, \bibinfo {author} {\bibfnamefont {S.~R.}\ \bibnamefont
  {Brozell}}, \bibinfo {author} {\bibfnamefont {D.~S.}\ \bibnamefont
  {Cerutti}}, \bibinfo {author} {\bibfnamefont {T.~E.}\ \bibnamefont
  {Cheatham}}, \bibinfo {author} {\bibnamefont {III}}, \bibinfo {author}
  {\bibfnamefont {V.~W.~D.}\ \bibnamefont {Cruzeiro}}, \bibinfo {author}
  {\bibfnamefont {T.~A.}\ \bibnamefont {Darden}}, \bibinfo {author}
  {\bibfnamefont {R.E.}\ \bibnamefont {Duke}}, \bibinfo {author} {\bibfnamefont
  {D.}~\bibnamefont {Ghoreishi}}, \bibinfo {author} {\bibfnamefont {M.~K.}\
  \bibnamefont {Gilson}}, \bibinfo {author} {\bibfnamefont {H.}~\bibnamefont
  {Gohlke}}, \bibinfo {author} {\bibfnamefont {A.~W.}\ \bibnamefont {Goetz}},
  \bibinfo {author} {\bibfnamefont {D.}~\bibnamefont {Greene}}, \bibinfo
  {author} {\bibfnamefont {R.}~\bibnamefont {Harris}}, \bibinfo {author}
  {\bibfnamefont {N.}~\bibnamefont {Homeyer}}, \bibinfo {author} {\bibfnamefont
  {S.}~\bibnamefont {Izadi}}, \bibinfo {author} {\bibfnamefont
  {A.}~\bibnamefont {Kovalenko}}, \bibinfo {author} {\bibfnamefont
  {T.}~\bibnamefont {Kurtzman}}, \bibinfo {author} {\bibfnamefont {T.~S.}\
  \bibnamefont {Lee}}, \bibinfo {author} {\bibfnamefont {S.}~\bibnamefont
  {LeGrand}}, \bibinfo {author} {\bibfnamefont {P.}~\bibnamefont {Li}},
  \bibinfo {author} {\bibfnamefont {C.}~\bibnamefont {Lin}}, \bibinfo {author}
  {\bibfnamefont {J.}~\bibnamefont {Liu}}, \bibinfo {author} {\bibfnamefont
  {T.}~\bibnamefont {Luchko}}, \bibinfo {author} {\bibfnamefont
  {R.}~\bibnamefont {Luo}}, \bibinfo {author} {\bibfnamefont {D.~J.}\
  \bibnamefont {Mermelstein}}, \bibinfo {author} {\bibfnamefont {K.~M.}\
  \bibnamefont {Merz}}, \bibinfo {author} {\bibfnamefont {Y.}~\bibnamefont
  {Miao}}, \bibinfo {author} {\bibfnamefont {G.}~\bibnamefont {Monard}},
  \bibinfo {author} {\bibfnamefont {C.}~\bibnamefont {Nguyen}}, \bibinfo
  {author} {\bibfnamefont {H.}~\bibnamefont {Nguyen}}, \bibinfo {author}
  {\bibfnamefont {I.}~\bibnamefont {Omelyan}}, \bibinfo {author} {\bibfnamefont
  {A.}~\bibnamefont {Onufriev}}, \bibinfo {author} {\bibfnamefont
  {F.}~\bibnamefont {Pan}}, \bibinfo {author} {\bibfnamefont {R.}~\bibnamefont
  {Qi}}, \bibinfo {author} {\bibfnamefont {D.~R.}\ \bibnamefont {Roe}},
  \bibinfo {author} {\bibfnamefont {A.}~\bibnamefont {Roitberg}}, \bibinfo
  {author} {\bibfnamefont {C.}~\bibnamefont {Sagui}}, \bibinfo {author}
  {\bibfnamefont {S.}~\bibnamefont {Schott-Verdugo}}, \bibinfo {author}
  {\bibfnamefont {J.}~\bibnamefont {Shen}}, \bibinfo {author} {\bibfnamefont
  {C.~L.}\ \bibnamefont {Simmerling}}, \bibinfo {author} {\bibfnamefont
  {J.}~\bibnamefont {Smith}}, \bibinfo {author} {\bibfnamefont
  {R.}~\bibnamefont {Salomon-Ferrer}}, \bibinfo {author} {\bibfnamefont
  {J.}~\bibnamefont {Swails}}, \bibinfo {author} {\bibfnamefont {R.~C.}\
  \bibnamefont {Walker}}, \bibinfo {author} {\bibfnamefont {J.}~\bibnamefont
  {Wang}}, \bibinfo {author} {\bibfnamefont {H.}~\bibnamefont {Wei}}, \bibinfo
  {author} {\bibfnamefont {R.~M.}\ \bibnamefont {Wolf}}, \bibinfo {author}
  {\bibfnamefont {X.}~\bibnamefont {Wu}}, \bibinfo {author} {\bibfnamefont
  {L.}~\bibnamefont {Xiao}}, \bibinfo {author} {\bibfnamefont {D.~M.}\
  \bibnamefont {York}}, \ and\ \bibinfo {author} {\bibfnamefont {P.~A.}\
  \bibnamefont {Kollman}},\ }\href@noop {} {\enquote {\bibinfo {title} {Amber
  18},}\ } (\bibinfo {year} {2018}),\ \bibinfo {note} {university of
  California, San Francisco}\BibitemShut {NoStop}%
\bibitem [{\citenamefont {Mukamel}(1995)}]{mukamel1995}%
  \BibitemOpen
  \bibfield  {author} {\bibinfo {author} {\bibfnamefont {Shaul}\ \bibnamefont
  {Mukamel}},\ }\href {https://cir.nii.ac.jp/crid/1130000797363520000} {\emph
  {\bibinfo {title} {Principles of nonlinear optical spectroscopy}}},\ Oxford
  series in optical and imaging sciences\ (\bibinfo  {publisher} {Oxford
  University Press},\ \bibinfo {year} {1995})\BibitemShut {NoStop}%
\bibitem [{\citenamefont {Loco}\ and\ \citenamefont
  {Cupellini}(2018)}]{Loco2018}%
  \BibitemOpen
  \bibfield  {author} {\bibinfo {author} {\bibfnamefont {Daniele}\ \bibnamefont
  {Loco}}\ and\ \bibinfo {author} {\bibfnamefont {Lorenzo}\ \bibnamefont
  {Cupellini}},\ }\bibfield  {title} {\enquote {\bibinfo {title} {Modeling the
  absorption lineshape of embedded systems from molecular dynamics: A tutorial
  review},}\ }\href {\doibase 10.1002/qua.25726} {\bibfield  {journal}
  {\bibinfo  {journal} {International Journal of Quantum Chemistry}\ }\textbf
  {\bibinfo {volume} {119}},\ \bibinfo {pages} {e25726} (\bibinfo {year}
  {2018})}\BibitemShut {NoStop}%
\bibitem [{\citenamefont {Cupellini}\ \emph {et~al.}(2020)\citenamefont
  {Cupellini}, \citenamefont {Viani},\ and\ \citenamefont
  {Mennucci}}]{specden2020}%
  \BibitemOpen
  \bibfield  {author} {\bibinfo {author} {\bibfnamefont {Lorenzo}\ \bibnamefont
  {Cupellini}}, \bibinfo {author} {\bibfnamefont {Lucas}\ \bibnamefont
  {Viani}}, \ and\ \bibinfo {author} {\bibfnamefont {Benedetta}\ \bibnamefont
  {Mennucci}},\ }\href {\doibase 10.5281/zenodo.3948106} {\enquote {\bibinfo
  {title} {{SPECDEN - Python tool to compute spectral densities from
  autocorrelation functions, and the corresponding vibronic spectrum.}}}\ }
  (\bibinfo {year} {2020})\BibitemShut {NoStop}%
\bibitem [{\citenamefont {Nigam}\ \emph {et~al.}(2020)\citenamefont {Nigam},
  \citenamefont {Pozdnyakov},\ and\ \citenamefont {Ceriotti}}]{niga+20jcp}%
  \BibitemOpen
  \bibfield  {author} {\bibinfo {author} {\bibfnamefont {Jigyasa}\ \bibnamefont
  {Nigam}}, \bibinfo {author} {\bibfnamefont {Sergey}\ \bibnamefont
  {Pozdnyakov}}, \ and\ \bibinfo {author} {\bibfnamefont {Michele}\
  \bibnamefont {Ceriotti}},\ }\bibfield  {title} {\enquote {\bibinfo {title}
  {Recursive evaluation and iterative contraction of {{{\emph{N}}}} -body
  equivariant features},}\ }\href {\doibase 10.1063/5.0021116} {\bibfield
  {journal} {\bibinfo  {journal} {The Journal of Chemical Physics}\ }\textbf
  {\bibinfo {volume} {153}},\ \bibinfo {pages} {121101} (\bibinfo {year}
  {2020})}\BibitemShut {NoStop}%
\bibitem [{\citenamefont {Nigam}\ \emph
  {et~al.}(2022{\natexlab{b}})\citenamefont {Nigam}, \citenamefont
  {Pozdnyakov}, \citenamefont {Fraux},\ and\ \citenamefont
  {Ceriotti}}]{niga+22jcp2}%
  \BibitemOpen
  \bibfield  {author} {\bibinfo {author} {\bibfnamefont {Jigyasa}\ \bibnamefont
  {Nigam}}, \bibinfo {author} {\bibfnamefont {Sergey}\ \bibnamefont
  {Pozdnyakov}}, \bibinfo {author} {\bibfnamefont {Guillaume}\ \bibnamefont
  {Fraux}}, \ and\ \bibinfo {author} {\bibfnamefont {Michele}\ \bibnamefont
  {Ceriotti}},\ }\bibfield  {title} {\enquote {\bibinfo {title} {Unified theory
  of atom-centered representations and message-passing machine-learning
  schemes},}\ }\href {\doibase 10.1063/5.0087042} {\bibfield  {journal}
  {\bibinfo  {journal} {The Journal of Chemical Physics}\ }\textbf {\bibinfo
  {volume} {156}},\ \bibinfo {pages} {204115} (\bibinfo {year}
  {2022}{\natexlab{b}})}\BibitemShut {NoStop}%
\bibitem [{\citenamefont {Musil}\ \emph
  {et~al.}(2021{\natexlab{b}})\citenamefont {Musil}, \citenamefont {Veit},
  \citenamefont {Goscinski}, \citenamefont {Fraux}, \citenamefont {Willatt},
  \citenamefont {Stricker},\ and\ \citenamefont {Ceriotti}}]{musi+21jcp}%
  \BibitemOpen
  \bibfield  {author} {\bibinfo {author} {\bibfnamefont {F{\'e}lix}\
  \bibnamefont {Musil}}, \bibinfo {author} {\bibfnamefont {Max}\ \bibnamefont
  {Veit}}, \bibinfo {author} {\bibfnamefont {Alexander}\ \bibnamefont
  {Goscinski}}, \bibinfo {author} {\bibfnamefont {Guillaume}\ \bibnamefont
  {Fraux}}, \bibinfo {author} {\bibfnamefont {Michael~J}\ \bibnamefont
  {Willatt}}, \bibinfo {author} {\bibfnamefont {Markus}\ \bibnamefont
  {Stricker}}, \ and\ \bibinfo {author} {\bibfnamefont {Michele}\ \bibnamefont
  {Ceriotti}},\ }\bibfield  {title} {\enquote {\bibinfo {title} {Efficient
  implementation of atom-density representations},}\ }\href {\doibase
  10.1063/5.0044689} {\bibfield  {journal} {\bibinfo  {journal} {The Journal of
  Chemical Physics}\ }\textbf {\bibinfo {volume} {154}},\ \bibinfo {pages}
  {114109} (\bibinfo {year} {2021}{\natexlab{b}})}\BibitemShut {NoStop}%
\bibitem [{\citenamefont {Yue}\ \emph {et~al.}(2021)\citenamefont {Yue},
  \citenamefont {Muniz}, \citenamefont {Andrade}, \citenamefont {Zhang},
  \citenamefont {Car},\ and\ \citenamefont {Panagiotopoulos}}]{Yue2021}%
  \BibitemOpen
  \bibfield  {author} {\bibinfo {author} {\bibfnamefont {Shuwen}\ \bibnamefont
  {Yue}}, \bibinfo {author} {\bibfnamefont {Maria~Carolina}\ \bibnamefont
  {Muniz}}, \bibinfo {author} {\bibfnamefont {Marcos F.~Calegari}\ \bibnamefont
  {Andrade}}, \bibinfo {author} {\bibfnamefont {Linfeng}\ \bibnamefont
  {Zhang}}, \bibinfo {author} {\bibfnamefont {Roberto}\ \bibnamefont {Car}}, \
  and\ \bibinfo {author} {\bibfnamefont {Athanassios~Z.}\ \bibnamefont
  {Panagiotopoulos}},\ }\bibfield  {title} {\enquote {\bibinfo {title} {When do
  short-range atomistic machine-learning models fall short?}}\ }\href {\doibase
  10.1063/5.0031215} {\bibfield  {journal} {\bibinfo  {journal} {The Journal of
  Chemical Physics}\ }\textbf {\bibinfo {volume} {154}} (\bibinfo {year}
  {2021}),\ 10.1063/5.0031215}\BibitemShut {NoStop}%
\bibitem [{\citenamefont {Anstine}\ and\ \citenamefont
  {Isayev}(2023)}]{Anstine2023}%
  \BibitemOpen
  \bibfield  {author} {\bibinfo {author} {\bibfnamefont {Dylan~M.}\
  \bibnamefont {Anstine}}\ and\ \bibinfo {author} {\bibfnamefont {Olexandr}\
  \bibnamefont {Isayev}},\ }\bibfield  {title} {\enquote {\bibinfo {title}
  {Machine learning interatomic potentials and long-range physics},}\ }\href
  {\doibase 10.1021/acs.jpca.2c06778} {\bibfield  {journal} {\bibinfo
  {journal} {The Journal of Physical Chemistry A}\ }\textbf {\bibinfo {volume}
  {127}},\ \bibinfo {pages} {2417--2431} (\bibinfo {year} {2023})}\BibitemShut
  {NoStop}%
\bibitem [{\citenamefont {Kabylda}\ \emph {et~al.}(2023)\citenamefont
  {Kabylda}, \citenamefont {Vassilev-Galindo}, \citenamefont {Chmiela},
  \citenamefont {Poltavsky},\ and\ \citenamefont {Tkatchenko}}]{Kabylda2023}%
  \BibitemOpen
  \bibfield  {author} {\bibinfo {author} {\bibfnamefont {Adil}\ \bibnamefont
  {Kabylda}}, \bibinfo {author} {\bibfnamefont {Valentin}\ \bibnamefont
  {Vassilev-Galindo}}, \bibinfo {author} {\bibfnamefont {Stefan}\ \bibnamefont
  {Chmiela}}, \bibinfo {author} {\bibfnamefont {Igor}\ \bibnamefont
  {Poltavsky}}, \ and\ \bibinfo {author} {\bibfnamefont {Alexandre}\
  \bibnamefont {Tkatchenko}},\ }\bibfield  {title} {\enquote {\bibinfo {title}
  {Efficient interatomic descriptors for accurate machine learning force fields
  of extended molecules},}\ }\href {\doibase 10.1038/s41467-023-39214-w}
  {\bibfield  {journal} {\bibinfo  {journal} {Nature Communications}\ }\textbf
  {\bibinfo {volume} {14}} (\bibinfo {year} {2023}),\
  10.1038/s41467-023-39214-w}\BibitemShut {NoStop}%
\bibitem [{\citenamefont {Grisafi}\ and\ \citenamefont
  {Ceriotti}(2019)}]{gris-ceri19jcp}%
  \BibitemOpen
  \bibfield  {author} {\bibinfo {author} {\bibfnamefont {Andrea}\ \bibnamefont
  {Grisafi}}\ and\ \bibinfo {author} {\bibfnamefont {Michele}\ \bibnamefont
  {Ceriotti}},\ }\bibfield  {title} {\enquote {\bibinfo {title} {Incorporating
  long-range physics in atomic-scale machine learning},}\ }\href {\doibase
  10.1063/1.5128375} {\bibfield  {journal} {\bibinfo  {journal} {The Journal of
  Chemical Physics}\ }\textbf {\bibinfo {volume} {151}},\ \bibinfo {pages}
  {204105} (\bibinfo {year} {2019})}\BibitemShut {NoStop}%
\end{thebibliography}
\end{document}